\documentclass[sigconf, screen]{acmart}
\AtBeginDocument{%
  \providecommand\BibTeX{{%
    \normalfont B\kern-0.5em{\scshape i\kern-0.25em b}\kern-0.8em\TeX}}}

\copyrightyear{2024}
\acmYear{2024}
\setcopyright{cc}
\setcctype[4.0]{by}
\acmConference[CHI '24]{Proceedings of the CHI Conference on Human Factors in Computing Systems}{May 11--16, 2024}{Honolulu, HI, USA}
\acmBooktitle{Proceedings of the CHI Conference on Human Factors in Computing Systems (CHI '24), May 11--16, 2024, Honolulu, HI, USA}
\acmDOI{10.1145/3613904.3642335}
\acmISBN{979-8-4007-0330-0/24/05}

\makeatletter

\newcommand\xlabel[2][]{\phantomsection\def\@currentlabelname{#1}\label{#2}}

\makeatother

\usepackage{subcaption}
\usepackage{enumitem}
\usepackage{rotating}
\usepackage{microtype}
\usepackage{bm}
\usepackage{soul}
\usepackage{wrapfig}
\usepackage{multirow}
\usepackage{longtable}
\usepackage{setspace}

\usepackage[varqu]{zi4}
\usepackage{units}
\usepackage{booktabs}
\usepackage{colortbl}

\usepackage{listings}
\usepackage{amsmath}
\usepackage{mathtools}

\usepackage{acmart-taps}

\definecolor{redOV}{RGB}{255, 235, 238}
\definecolor{redI}{RGB}{255, 205, 210}
\definecolor{redII}{RGB}{239, 154, 154}
\definecolor{redIII}{RGB}{229, 115, 115}
\definecolor{redIV}{RGB}{239, 83, 80}
\definecolor{redV}{RGB}{244, 67, 54}
\definecolor{redVI}{RGB}{229, 57, 53}
\definecolor{redVII}{RGB}{211, 47, 47}
\definecolor{redVIII}{RGB}{198, 40, 40}
\definecolor{redIX}{RGB}{183, 28, 28}
\definecolor{redAI}{RGB}{255, 138, 128}
\definecolor{redAII}{RGB}{255, 82, 82}
\definecolor{redAIV}{RGB}{255, 23, 68}
\definecolor{redAVII}{RGB}{213, 0, 0}

\definecolor{pinkOV}{RGB}{252, 228, 236}
\definecolor{pinkI}{RGB}{248, 187, 208}
\definecolor{pinkII}{RGB}{244, 143, 177}
\definecolor{pinkIII}{RGB}{240, 98, 146}
\definecolor{pinkIV}{RGB}{236, 64, 122}
\definecolor{pinkV}{RGB}{233, 30, 99}
\definecolor{pinkVI}{RGB}{216, 27, 96}
\definecolor{pinkVII}{RGB}{194, 24, 91}
\definecolor{pinkVIII}{RGB}{173, 20, 87}
\definecolor{pinkIX}{RGB}{136, 14, 79}
\definecolor{pinkAI}{RGB}{255, 128, 171}
\definecolor{pinkAII}{RGB}{255, 64, 129}
\definecolor{pinkAIV}{RGB}{245, 0, 87}
\definecolor{pinkAVII}{RGB}{197, 17, 98}

\definecolor{purpleOV}{RGB}{243, 229, 245}
\definecolor{purpleI}{RGB}{225, 190, 231}
\definecolor{purpleII}{RGB}{206, 147, 216}
\definecolor{purpleIII}{RGB}{186, 104, 200}
\definecolor{purpleIV}{RGB}{171, 71, 188}
\definecolor{purpleV}{RGB}{156, 39, 176}
\definecolor{purpleVI}{RGB}{142, 36, 170}
\definecolor{purpleVII}{RGB}{123, 31, 162}
\definecolor{purpleVIII}{RGB}{106, 27, 154}
\definecolor{purpleIX}{RGB}{74, 20, 140}
\definecolor{purpleAI}{RGB}{234, 128, 252}
\definecolor{purpleAII}{RGB}{224, 64, 251}
\definecolor{purpleAIV}{RGB}{213, 0, 249}
\definecolor{purpleAVII}{RGB}{170, 0, 255}

\definecolor{deeppurpleOV}{RGB}{237, 231, 246}
\definecolor{deeppurpleI}{RGB}{209, 196, 233}
\definecolor{deeppurpleII}{RGB}{179, 157, 219}
\definecolor{deeppurpleIII}{RGB}{149, 117, 205}
\definecolor{deeppurpleIV}{RGB}{126, 87, 194}
\definecolor{deeppurpleV}{RGB}{103, 58, 183}
\definecolor{deeppurpleVI}{RGB}{94, 53, 177}
\definecolor{deeppurpleVII}{RGB}{81, 45, 168}
\definecolor{deeppurpleVIII}{RGB}{69, 39, 160}
\definecolor{deeppurpleIX}{RGB}{49, 27, 146}
\definecolor{deeppurpleAI}{RGB}{179, 136, 255}
\definecolor{deeppurpleAII}{RGB}{124, 77, 255}
\definecolor{deeppurpleAIV}{RGB}{101, 31, 255}
\definecolor{deeppurpleAVII}{RGB}{98, 0, 234}

\definecolor{indigoOV}{RGB}{232, 234, 246}
\definecolor{indigoI}{RGB}{197, 202, 233}
\definecolor{indigoII}{RGB}{159, 168, 218}
\definecolor{indigoIII}{RGB}{121, 134, 203}
\definecolor{indigoIV}{RGB}{92, 107, 192}
\definecolor{indigoV}{RGB}{63, 81, 181}
\definecolor{indigoVI}{RGB}{57, 73, 171}
\definecolor{indigoVII}{RGB}{48, 63, 159}
\definecolor{indigoVIII}{RGB}{40, 53, 147}
\definecolor{indigoIX}{RGB}{26, 35, 126}
\definecolor{indigoAI}{RGB}{140, 158, 255}
\definecolor{indigoAII}{RGB}{83, 109, 254}
\definecolor{indigoAIV}{RGB}{61, 90, 254}
\definecolor{indigoAVII}{RGB}{48, 79, 254}

\definecolor{blueOV}{RGB}{227, 242, 253}
\definecolor{blueI}{RGB}{187, 222, 251}
\definecolor{blueII}{RGB}{144, 202, 249}
\definecolor{blueIII}{RGB}{100, 181, 246}
\definecolor{blueIV}{RGB}{66, 165, 245}
\definecolor{blueV}{RGB}{33, 150, 243}
\definecolor{blueVI}{RGB}{30, 136, 229}
\definecolor{blueVII}{RGB}{25, 118, 210}
\definecolor{blueVIII}{RGB}{21, 101, 192}
\definecolor{blueIX}{RGB}{13, 71, 161}
\definecolor{blueAI}{RGB}{130, 177, 255}
\definecolor{blueAII}{RGB}{68, 138, 255}
\definecolor{blueAIV}{RGB}{41, 121, 255}
\definecolor{blueAVII}{RGB}{41, 98, 255}

\definecolor{lightblueOV}{RGB}{225, 245, 254}
\definecolor{lightblueI}{RGB}{179, 229, 252}
\definecolor{lightblueII}{RGB}{129, 212, 250}
\definecolor{lightblueIII}{RGB}{79, 195, 247}
\definecolor{lightblueIV}{RGB}{41, 182, 246}
\definecolor{lightblueV}{RGB}{3, 169, 244}
\definecolor{lightblueVI}{RGB}{3, 155, 229}
\definecolor{lightblueVII}{RGB}{2, 136, 209}
\definecolor{lightblueVIII}{RGB}{2, 119, 189}
\definecolor{lightblueIX}{RGB}{1, 87, 155}
\definecolor{lightblueAI}{RGB}{128, 216, 255}
\definecolor{lightblueAII}{RGB}{64, 196, 255}
\definecolor{lightblueAIV}{RGB}{0, 176, 255}
\definecolor{lightblueAVII}{RGB}{0, 145, 234}

\definecolor{cyanOV}{RGB}{224, 247, 250}
\definecolor{cyanI}{RGB}{178, 235, 242}
\definecolor{cyanII}{RGB}{128, 222, 234}
\definecolor{cyanIII}{RGB}{77, 208, 225}
\definecolor{cyanIV}{RGB}{38, 198, 218}
\definecolor{cyanV}{RGB}{0, 188, 212}
\definecolor{cyanVI}{RGB}{0, 172, 193}
\definecolor{cyanVII}{RGB}{0, 151, 167}
\definecolor{cyanVIII}{RGB}{0, 131, 143}
\definecolor{cyanIX}{RGB}{0, 96, 100}
\definecolor{cyanAI}{RGB}{132, 255, 255}
\definecolor{cyanAII}{RGB}{24, 255, 255}
\definecolor{cyanAIV}{RGB}{0, 229, 255}
\definecolor{cyanAVII}{RGB}{0, 184, 212}

\definecolor{tealOV}{RGB}{224, 242, 241}
\definecolor{tealI}{RGB}{178, 223, 219}
\definecolor{tealII}{RGB}{128, 203, 196}
\definecolor{tealIII}{RGB}{77, 182, 172}
\definecolor{tealIV}{RGB}{38, 166, 154}
\definecolor{tealV}{RGB}{0, 150, 136}
\definecolor{tealVI}{RGB}{0, 137, 123}
\definecolor{tealVII}{RGB}{0, 121, 107}
\definecolor{tealVIII}{RGB}{0, 105, 92}
\definecolor{tealIX}{RGB}{0, 77, 64}
\definecolor{tealAI}{RGB}{167, 255, 235}
\definecolor{tealAII}{RGB}{100, 255, 218}
\definecolor{tealAIV}{RGB}{29, 233, 182}
\definecolor{tealAVII}{RGB}{0, 191, 165}

\definecolor{greenOV}{RGB}{232, 245, 233}
\definecolor{greenI}{RGB}{200, 230, 201}
\definecolor{greenII}{RGB}{165, 214, 167}
\definecolor{greenIII}{RGB}{129, 199, 132}
\definecolor{greenIV}{RGB}{102, 187, 106}
\definecolor{greenV}{RGB}{76, 175, 80}
\definecolor{greenVI}{RGB}{67, 160, 71}
\definecolor{greenVII}{RGB}{56, 142, 60}
\definecolor{greenVIII}{RGB}{46, 125, 50}
\definecolor{greenIX}{RGB}{27, 94, 32}
\definecolor{greenAI}{RGB}{185, 246, 202}
\definecolor{greenAII}{RGB}{105, 240, 174}
\definecolor{greenAIV}{RGB}{0, 230, 118}
\definecolor{greenAVII}{RGB}{0, 200, 83}

\definecolor{lightgreenOV}{RGB}{241, 248, 233}
\definecolor{lightgreenI}{RGB}{220, 237, 200}
\definecolor{lightgreenII}{RGB}{197, 225, 165}
\definecolor{lightgreenIII}{RGB}{174, 213, 129}
\definecolor{lightgreenIV}{RGB}{156, 204, 101}
\definecolor{lightgreenV}{RGB}{139, 195, 74}
\definecolor{lightgreenVI}{RGB}{124, 179, 66}
\definecolor{lightgreenVII}{RGB}{104, 159, 56}
\definecolor{lightgreenVIII}{RGB}{85, 139, 47}
\definecolor{lightgreenIX}{RGB}{51, 105, 30}
\definecolor{lightgreenAI}{RGB}{204, 255, 144}
\definecolor{lightgreenAII}{RGB}{178, 255, 89}
\definecolor{lightgreenAIV}{RGB}{118, 255, 3}
\definecolor{lightgreenAVII}{RGB}{100, 221, 23}

\definecolor{limeOV}{RGB}{249, 251, 231}
\definecolor{limeI}{RGB}{240, 244, 195}
\definecolor{limeII}{RGB}{230, 238, 156}
\definecolor{limeIII}{RGB}{220, 231, 117}
\definecolor{limeIV}{RGB}{212, 225, 87}
\definecolor{limeV}{RGB}{205, 220, 57}
\definecolor{limeVI}{RGB}{192, 202, 51}
\definecolor{limeVII}{RGB}{175, 180, 43}
\definecolor{limeVIII}{RGB}{158, 157, 36}
\definecolor{limeIX}{RGB}{130, 119, 23}
\definecolor{limeAI}{RGB}{244, 255, 129}
\definecolor{limeAII}{RGB}{238, 255, 65}
\definecolor{limeAIV}{RGB}{198, 255, 0}
\definecolor{limeAVII}{RGB}{174, 234, 0}

\definecolor{yellowOV}{RGB}{255, 253, 231}
\definecolor{yellowI}{RGB}{255, 249, 196}
\definecolor{yellowII}{RGB}{255, 245, 157}
\definecolor{yellowIII}{RGB}{255, 241, 118}
\definecolor{yellowIV}{RGB}{255, 238, 88}
\definecolor{yellowV}{RGB}{255, 235, 59}
\definecolor{yellowVI}{RGB}{253, 216, 53}
\definecolor{yellowVII}{RGB}{251, 192, 45}
\definecolor{yellowVIII}{RGB}{249, 168, 37}
\definecolor{yellowIX}{RGB}{245, 127, 23}
\definecolor{yellowAI}{RGB}{255, 255, 141}
\definecolor{yellowAII}{RGB}{255, 255, 0}
\definecolor{yellowAIV}{RGB}{255, 234, 0}
\definecolor{yellowAVII}{RGB}{255, 214, 0}

\definecolor{amberOV}{RGB}{255, 248, 225}
\definecolor{amberI}{RGB}{255, 236, 179}
\definecolor{amberII}{RGB}{255, 224, 130}
\definecolor{amberIII}{RGB}{255, 213, 79}
\definecolor{amberIV}{RGB}{255, 202, 40}
\definecolor{amberV}{RGB}{255, 193, 7}
\definecolor{amberVI}{RGB}{255, 179, 0}
\definecolor{amberVII}{RGB}{255, 160, 0}
\definecolor{amberVIII}{RGB}{255, 143, 0}
\definecolor{amberIX}{RGB}{255, 111, 0}
\definecolor{amberAI}{RGB}{255, 229, 127}
\definecolor{amberAII}{RGB}{255, 215, 64}
\definecolor{amberAIV}{RGB}{255, 196, 0}
\definecolor{amberAVII}{RGB}{255, 171, 0}

\definecolor{orangeOV}{RGB}{255, 243, 224}
\definecolor{orangeI}{RGB}{255, 224, 178}
\definecolor{orangeII}{RGB}{255, 204, 128}
\definecolor{orangeIII}{RGB}{255, 183, 77}
\definecolor{orangeIV}{RGB}{255, 167, 38}
\definecolor{orangeV}{RGB}{255, 152, 0}
\definecolor{orangeVI}{RGB}{251, 140, 0}
\definecolor{orangeVII}{RGB}{245, 124, 0}
\definecolor{orangeVIII}{RGB}{239, 108, 0}
\definecolor{orangeIX}{RGB}{230, 81, 0}
\definecolor{orangeAI}{RGB}{255, 209, 128}
\definecolor{orangeAII}{RGB}{255, 171, 64}
\definecolor{orangeAIV}{RGB}{255, 145, 0}
\definecolor{orangeAVII}{RGB}{255, 109, 0}

\definecolor{deeporangeOV}{RGB}{251, 233, 231}
\definecolor{deeporangeI}{RGB}{255, 204, 188}
\definecolor{deeporangeII}{RGB}{255, 171, 145}
\definecolor{deeporangeIII}{RGB}{255, 138, 101}
\definecolor{deeporangeIV}{RGB}{255, 112, 67}
\definecolor{deeporangeV}{RGB}{255, 87, 34}
\definecolor{deeporangeVI}{RGB}{244, 81, 30}
\definecolor{deeporangeVII}{RGB}{230, 74, 25}
\definecolor{deeporangeVIII}{RGB}{216, 67, 21}
\definecolor{deeporangeIX}{RGB}{191, 54, 12}
\definecolor{deeporangeAI}{RGB}{255, 158, 128}
\definecolor{deeporangeAII}{RGB}{255, 110, 64}
\definecolor{deeporangeAIV}{RGB}{255, 61, 0}
\definecolor{deeporangeAVII}{RGB}{221, 44, 0}

\definecolor{brownOV}{RGB}{239, 235, 233}
\definecolor{brownI}{RGB}{215, 204, 200}
\definecolor{brownII}{RGB}{188, 170, 164}
\definecolor{brownIII}{RGB}{161, 136, 127}
\definecolor{brownIV}{RGB}{141, 110, 99}
\definecolor{brownV}{RGB}{121, 85, 72}
\definecolor{brownVI}{RGB}{109, 76, 65}
\definecolor{brownVII}{RGB}{93, 64, 55}
\definecolor{brownVIII}{RGB}{78, 52, 46}
\definecolor{brownIX}{RGB}{62, 39, 35}

\definecolor{grayOV}{RGB}{250, 250, 250}
\definecolor{grayI}{RGB}{245, 245, 245}
\definecolor{grayII}{RGB}{238, 238, 238}
\definecolor{grayIII}{RGB}{224, 224, 224}
\definecolor{grayIV}{RGB}{189, 189, 189}
\definecolor{grayV}{RGB}{158, 158, 158}
\definecolor{grayVI}{RGB}{117, 117, 117}
\definecolor{grayVII}{RGB}{97, 97, 97}
\definecolor{grayVIII}{RGB}{66, 66, 66}
\definecolor{grayIX}{RGB}{33, 33, 33}

\definecolor{bluegrayOV}{RGB}{236, 239, 241}
\definecolor{bluegrayI}{RGB}{207, 216, 220}
\definecolor{bluegrayII}{RGB}{176, 190, 197}
\definecolor{bluegrayIII}{RGB}{144, 164, 174}
\definecolor{bluegrayIV}{RGB}{120, 144, 156}
\definecolor{bluegrayV}{RGB}{96, 125, 139}
\definecolor{bluegrayVI}{RGB}{84, 110, 122}
\definecolor{bluegrayVII}{RGB}{69, 90, 100}
\definecolor{bluegrayVIII}{RGB}{55, 71, 79}
\definecolor{bluegrayIX}{RGB}{38, 50, 56}
\definecolor{bluegrayX}{RGB}{17, 23, 26}

\definecolor{myACMBlue}{cmyk}{1,0.1,0,0.1}
\definecolor{myACMYellow}{cmyk}{0,0.16,1,0}
\definecolor{myACMOrange}{cmyk}{0,0.42,1,0.01}
\definecolor{myACMRed}{cmyk}{0,0.90,0.86,0}
\definecolor{myACMLightBlue}{cmyk}{0.49,0.01,0,0}
\definecolor{myACMGreen}{cmyk}{0.20,0,1,0.19}
\definecolor{myACMPurple}{cmyk}{0.55,1,0,0.15}
\definecolor{myACMDarkBlue}{cmyk}{1,0.58,0,0.21}

\newcommand{\nnlink}[1]{{\href{#1}{\color{blueVI}\textbf{\texttt{#1}}}}}
\newcommand{\myhref}[2]{{\href{#1}{\color{blueVI}\textbf{#2}}}}

\newcommand{\mypar}[1]{\vspace{3pt}\textbf{{#1}}}
\newcommand{\myref}[1]{\mbox{\nameref{#1}}}

\newcommand{\figpart}[1]{\textcolor{myACMPurple}{#1}}

\newcommand{\hone}{\myref{sec:harm-envision-one}}
\newcommand{\htwo}{\myref{sec:harm-envision-two}}
\newcommand{\hthree}{\myref{sec:harm-envision-three}}
\newcommand{\hfour}{\myref{sec:harm-envision-four}}

\aptLtoX[graphic=no, type=html]{}{
  \setulcolor{grayIII}
  \setul{0.45ex}{0.7pt}
}

\newcommand{\tool}{\textsc{Farsight}}
\newcommand{\lite}{\textsc{Farsight Lite}}
\newcommand{\guide}{\textsc{Envisioning Guide}}

\newcommand{\symbolview}{\textit{Alert Symbol}}
\newcommand{\sidebarview}{\textit{Awareness Sidebar}}
\newcommand{\windowview}{\textit{Harm Envisioner}}

\newcommand{\symbolviewnormal}{Alert Symbol}
\newcommand{\sidebarviewnormal}{Awareness Sidebar}
\newcommand{\windowviewnormal}{Harm Envisioner}

\newcommand{\incidentpanel}{\textit{Incident Panel}}
\newcommand{\usecasepanel}{\textit{Use Case Panel}}

\newcommand{\riskscore}{alert level}
\newcommand{\riskscoretitle}{Alert Level}

\newcommand{\incidentpanelnormal}{Incident Panel}
\newcommand{\usecasepanelnormal}{Use Case Panel}

\definecolor{farsightcolor}{HTML}{2D8C72}
\definecolor{litecolor}{HTML}{5B6D95}
\definecolor{groupcolor}{HTML}{DD734A}

\definecolor{raicolor}{HTML}{64B5F6}
\definecolor{promptingcolor}{HTML}{42A299}

\newcommand{\toolc}{\textcolor{farsightcolor}{{\tool{}}}}
\newcommand{\litec}{\textcolor{litecolor}{{\lite{}}}}
\newcommand{\guidec}{\textcolor{groupcolor}{{\guide{}}}}

\newcommand{\toolcolor}[1]{\textcolor{farsightcolor}{#1}}
\newcommand{\litecolor}[1]{\textcolor{litecolor}{#1}}
\newcommand{\guidecolor}[1]{\textcolor{groupcolor}{#1}}

\newcommand{\conditionfg}{$C_{FG}$}
\newcommand{\conditionf}{$C_{F}$}
\newcommand{\conditionlg}{$C_{LG}$}
\newcommand{\conditionl}{$C_{L}$}
\newcommand{\conditiongf}{$C_{GF}$}
\newcommand{\conditiongl}{$C_{GL}$}

\definecolor{soulorange}{RGB}{255, 212, 153}
\definecolor{soulgray}{RGB}{220, 220, 220}
\definecolor{soulgraylight}{RGB}{235, 235, 235}
\definecolor{soulred}{RGB}{252, 217, 218}
\definecolor{soulbluelight}{RGB}{208, 233, 253}
\definecolor{souldorangelight}{RGB}{254, 234, 212}
\colorlet{soulblue}{blueV!30}

\newcommand*{\vcenteredhbox}[1]{\begingroup\setbox0=\hbox{#1}\parbox{\wd0}{\box0}\endgroup}
\definecolor{tagbordercolor}{rgb}{0.8, 0.8, 0.8}
\definecolor{tagbgcolor}{rgb}{0.9, 0.9, 0.9}
\definecolor{lightgray}{RGB}{247, 247, 247}
\definecolor{midgray}{RGB}{179, 179, 179}

\newcommand*\myquote[1]{``\textit{#1}''}

\definecolor{tagbgcolor}{rgb}{1, 1, 1}
\definecolor{boxyellow}{RGB}{206, 171, 1}
\definecolor{boxgreen}{RGB}{14, 152, 136}
\definecolor{boxblue}{RGB}{77, 167, 223}

\begin{document}

\title{\tool{}: Fostering Responsible AI Awareness During AI Application Prototyping}

\author{Zijie J. Wang}
\authornote{The work was done when the authors were at Google Research.}
\orcid{0000-0003-4360-1423}
\affiliation{%
  \institution{Georgia Tech}
  \city{Atlanta}
  \state{Georgia}
  \country{USA}
}

\author{Chinmay Kulkarni}
\authornotemark[1]
\orcid{0000-0003-4922-6601}
\affiliation{%
  \institution{Emory University}
  \city{Atlanta}
  \state{Georgia}
  \country{USA}
}

\author{Lauren Wilcox}
\authornotemark[1]
\orcid{0000-0001-6598-1733}
\affiliation{%
  \institution{eBay}
  \city{San Jose}
  \state{California}
  \country{USA}
}

\author{Michael Terry}
\orcid{0000-0003-1941-939X}
\affiliation{%
  \institution{Google Research}
  \city{Cambridge}
  \state{Massachusetts}
  \country{USA}
}

\author{Michael Madaio}
\orcid{0000-0001-5772-0488}
\affiliation{%
  \institution{Google Research}
  \city{New York}
  \state{New York}
  \country{USA}
}
\renewcommand{\shortauthors}{Zijie J. Wang, et al.}

\begin{abstract}
  Prompt-based interfaces for Large Language Models (LLMs) have made prototyping and building AI-powered applications easier than ever before.
  However, identifying potential harms that may arise from AI applications remains a challenge, particularly during prompt-based prototyping.
  To address this, we present \tool{}, a novel \textit{in situ} interactive tool that helps people identify potential harms from the AI applications they are prototyping.
  Based on a user's prompt, \tool{} highlights news articles about relevant AI incidents and allows users to explore and edit LLM-generated use cases, stakeholders, and harms.
  We report design insights from a co-design study with 10 AI prototypers and findings from a user study with 42 AI prototypers.
  After using \tool{}, AI prototypers in our user study are better able to independently identify potential harms associated with a prompt and find our tool more useful and usable than existing resources.
  Their qualitative feedback also highlights that \tool{} encourages them to focus on end-users and think beyond immediate harms.
  We discuss these findings and reflect on their implications for designing AI prototyping experiences that meaningfully engage with AI harms.
  \tool{} is publicly accessible at: \nnlink{https://pair-code.github.io/farsight}.
\end{abstract} 

\ccsdesc[500]{Human-centered computing~Interactive systems and tools}
\ccsdesc[500]{Computing methodologies~Machine learning}
\keywords{Responsible AI, Human-AI Collaboration, Large Language Models}

\begin{teaserfigure}
  \centering
  \includegraphics[width=500pt]{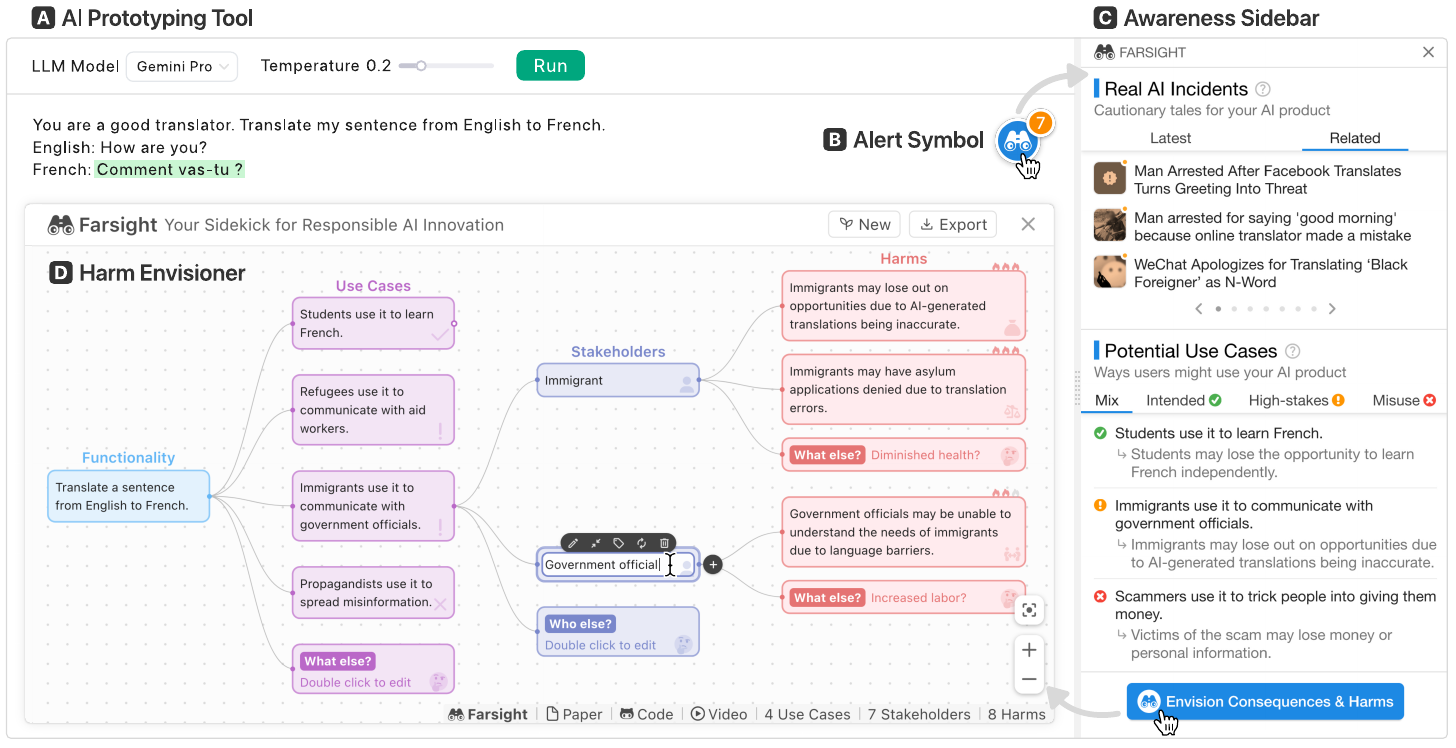}
  \vspace{-11pt}
  \caption{With \textit{in situ} interfaces and novel techniques, \tool{} empowers AI prototypers to envision potential harms that may arise from their large language models (LLMs)-powered AI applications during early prototyping.
    \textbf{(A)} In this example, an AI prototyper is creating a prompt for an English-to-French translator in a web-based AI prototyping tool.
    \textbf{(B)} The \symbolview{} from \tool{} warns the user of potential risks associated with their AI application.
    \textbf{(C)} Clicking the symbol expands the \sidebarview{}, highlighting news articles relevant to the user's prompt (top), and LLM-generated potential use cases and harms (bottom).
    \textbf{(D)} Clicking the blue button opens the \windowview{} that allows the user to interactively envision, assess, and reflect on the potential use cases, stakeholders, and harms of their AI application with the assistance of an LLM.}
  \Description{Teaser figure with four components. Component A shows an AI prototyper is writing a prompt for an English-to-French translator within a web-based AI prototyping tool. Component B shows the alert symbol. Component C shows the awareness sidebar. Component D shows the harm envisioner.}
  \label{fig:teaser}
\end{teaserfigure}

\maketitle

\section{Introduction}

\setlength{\belowcaptionskip}{0pt}
\setlength{\abovecaptionskip}{5pt}
\begin{figure}[tb]
      \includegraphics[width=1\linewidth]{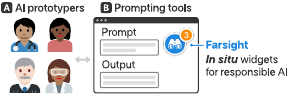}
      \caption[]{
            \textbf{(A)} Many AI prototypers from diverse backgrounds and roles use \textbf{(B)} prompting tools to prototype AI applications.
            \tool{} provides a range of \textit{in situ} widgets for these tools, helping AI prototypers envision the potential harms of their AI applications during an early prototyping stage.
      }
      \Description{An illustration divided into two sections labeled A and B. Section A is titled "AI prototypers" and shows four cartoon avatars representing diverse professionals, including what appears to be a doctor, two people with casual attire, and one person wearing lab goggles. Section B is titled "Prompting tools" and displays a graphical user interface with a text box labeled "Prompt" above another box labeled "Output", with a Farsight alert symbol attached to the "Prompt" box, with a text: "Farsight: in situ widgets for responsible AI."}
      \label{fig:diagram}
\end{figure}
\setlength{\belowcaptionskip}{0pt}
\setlength{\abovecaptionskip}{12pt}

\setlength{\belowcaptionskip}{6pt}
\setlength{\abovecaptionskip}{6pt}
\begin{figure*}[tb]
      \includegraphics[width=1\linewidth]{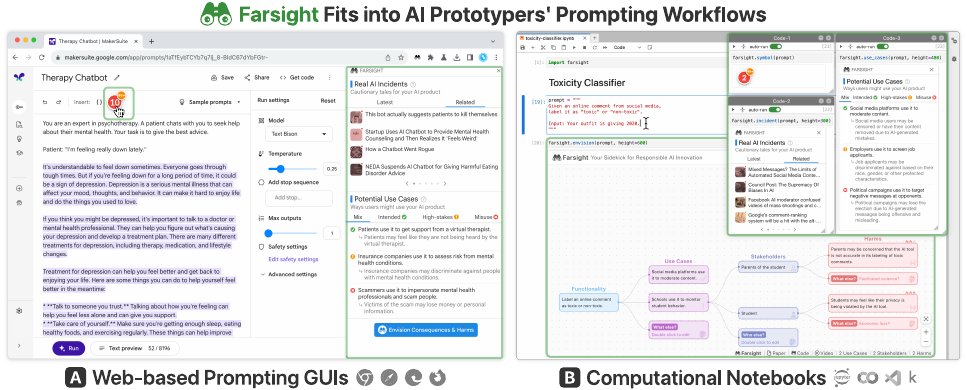}
      \caption[]{
            \tool{} fits into AI prototypers' diverse prompting workflows including prompting GUIs and computational notebooks.
            For example, (A) when an AI prototyper writes prompts for a therapy chatbot in Google AI Studio~\cite{googleGoogleAiStudio2023}, \tool{}'s Chrome extension alerts the user about related accidents and potential harms.
            (B) When an AI prototyper writes prompts for a toxicity classifier in Jupyter Notebook~\cite{kluyverJupyterNotebooksaPublishing2016,wangStickyLandBreakingLinear2022a}, \tool{}'s Python library shows potential negative consequences of this classifier.
      }

      \Description{A figure with two components, showing the use case of Farsight in (A) web-based prompting GUIs and (B) computational notebooks.
            Component A shows a screenshot of a user prompting a therapy chatbot with Google AI Studio in a Chrome browser, and Farsight shows related accident reports and potential harms that may arise for the therapy chatbot.
            Component B shows a screenshot of a user prompting a toxicity classifier in Jupyter notebooks, and Farsight shows related accident reports and potential harms associated with a toxicity classifier.}
      \label{fig:use-cases}
\end{figure*}
\setlength{\belowcaptionskip}{0pt}
\setlength{\abovecaptionskip}{12pt}

As artificial intelligence (AI) becomes increasingly integrated into our everyday lives, mitigating the societal harms posed by AI technologies has never been more important.
In response to the demand for accountable and safe AI, there have been growing efforts from both industry and academia towards responsible design and development of AI~\cite{rakovaWhereResponsibleAI2021,wangDesigningResponsibleAI2023}.
The majority of these endeavors focus on machine learning (ML) experts, such as ML developers and other AI practitioners.
For example, researchers have introduced techniques that help ML developers interpret ML models~\cite{lundbergUnifiedApproachInterpreting2017, ribeiroWhyShouldTrust2016,noriInterpretMLUnifiedFramework2019} and assess model fairness~\cite{chouldechovaFairPredictionDisparate2017, weertsFairlearnAssessingImproving2023,kleinbergInherentTradeOffsFair2016}.
Additionally, researchers have also proposed frameworks that target ML developers' workflows, such as improving data collection and annotation practices~\cite{berettaDetectingDiscriminatoryRisk2021, miceliSubjectivityImpositionPower2020, mostafazadehdavaniDealingDisagreementsLooking2022}, documenting training data and models~\cite{mitchellModelCardsModel2019, gebruDatasheetsDatasets2020,diazCrowdWorkSheetsAccountingIndividual2022}, and anticipating an ML product's potentials for harms~\cite{microsoftHarmsModelingAzure2022,doteveryoneConsequenceScanningAgile2019}.

However, more recently, we have witnessed a rapid advancement of large language models (LLMs) such as Gemini~\cite{teamGeminiFamilyHighly2023} and GPT-4~\cite{openaiGPT4TechnicalReport2023}, alongside the emergence of prompt-based interfaces like Google AI Studio~\cite{googleGoogleAiStudio2023}, GPT Playground~\cite{openaiOpenAIPlayground2023}, AI Chains~\cite{wuAIChainsTransparent2022}, and Wordflow~\cite{wangWordflowSocialPrompt2024}~(\autoref{fig:diagram}\figpart{B}).
These general-purpose models and easy-to-use interfaces have significantly increased access to the process of prototyping and building diverse AI-powered applications---leading to a paradigm shift in AI development workflows that poses unique challenges to responsible AI, including introducing new potential harms to avoid~\cite{weidingerEthicalSocialRisks2021}, as well as challenges applying existing responsible AI practices~\cite{liaoAITransparencyAge2023}.

Many people who use prompts to create AI applications now encompass a broader spectrum of roles beyond traditional ML experts~(\autoref{fig:diagram}\figpart{A}), such as designers, writers, lawyers, and everyday users~\cite{fiannacaProgrammingProgrammingLanguage2023,jiangPromptMakerPromptbasedPrototyping2022, zamfirescu-pereiraWhyJohnnyCan2023, weiserChatGPTLawyerExplains2023}, whereas existing responsible AI research often targets ML experts such as ML engineers and data scientists~\cite{holsteinImprovingFairnessMachine2019,wongSeeingToolkitHow2023}.
Many users of AI prompt-based prototyping interfaces \cite[e.g.,][]{googleGoogleAiStudio2023,wuAIChainsTransparent2022,openaiOpenAIPlayground2023, wangWordflowSocialPrompt2024}, or ``AI prototypers'' \cite[cf.][]{jiangPromptMakerPromptbasedPrototyping2022} do not have experience in AI or computer science, which can lead to challenges in anticipating the consequences of their AI applications~\cite{rakovaWhereResponsibleAI2021}---a difficult task even for computer science faculty and AI researchers~\cite{doThatImportantHow2023, boyarskayaOvercomingFailuresImagination2020}.
Furthermore, LLMs demonstrate a wide range of capabilities that are continually being discovered across various contexts, including tasks such as summarization, classification, and translation~\cite{bommasaniOpportunitiesRisksFoundation2022, srivastavaImitationGameQuantifying2022a}.
This characteristic of LLMs gives rise to \textit{complex} and \textit{uncertain} impacts of LLM-powered applications~\cite{ganguliPredictabilitySurpriseLarge2022}, presenting a significant departure from the classical ML models targeted by existing responsible AI endeavors~\cite{liaoAITransparencyAge2023,weidingerEthicalSocialRisks2021} and introducing a new layer of complexity for responsible AI researchers to help AI developers anticipate downstream consequences.

To help address these challenges in applying responsible AI practices to LLM-powered AI applications,
we present \tool{}~(\autoref{fig:teaser}, \autoref{fig:diagram}\figpart{B}), an interactive tool to help AI prototypers identify potential harms of their LLM-powered applications---a key early step in harm prevention and mitigation~\cite{oneilNearTermArtificialIntelligence2020, sureshFrameworkUnderstandingSources2021,tabassiAIRiskManagement2023,madaioCoDesigningChecklistsUnderstand2020,microsoftHarmsModelingAzure2022}---during the prototyping stage.
Using \tool{} as a probe, we conduct multiple mixed-method user studies to investigate (1) how an early-stage intervention changes AI prototypers' awareness of and approach to identifying harms, (2) the effectiveness of our tool in helping people envision harms, and (3) the challenges AI prototypers face during this harm envisioning process.
\textbf{We contribute:}

\aptLtoX[graphic=no,type=html]{
      \begin{itemize}
            \item \textbf{\tool{}, the first \textit{in situ} interactive harm envisioning tool that empowers AI prototypers} to identify potential harms that may arise from their prompt-based AI applications, directly within their prompting environments~(\autoref{fig:teaser}, \autoref{fig:diagram}).
                  Inspired by prior harm envisioning frameworks~\cite{microsoftHarmsModelingAzure2022,bucincaAHAFacilitatingAI2023,doteveryoneConsequenceScanningAgile2019} and \textit{in situ} security alert tools~\cite{reinheimerInvestigationPhishingAwareness2020, maurerUsingDataType2011, mylonasDelegateSmartphoneUser2013}, \tool{} overcomes unique design challenges identified from a literature review~(\autoref{sec:related}) and a co-design user study with 10 AI prototypers~(\autoref{sec:formative}).

            \item \textbf{Novel techniques and interactive system designs} to foster responsible AI awareness among AI prototypers.
                  Given a user's prompt, \tool{} leverages embedding similarities to surface news articles about relevant AI incidents from the AI Incident Database~\cite{mcgregorPreventingRepeatedReal2020} and uses LLMs to generate potential use cases, impacted stakeholders, and harms for AI prototypers to review, edit, and add to.
                  Applying a progressive disclosure design~\cite{normanUserCenteredSystem1986}, our tool fits into users' diverse prompting workflows.
                  With a novel adaptation of node-link diagrams~\cite{reingoldTidierDrawingsTrees1981}, \tool{} enables users to interactively visualize, generate, and edit use cases, stakeholders, and harms~(\autoref{sec:system}).

            \item \textbf{Empirical findings about harm envisioning processes from a co-design study and an evaluation study.}
                  During our design of \tool{}, we conducted a co-design study with 10 AI prototypers to evaluate our design ideas and generate new ideas~(\autoref{sec:formative}).
                  After developing \tool{}, we conducted an evaluation user study with 42 AI prototypers to examine the effectiveness of \tool{} in aiding users to brainstorm harms and improving their ability to independently identify harms.
                  Our mixed-method analysis highlights that, after using \tool{}, AI prototypers are better able to independently identify potential harms that might arise from an application developed with a given prompt, and participants report that our tool is more useful and usable than existing resources.
                  In particular, \tool{} encourages users to shift their focus from the AI model to the end-users, providing them with a broader perspective to consider indirect stakeholders and cascading harms~(\autoref{sec:evaluation}).

            \item \textbf{An open-source, web-based implementation} that lowers the barrier to applying responsible AI practices.
                  We develop \tool{} with cutting-edge web technologies, such as Web Components~\cite{mdnWebComponentsWeb2021} and WebGL~\cite{mdnWebGL2D3D2011}, so that it can be easily integrated into any web-based prompt development environments, such as Google AI Studio and Jupyter Notebook~(\autoref{fig:use-cases}).
                  We open source\footnote{\tool{} code: \nnlink{https://github.com/PAIR-code/farsight}} \tool{} as a collection of reusable interactive components that future researchers and designers can easily adopt~(\autoref{sec:system:implementation}).
                  To see a demo video of \tool{}, visit \nnlink{https://youtu.be/BlSFbGkOlHk}.

      \end{itemize}
}{
      \begin{itemize}[topsep=5pt, itemsep=0mm, parsep=1mm, leftmargin=10pt]
            \item \textbf{\tool{}, the first \textit{in situ} interactive harm envisioning tool that empowers AI prototypers} to identify potential harms that may arise from their prompt-based AI applications, directly within their prompting environments~(\autoref{fig:teaser}, \autoref{fig:diagram}).
                  Inspired by prior harm envisioning frameworks~\cite{microsoftHarmsModelingAzure2022,bucincaAHAFacilitatingAI2023,doteveryoneConsequenceScanningAgile2019} and \textit{in situ} security alert tools~\cite{reinheimerInvestigationPhishingAwareness2020, maurerUsingDataType2011, mylonasDelegateSmartphoneUser2013}, \tool{} overcomes unique design challenges identified from a literature review~(\autoref{sec:related}) and a co-design user study with 10 AI prototypers~(\autoref{sec:formative}).

            \item \textbf{Novel techniques and interactive system designs} to foster responsible AI awareness among AI prototypers.
                  Given a user's prompt, \tool{} leverages embedding similarities to surface news articles about relevant AI incidents from the AI Incident Database~\cite{mcgregorPreventingRepeatedReal2020} and uses LLMs to generate potential use cases, impacted stakeholders, and harms for AI prototypers to review, edit, and add to.
                  Applying a progressive disclosure design~\cite{normanUserCenteredSystem1986}, our tool fits into users' diverse prompting workflows.
                  With a novel adaptation of node-link diagrams~\cite{reingoldTidierDrawingsTrees1981}, \tool{} enables users to interactively visualize, generate, and edit use cases, stakeholders, and harms~(\autoref{sec:system}).

            \item \textbf{Empirical findings about harm envisioning processes from a co-design study and an evaluation study.}
                  During our design of \tool{}, we conducted a co-design study with 10 AI prototypers to evaluate our design ideas and generate new ideas~(\autoref{sec:formative}).
                  After developing \tool{}, we conducted an evaluation user study with 42 AI prototypers to examine the effectiveness of \tool{} in aiding users to brainstorm harms and improving their ability to independently identify harms.
                  Our mixed-method analysis highlights that, after using \tool{}, AI prototypers are better able to independently identify potential harms that might arise from an application developed with a given prompt, and participants report that our tool is more useful and usable than existing resources.
                  In particular, \tool{} encourages users to shift their focus from the AI model to the end-users, providing them with a broader perspective to consider indirect stakeholders and cascading harms~(\autoref{sec:evaluation}).

            \item \textbf{An open-source, web-based implementation} that lowers the barrier to applying responsible AI practices.
                  We develop \tool{} with cutting-edge web technologies, such as Web Components~\cite{mdnWebComponentsWeb2021} and WebGL~\cite{mdnWebGL2D3D2011}, so that it can be easily integrated into any web-based prompt development environments, such as Google AI Studio and Jupyter Notebook~(\autoref{fig:use-cases}).
                  We open source\footnote{\tool{} code: \nnlink{https://github.com/PAIR-code/farsight}} \tool{} as a collection of reusable interactive components that future researchers and designers can easily adopt~(\autoref{sec:system:implementation}).
                  To see a demo video of \tool{}, visit \nnlink{https://youtu.be/BlSFbGkOlHk}.

      \end{itemize}
} %
\section{Related Work}
\label{sec:related}

\subsection{Anticipating Technology's Negative Impacts}
Various design methods and approaches have been developed to support ideation about potential downstream impacts of technology, including anticipatory tech ethics \cite{breyAnticipatoryEthicsEmerging2012,nanayakkaraAnticipatoryEthicsRole2020}, speculative design \cite{wongSpeculativeDesignHCI2018,augerSpeculativeDesignCrafting2013,dunneSpeculativeEverythingDesign2013}, and value-sensitive design \cite{friedmanValuesensitiveDesign1996,friedmanValueSensitiveDesign2002,friedmanSurveyValueSensitive2017} among others.
To support designers with this, prior research has developed design toolkits \cite[e.g.,][]{chivukulaSurveyingLandscapeEthicsfocused2021} and resources, such as Envisioning Cards \cite{friedmanEnvisioningCardsToolkit2012}, Value Cards \cite{shenValueCardsEducational2021}, Timelines \cite{wongTimelinesWorldbuildingActivity2021}, and the Black Mirror Writers' Room \cite{klassenRunWildLittle2022}, among others \cite[e.g.,][]{ballardJudgmentCallGame2019,doteveryoneConsequenceScanningAgile2019}.
Such resources are intended to be used by designers of technology early in the design process, but they may not fit neatly into existing product design and development processes, particularly for AI-powered application design paradigms, where large pre-trained models are used for many downstream tasks \cite{wangDesigningResponsibleAI2023}.

In addition to technology designers, computing researchers have called for the computer science field to consider the negative impacts of their work in addition to the positive impacts \cite{hechtItTimeSomething2021}. In AI research, conferences such as NeurIPS have begun requiring that researchers articulate potential negative broader impacts of their work in statements at the ends of their papers \cite{prunklInstitutionalizingEthicsAI2021} to avoid the ``failures of imagination'' \cite{boyarskayaOvercomingFailuresImagination2020} that may lead to downstream harms. Prior work analyzed these broader impacts statements, finding convergence around a set of topics such as risks to privacy and bias, but often lacking concrete specifics or strategies for mitigation \cite{ashurstAiEthicsStatements2022,nanayakkaraUnpackingExpressedConsequences2021,liuExaminingResponsibilityDeliberation2022,simThinkingWritingResearch2021}. However, prior work suggests that many CS researchers may not have the training, resources, or inclination to engage in this type of anticipatory work \cite{sturdeeConsequencesSchmonsequencesConsidering2021,doThatImportantHow2023}, suggesting that new tools, training, and processes, are needed to support researchers and developers in engaging in anticipatory work in ways that are integrated into their research practices.
More recently, researchers have proposed a framework that uses LLMs to anticipate harms for classifiers by generating stakeholders and vignettes for a given scenario~\cite{bucincaAHAFacilitatingAI2023}, evaluating this framework through interviews with responsible AI researchers.
\tool{} builds upon this framework and extends it to (1) target an early prototyping stage through \textit{in situ} and interactive interfaces that promote user engagement in the harm envisioning process, (2) support LLM-powered applications with diverse tasks beyond classification, and (3) evaluates its effectiveness through a user study with 42 AI prototypers.

\subsection{Identifying and Mitigating LLM Harms}
\label{sec:related:llm}
More recently, there has been a growing body of research that specifically focuses on identifying and mitigating the harms of LLMs.
Researchers have introduced harm taxonomies specifically for LLMs, which identify known risks~(i.e., informed by observed instances of harm)~\cite{bommasaniOpportunitiesRisksFoundation2022, weidingerEthicalSocialRisks2021, liuTrustworthyLLMsSurvey2023} and emerging risks of LLMs~(anticipated risks based on foreseeable capabilities of LLMs)~\cite{matzPotentialGenerativeAI2023, shevlaneModelEvaluationExtreme2023}.
Since LLMs can be used for a wide range of tasks associated with many different categories of harms, researchers have presented frameworks and evaluation methods to assess a particular type of LLM harm, including misinformation~\cite{panRiskMisinformationPollution2023, hanleyMachineMadeMediaMonitoring2023}, representation and toxicity~\cite{gehmanRealToxicityPromptsEvaluatingNeural2020, deshpandeToxicityChatGPTAnalyzing2023}, human autonomy~\cite{glaeseImprovingAlignmentDialogue2022, simmonsMoralMimicryLarge2023}, malicious use~\cite{dengMasterKeyAutomatedJailbreak2023, royGeneratingPhishingAttacks2023}, and data privacy~\cite{liMultistepJailbreakingPrivacy2023, kimProPILEProbingPrivacy2023}.
The popular methods to identify these harms include benchmarking~\cite{carliniQuantifyingMemorizationNeural2023, carliniExtractingTrainingData2021}, user research~\cite{malinkaEducationalImpactChatGPT2023, longoniNewsGenerativeArtificial2022}, and adversarial testing~\cite{perezRedTeamingLanguage2022, dernerSafeguardsExploringSecurity2023}.
Based on existing benchmarks and harm taxonomies of LLM risks, \citet{weidingerSociotechnicalSafetyEvaluation2023} introduce a sociotechnical evaluation framework that identifies three AI actors with LLM safety responsibilities: AI model developers, AI application developers, and third-party stakeholders.

The mitigation strategies for these harms depend on the use cases and context.
Popular strategies include algorithmic and sociotechnical approaches~\cite{weidingerTaxonomyRisksPosed2022}, such as improving the training data to mitigate social stereotypes and biases~\cite{solaimanProcessAdaptingLanguage2021}; fine-tuning LLM models on curated datasets~\cite{gehmanRealToxicityPromptsEvaluatingNeural2020}; filtering LLM outputs~\cite{welblChallengesDetoxifyingLanguage2021, xuDetoxifyingLanguageModels2021}; employing special decoding techniques~\cite{krauseGeDiGenerativeDiscriminator2021, schickSelfDiagnosisSelfDebiasingProposal2021}, adding instructions in prompts~\cite{askellGeneralLanguageAssistant2021}, monitoring the use of LLMs~\cite{weidingerTaxonomyRisksPosed2022}; as well as inclusive product design and development from the beginning~\cite{aiQueerAICase2023, delgadoStakeholderParticipationAI2021,harringtonDeconstructingCommunityBasedCollaborative2019,cooperSystematicReviewThematic2022}.
Building on this prior work, \tool{} introduces a novel framework that leverages human-AI collaboration to help AI prototypers identify the potential harms of LLMs.
Specifically targeting AI prototypers as one subset of AI application developers~\cite{weidingerSociotechnicalSafetyEvaluation2023,wangDesigningResponsibleAI2023}, \tool{} introduces novel techniques and \textit{in situ} interfaces to foster responsible AI awareness during AI prototyping, although the current version of \tool{} does not assist AI prototypers in mitigating potential LLM harms.

\subsection{Responsible AI Tools and Practices}

Despite the increasing emphasis on responsible AI from the technology industry~\cite{wrightComparativeAnalysisIndustry2020, amershiGuidelinesHumanAIInteraction2019, googlepairPeopleAIGuidebook2019,appleHumanInterfaceGuidelines2023},
academia~\cite{dilhacReportMontrealDeclaration2018,liAnnualReport20222022}, and policymakers~\cite{madiegaArtificialIntelligenceAct2021, thewhitehouseBlueprintAIBill2022,tabassiAIRiskManagement2023}, incorporating responsible AI practices into AI product development remains a challenge~\cite[e.g.,][]{wangDesigningResponsibleAI2023,varanasiItCurrentlyHodgepodge2023, aliWalkingWalkAI2023}---
in part due to practitioners' insufficient knowledge of responsible AI~\cite[e.g.,][]{schiffPrinciplesPracticesResponsible2020, rajiYouCanSit2021,garrettMoreIfTime2020}, lack of engagement with direct stakeholders or domain experts~\cite{madaioAssessingFairnessAI2022, hongHumanFactorsModel2020}, and organizational culture and structure~\cite{rakovaWhereResponsibleAI2021,madaioCoDesigningChecklistsUnderstand2020}.

To address these challenges and facilitate the adoption of responsible AI practices, researchers have proposed several approaches.
These include incorporating ethics into AI education~\cite{fieslerWhatWeTeach2020,smithIncorporatingEthicsComputing2023, shenValueCardsEducational2021}, providing engaging playbooks or design activities~\cite{yildirimInvestigatingHowPractitioners2023,hongPlanningNaturalLanguage2021, ballardJudgmentCallGame2019}, and fostering ethical norms in AI research and development~\cite{liuExaminingResponsibilityDeliberation2022, smithREALMLRecognizing2022, rajiClosingAIAccountability2020}.
In addition, researchers have also proposed a wide range of tools to operationalize responsible AI practices~\cite{leeLandscapeGapsOpen2020, wongSeeingToolkitHow2023}.
These tools encompass libraries and frameworks that cover various dimensions of responsible AI, including fairness~\cite{weertsFairlearnAssessingImproving2023, saleiroAequitasBiasFairness2019, bellamyAIFairness3602018}, explainability~\cite{ribeiroWhyShouldTrust2016,noriInterpretMLUnifiedFramework2019}, testing and error analysis~\cite{vilarErrorAnalysisStatistical2006, microsoftResponsibleAIToolbox2020, ribeiroAccuracyBehavioralTesting2020}, and model development documentation~\cite{rajiClosingAIAccountability2020, mitchellModelCardsModel2019, gebruDatasheetsDatasets2020}.\looseness=-1

Moreover, alongside these advancements, there has been a rise in the research and development of easy-to-use interactive visualization tools to further facilitate the operationalization of responsible AI.
For example, tools like \textsc{What-If Tool}~\cite{wexlerWhatIfToolInteractive2019}, \textsc{FairVis}~\cite{cabreraFAIRVISVisualAnalytics2019}, and \textsc{Visual Auditor}~\cite{munechikaVisualAuditorInteractive2022} enable ML developers to visually assess the fairness of ML models across a diverse range of inputs.
Visual analytics systems such as \textsc{Summit}~\cite{hohmanSUMMITScalingDeep2019}, \textsc{LIT}~\cite{tenneyLanguageInterpretabilityTool2020}, and \textsc{GAM Changer}~\cite{wangInterpretabilityThenWhat2022} empower ML developers to interpret their models and fix problematic behaviors.
Interactive visual testing tools like \textsc{Errudite}~\cite{wuErruditeScalableReproducible2019}, \textsc{Angler}~\cite{robertsonAnglerHelpingMachine2023}, and \textsc{AdaTest}~\cite{ribeiroAdaptiveTestingDebugging2022} help ML developers surface weaknesses in their models.\looseness=-1

Inspired by these tools, \tool{} joins the body of research of interactive visualization tools for responsible AI by visualizing use cases, stakeholders, and harms~(\autoref{sec:system:window}).
In contrast to existing tools that target traditional ML models after they have been trained, \tool{} focuses on diverse LLM-powered applications in an early prototyping stage.
During this stage, AI prototypers have greater flexibility to iterate on the design and objectives of their applications and implement early mitigation strategies such as engaging with stakeholders and improving data collection~\cite{amershiSoftwareEngineeringMachine2019}.

\vspace{-2pt}
\subsection{\textit{In Situ} Alerting Tools}
\label{sec:related:alert}

Although \textit{in situ} responsible AI tools are relatively nascent, there is a large body of research in designing in-context warning tools and interfaces.
For example, security and HCI researchers study how to best present warnings to raise people's online security awareness~\cite[e.g.,][]{reinheimerInvestigationPhishingAwareness2020, maurerUsingDataType2011, mylonasDelegateSmartphoneUser2013} and protect people from malware and phishing attacks~\cite[e.g.,][]{egelmanImportanceBeingEarnest2013, reederExperienceSamplingStudy2018, feltImprovingSSLWarnings2015}.
The key challenges when designing effective warning interfaces include the presentation of comprehensible messages and supporting evidence~\cite{biddleBrowserInterfacesExtended2009, feltExperimentingScaleGoogle2014}, engaging users~\cite{wuSecurityToolbarsActually2006, egelmanYouVeBeen2008}, and preventing alert fatigue and habituation~\cite{andersonHowPolymorphicWarnings2015, bohmeTrainedAcceptField2010}.
To address these challenges, researchers recommend designing simple interfaces~\cite{goodStoppingSpywareGate2005, goodNoticingNoticeLargescale2007}, considering the trade-off between blocking and non-blocking warnings~\cite{egelmanYouVeBeen2008}, varying interfaces~\cite{andersonHowPolymorphicWarnings2015}, and requiring user input~\cite{brustoloniImprovingSecurityDecisions2007}.

Using in-context warnings to improve users' safety awareness and encourage users to take protection measures can be considered a form of ``digital nudging''~\cite{schneiderDigitalNudgingGuiding2018,caraban23WaysNudge2019}.
More recently, researchers have also adapted in-context security warnings to nudge social media users to recognize and avoid online disinformation~\cite{kaiserAdaptingSecurityWarnings2021, sharevskiMeaningfulContextRed2022} and reflect before posting potentially harmful content~\cite{simonOpenWebTestsImpact2020, kiskolaOnlineSurveyNovel2022, wrightRECASTEnablingUser2021}.
Beyond platform-initiated integration of warnings, end-users also voluntarily seek in-context alert interfaces for productivity improvement.
For example, writers use grammar checker tools like Grammarly~\cite{grammarlyGrammarlyFreeWriting2023}, which offer in-context warnings and scores to improve their writing.
Similarly, software developers use accessibility developer tools~\cite{googleLighthouse2016, dequeAxeDevToolsDigital2023} to detect potential accessibility issues during the development process.
However, there has been little work in designing and evaluating \textit{in situ} warnings for developing AI applications, particularly for responsible AI.
\tool{}'s design draws inspiration from many existing warning interfaces~(\autoref{sec:formative}).
Our work advances the landscape of \textit{in situ} alerting research by addressing responsible AI for modern AI application development.

\section{Formative Study \& Design Goals}
\label{sec:formative}

To identify the needs and potential challenges faced by users in envisioning harms, we conducted a formative co-design study to investigate (1) how AI prototypers envision harms (if they do), (2) what design ideas are most helpful for them, and (3) how to motivate users to think about potential risks when prototyping an AI application.
In this section, we report our findings from the formative co-design study, and in~\autoref{sec:evaluation}, we report on our findings from a subsequent evaluation user study.

\subsection{Co-design Study}
\label{sec:formative-study}

\mypar{Participants.}
To inform our tool's design, we conducted a co-design user study with 10 AI prototypers based in the United States.
These participants were recruited from Google through internal mailing lists.
Our recruitment criteria required participants to have experience using an internal prompt-crafting tool, PromptMaker \cite{jiangPromptMakerPromptbasedPrototyping2022}, which is similar to Google AI Studio~\cite{googleGoogleAiStudio2023} and GPT Playground~\cite{openaiOpenAIPlayground2023}.
Each session was 60 minutes, and each participant received an average of \$50 USD in their choice of a gift card or a donation to their preferred charity.
Among the 10 participants (U1–U10), 6 identified as men, 3 identified as women, and 1 identified as non-binary.
Four participants self-reported having expertise in responsible AI.
Information about participants' job roles is listed in \autoref{table:design-participant}.
All participants are our targeted users (AI prototypers).

\begin{table}
      \caption[]{
            The co-design user study includes 10 participants with diverse roles. All participants have experience in prompting LLMs. Four participants who self-reported having expertise in responsible AI are marked with asterisks ($\ast$).
      }
      \label{table:design-participant}
      \begin{tabular}{l l}
            \textbf{Participant Roles} & \textbf{Participant IDs} \\
            \midrule
            Software Engineer          & 1$\ast$, 4, 5, 6, 7, 10  \\
            Research Scientist         & 3$\ast$, 8$\ast$         \\
            Technical Writer           & 2                        \\
            Program Manager            & 9$\ast$                  \\
      \end{tabular}
      \Description{
            This table lists participant roles and their IDs in the co-design user study. The study includes 10 participants with diverse roles, all experienced in prompting Large Language Models (LLMs). The table has two columns: "Participant Roles" and "Participant IDs." The roles listed are Software Engineer, Research Scientist, Technical Writer, and Program Manager. Participants with expertise in responsible AI are marked with asterisks. Participant IDs for Software Engineer are 1* (marked with an asterisk indicating expertise in responsible AI), 4, 5, 6, 7, and 10. Research Scientists are numbered 3* and 8*, both marked with asterisks. The Technical Writer is assigned the number 2, and the Program Manager is labeled 9* with an asterisk. The asterisks denote those with self-reported expertise in responsible AI.
      }
\end{table}

\mypar{Procedure.}
We structured our study as a ``during-design co-design study''~\cite{sarmientoParticipatoryCoDesignLearning2022}.
Participants were asked to bring a recent prompt that they had written to the study.
The study started with a semi-structured interview regarding participants' prompting workflows and their experience in thinking about potential harms linked to their applications~(\autoref{appendix:evaluation-co-design-interview-1}).
Then participants were asked to use our very early-stage design prototypes~(\autoref{appendix:evaluation-co-design-interview-1}) to envision potential harms associated with their application while thinking aloud.
Participants were also presented with low-fidelity sketches for our other design ideas.
These prototypes and sketches can be found in~\autoref{fig:appendix-prototypes}.
Finally, we asked participants to rate and provide feedback on all of our design ideas and generate their own design suggestions~(\autoref{appendix:evaluation-co-design-interview-2}).

\setlength{\belowcaptionskip}{0pt}
\setlength{\abovecaptionskip}{5pt}
\begin{figure}[tb]
      \centering
      \includegraphics[width=1\linewidth]{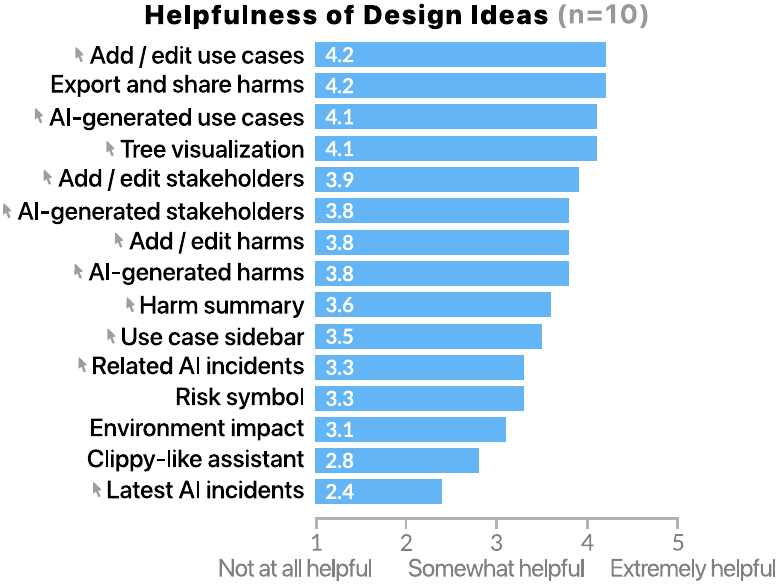}
      \caption[]{
            Average ratings on our design ideas from 10 AI prototypers.
            Features marked with \protect\vcenteredhbox{\includegraphics[height=6pt]{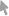}} were presented to participants as early-stage prototypes, while other features were presented as sketches~(see details in \autoref{fig:appendix-prototypes}).
      }
      \Description{Average ratings on our design ideas, sorted by add/edit use cases, export and share harms, AI-generated use cases, tree visualization, add/edit stakeholders, AI-generated stakeholders, add/edit harms, AI-generated harms, harm summary, use case sidebar, related AI incidents, risk symbol, environment impact, slippy-like assistant, latest AI incidents.}
      \label{fig:formative-feature}
\end{figure}
\setlength{\belowcaptionskip}{0pt}
\setlength{\abovecaptionskip}{12pt}

\mypar{Design feedback.}
Interestingly, although perhaps not surprisingly \cite[cf.][]{holsteinImprovingFairnessMachine2019}, none of the 6 participants without expertise in responsible AI reported that they typically considered the potential harms of their AI prototypes when writing prompts, while 3 of the 4 participants with expertise in responsible AI did report typically anticipating harms during the prototyping process.
Participants' ratings were shown in \autoref{fig:formative-feature}.
Overall, participants favored using AI to generate use cases of their AI prototypes, potential stakeholders, and potential harms.
Many participants also highlighted the importance of being able to edit AI-generated content and control the generation direction~(U4, U8).
On the other hand, participants were less in favor of more distracting design ideas (e.g., an anthropomorphized assistant tool similar to Microsoft Office's Clippy) or irrelevant content (e.g., the latest, rather than the most relevant AI incidents).
Participants also provided us with helpful usability feedback that we integrated into our final design of \tool{}~(\autoref{sec:system}).

\mypar{New design ideas.}
Participants generated many interesting design ideas to help raise responsible awareness among AI prototypers.
For example, participants recommended categorizing AI-generated harms~(U1, U5), allowing users to rate the severity of harms~(U6), and using users' input to steer AI generation~(U10).
We integrated these design ideas into the final design of \tool{}~(\autoref{sec:system}).
Some other interesting design ideas include designing a game-like reward system to incentivize users to anticipate harms~(U5), building online communities to allow users to share their envisioned harms using \tool{} and seek support~(U2), allowing real-time collaborative harm envisioning similar to Google Slides~(U1, U4), and automatically revising a user's prompt to address identified harms~(U4).
We discuss the implications of these design ideas in user motivation~(\autoref{sec:discussion-motivation}) and mitigation strategies~(\autoref{sec:discussion-action}).

\subsection{Design Goals}
\label{sec:formative-goals}

Based on our literature review~(\autoref{sec:related}) and findings and early feedback from the co-design user study, we identify the following five design goals~(\aptLtoX[graphic=no,type=html]{\textbf{G1}--\textbf{G5}}{\ref{item:goal-use-case}--\ref{item:goal-open-source}}) for \tool{}.

\aptLtoX[graphic=no,type=html]{
      \begin{itemize}
            \item [\textbf{G1.}] \textbf{Guide users in imagining use cases.}
                  Existing research highlights the challenges faced by ML practitioners when attempting to anticipate the uses of their ML-powered applications and how different individuals or groups may be affected~\cite{boyarskayaOvercomingFailuresImagination2020, smithREALMLRecognizing2022,doThatImportantHow2023, madaioAssessingFairnessAI2022}.
                  Confirming this, software engineer U6 noted \myquote{You don't really know how your tool could be used, so it's really hard to envision what harms would be.}
                  The availability of LLMs and prompt-crafting tools has broadened the spectrum of AI prototypers to include people without prior technology development experience~\cite{fiannacaProgrammingProgrammingLanguage2023,jiangPromptMakerPromptbasedPrototyping2022}, which can further magnify the challenges associated with envisioning diverse use cases for AI applications.
                  Therefore, we design \tool{} to help AI prototypers with diverse backgrounds to brainstorm a wide array of use cases for their AI applications.

            \item [\textbf{G2.}] \textbf{Help users understand, organize, and prioritize harms.}
                  Depending on an AI application's goal, implementation, and context, some harms are more salient than others~\cite{bucincaAHAFacilitatingAI2023, microsoftMicrosoftResponsibleAI2022}.
                  To help AI prototypers assess harms, \tool{} should first help them understand \textit{where} and \textit{how} harms might occur and \textit{who} might be impacted, by connecting harms to use cases and stakeholders~\cite{madaioAssessingFairnessAI2022,barocasDesigningDisaggregatedEvaluations2021,microsoftHarmsModelingAzure2022}.
                  Participants expressed a desire for the ability to categorize~(U1, U5) and rate the severity~(U6) of harms.
                  To meet these needs, we aim to design an easy-to-use interface that empowers users to navigate, comprehend, and label harms within diverse potential harm scenarios.

            \item [\textbf{G3.}] \textbf{Fit into current workflows and mitigate habituation.}
                  In our co-design study, none of the 6 participants without expertise in responsible AI had previously thought about harms when writing prompts.
                  We also found some participants were not incentivized to anticipate harm on their own; for example, U6 explained \myquote{To be honest, as a software engineer, I don't use policy tools [compliance tools like checklists] unless I have to.}
                  Thus, to make \tool{} easy to adopt, we aim to take inspiration from \textit{in situ} warning tools~\cite[e.g.,][]{egelmanImportanceBeingEarnest2013, reederExperienceSamplingStudy2018, feltImprovingSSLWarnings2015} to design it in a way that fits into AI prototypers' existing workflows instead of introducing new workflows.
                  In addition, we aim to apply strategies like varying content~\cite{andersonHowPolymorphicWarnings2015} and promoting user input~\cite{brustoloniImprovingSecurityDecisions2007} to mitigate habituation---a common pitfall of in-context warning designs~\cite{andersonHowPolymorphicWarnings2015,bohmeTrainedAcceptField2010}.

            \item [\textbf{G4.}] \textbf{Promote user engagement and provide compelling examples.}
                  Prior research highlights that the effectiveness of warning tools depends on their clarity and persuasiveness~\cite{biddleBrowserInterfacesExtended2009, feltExperimentingScaleGoogle2014}.
                  As we are targeting AI prototypers with diverse experience in AI and responsible AI, \tool{} should be easy to use and understand.
                  When asked what would help them envision potential harms for their AI applications, many participants mentioned referring to prior examples of AI harms~(U1, U2, U8).
                  For instance, U2 said \myquote{Giving some specific real [harm] examples for different types of seemingly innocuous applications would help alert people [to consider harms].}
                  Therefore, we aimed to integrate real examples in \tool{} to motivate and help users understand the potential risk of their applications.
                  Participants like being able to control the harm envisioning process~(\autoref{fig:formative-feature}), and active participation is a key factor in learning \cite{koedingerLearningNotSpectator2015}---essential to foster AI prototypers' ability to independently identify harms.
                  Thus, \tool{} is designed to provide users with human agency and encourage users to actively and critically think about harms.

            \item [\textbf{G5.}] \textbf{Open-source and adaptable implementation.}
                  Given the ever-expanding array of LLMs and prompt-crafting tools~\cite{chowAIArmsRace2023}, our approach in designing \tool{} is to ensure it remains adaptable to this dynamically evolving landscape.
                  We aimed to design \tool{} to be model-agnostic and environment-agnostic, thereby making it accessible to users of different LLM models (e.g., Gemini~\cite{teamGeminiFamilyHighly2023}, GPT-4~\cite{openaiGPT4TechnicalReport2023}, Llama 2~\cite{touvronLlamaOpenFoundation2023}) and prompt-crafting interfaces (e.g., GPT Playground~\cite{openaiOpenAIPlayground2023}, Google AI Studio~\cite{googleGoogleAiStudio2023}, Wordflow~\cite{wangWordflowSocialPrompt2024}).
                  Furthermore, we open source our implementation to foster future advancements in the design, research, and development of responsible AI tools.

      \end{itemize}
}{
      \begin{enumerate}[topsep=5pt, itemsep=0mm, parsep=1mm, leftmargin=18pt, label=\textbf{G\arabic*.}, ref=G\arabic*]
            \item \label{item:goal-use-case} \textbf{Guide users in imagining use cases.}
                  Existing research highlights the challenges faced by ML practitioners when attempting to anticipate the uses of their ML-powered applications and how different individuals or groups may be affected~\cite{boyarskayaOvercomingFailuresImagination2020, smithREALMLRecognizing2022,doThatImportantHow2023, madaioAssessingFairnessAI2022}.
                  Confirming this, software engineer U6 noted \myquote{You don't really know how your tool could be used, so it's really hard to envision what harms would be.}
                  The availability of LLMs and prompt-crafting tools has broadened the spectrum of AI prototypers to include people without prior technology development experience~\cite{fiannacaProgrammingProgrammingLanguage2023,jiangPromptMakerPromptbasedPrototyping2022}, which can further magnify the challenges associated with envisioning diverse use cases for AI applications.
                  Therefore, we design \tool{} to help AI prototypers with diverse backgrounds to brainstorm a wide array of use cases for their AI applications.

            \item \label{item:goal-explore} \textbf{Help users understand, organize, and prioritize harms.}
                  Depending on an AI application's goal, implementation, and context, some harms are more salient than others~\cite{bucincaAHAFacilitatingAI2023, microsoftMicrosoftResponsibleAI2022}.
                  To help AI prototypers assess harms, \tool{} should first help them understand \textit{where} and \textit{how} harms might occur and \textit{who} might be impacted, by connecting harms to use cases and stakeholders~\cite{madaioAssessingFairnessAI2022,barocasDesigningDisaggregatedEvaluations2021,microsoftHarmsModelingAzure2022}.
                  Participants expressed a desire for the ability to categorize~(U1, U5) and rate the severity~(U6) of harms.
                  To meet these needs, we aim to design an easy-to-use interface that empowers users to navigate, comprehend, and label harms within diverse potential harm scenarios.

            \item \label{item:goal-workflow} \textbf{Fit into current workflows and mitigate habituation.}
                  In our co-design study, none of the 6 participants without expertise in responsible AI had previously thought about harms when writing prompts.
                  We also found some participants were not incentivized to anticipate harm on their own; for example, U6 explained \myquote{To be honest, as a software engineer, I don't use policy tools [compliance tools like checklists] unless I have to.}
                  Thus, to make \tool{} easy to adopt, we aim to take inspiration from \textit{in situ} warning tools~\cite[e.g.,][]{egelmanImportanceBeingEarnest2013, reederExperienceSamplingStudy2018, feltImprovingSSLWarnings2015} to design it in a way that fits into AI prototypers' existing workflows instead of introducing new workflows.
                  In addition, we aim to apply strategies like varying content~\cite{andersonHowPolymorphicWarnings2015} and promoting user input~\cite{brustoloniImprovingSecurityDecisions2007} to mitigate habituation---a common pitfall of in-context warning designs~\cite{andersonHowPolymorphicWarnings2015,bohmeTrainedAcceptField2010}.

            \item \label{item:goal-engage} \textbf{Promote user engagement and provide compelling examples.}
                  Prior research highlights that the effectiveness of warning tools depends on their clarity and persuasiveness~\cite{biddleBrowserInterfacesExtended2009, feltExperimentingScaleGoogle2014}.
                  As we are targeting AI prototypers with diverse experience in AI and responsible AI, \tool{} should be easy to use and understand.
                  When asked what would help them envision potential harms for their AI applications, many participants mentioned referring to prior examples of AI harms~(U1, U2, U8).
                  For instance, U2 said \myquote{Giving some specific real [harm] examples for different types of seemingly innocuous applications would help alert people [to consider harms].}
                  Therefore, we aimed to integrate real examples in \tool{} to motivate and help users understand the potential risk of their applications.
                  Participants like being able to control the harm envisioning process~(\autoref{fig:formative-feature}), and active participation is a key factor in learning \cite{koedingerLearningNotSpectator2015}---essential to foster AI prototypers' ability to independently identify harms.
                  Thus, \tool{} is designed to provide users with human agency and encourage users to actively and critically think about harms.

            \item \label{item:goal-open-source} \textbf{Open-source and adaptable implementation.}
                  Given the ever-expanding array of LLMs and prompt-crafting tools~\cite{chowAIArmsRace2023}, our approach in designing \tool{} is to ensure it remains adaptable to this dynamically evolving landscape.
                  We aimed to design \tool{} to be model-agnostic and environment-agnostic, thereby making it accessible to users of different LLM models (e.g., Gemini~\cite{teamGeminiFamilyHighly2023}, GPT-4~\cite{openaiGPT4TechnicalReport2023}, Llama 2~\cite{touvronLlamaOpenFoundation2023}) and prompt-crafting interfaces (e.g., GPT Playground~\cite{openaiOpenAIPlayground2023}, Google AI Studio~\cite{googleGoogleAiStudio2023}, Wordflow~\cite{wangWordflowSocialPrompt2024}).
                  Furthermore, we open source our implementation to foster future advancements in the design, research, and development of responsible AI tools.

      \end{enumerate}
} %
\section{User Interface}
\label{sec:system}

Following the five design goals~(\aptLtoX[graphic=no,type=html]{\textbf{G1}--\textbf{G5}}{\ref{item:goal-use-case}--\ref{item:goal-open-source}}), we present \tool{}, the first \textit{in situ} interactive tool that aims to foster responsible AI awareness among AI prototypers during the AI prototyping process.
\tool{} is designed to be a plugin of any web-based prompt-crafting tools.
\tool{}'s interface employs progressive disclosure~\cite{normanUserCenteredSystem1986}, enabling users to smoothly transition across three main components, with each phase increasing the level of user engagement.
The \symbolview{}~(\autoref{sec:system:symbol}) presents an always-on symbol that shows the approximated \riskscore{} of a user's current prompt; the \sidebarview{}~(\autoref{sec:system:sidebar}) highlights news articles about related AI incidents and LLM-generated use cases and harms; and the \windowview{}~(\autoref{sec:system:window}) visualizes LLM-generated harms and allows users to edit, add, and share harms.
Examples in this section use PaLM 2 model through its APIs; we chose this model because it provided free API access to the public during our design process.
Researchers and designers can easily replace PaLM 2 model with other LLMs by changing the API endpoints in \tool{}.

\subsection{\symbolviewnormal{}}
\label{sec:system:symbol}

\setlength{\columnsep}{10pt}%
\setlength{\intextsep}{-5pt}%
\begin{wrapfigure}{R}{100pt}
  \vspace{0pt}
  \centering
  \includegraphics[width=100pt]{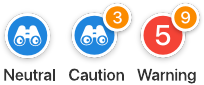}
  \vspace{-20pt}
  \caption[]{
    Three alert modes of the \symbolview{}.
  }
  \Description{
    Three alert modes of the Alert Symbol.
  }
  \vspace{0pt}
  \label{fig:system-risk-symbol}
\end{wrapfigure}
The \symbolview{} is an always-on display on top of the AI prototyping tool, displaying the \riskscore{} of a user's prompt~(\autoref{fig:system-risk-symbol}).
Every time the user runs their prompt, the \symbolview{} updates the \riskscore{} using the new prompt.
Based on the computed \riskscore{}, there are three modes~(\autoref{fig:system-risk-symbol}), each characterized by a progressively more attention-grabbing style.
Thus, \tool{} only disrupts AI prototypers' flow when their prompts require more caution~(\aptLtoX[graphic=no,type=html]{\textbf{G3}}{\ref{item:goal-workflow}}).\looseness=-1

\mypar{Calculating the \riskscoretitle{}.}
Auditing and quantifying the societal risk of LLM-powered applications is an open research problem~\cite{rastogiSupportingHumanAICollaboration2023}.
To categorize the potential harms that might arise from users' prompts, we propose a novel technique that uses the similarity between the prompt and previously documented AI incident reports as a proxy for the prompt's \riskscore{}.
First, we use an LLM to extract high-dimensional latent representations (embeddings) of all AI incident reports indexed in the AI Incident Database~\cite{mcgregorPreventingRepeatedReal2020}, which includes more than 3k community-curated news reports about AI failures and harms.
Then, we extract the embedding of the user's prompt and compute pairwise cosine distances between the prompt embedding and AI incident report embeddings.
We label each incident report as \vcenteredhbox{\includegraphics[height=9pt]{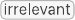}}, \vcenteredhbox{\includegraphics[height=9pt]{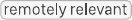}}, \vcenteredhbox{\includegraphics[height=9pt]{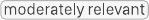}} based on two distance thresholds \texttt{0.69} and \texttt{0.75}.
We determine these two thresholds from an experiment with 1k random prompts (see \autoref{appendix:system-relevancy-thresholds} and \autoref{fig:wizmap} for details).
Researchers can easily adjust these two thresholds (between \texttt{0} and \texttt{1}) to calibrate an article's relevancy.

Finally, we show the numbers of AI incidents that are classified as \vcenteredhbox{\includegraphics[height=9pt]{figures/tag-remotely-relevant}} in an orange circle and \vcenteredhbox{\includegraphics[height=9pt]{figures/tag-moderately-relevant}} in a red circle~(\autoref{fig:system-risk-symbol}) as a proxy of the prompt's potential risk.
In other words, we consider a prompt to have a higher risk if many AI incident reports are semantically and syntactically similar to it.

\subsection{\sidebarviewnormal{}}
\label{sec:system:sidebar}

\setlength{\belowcaptionskip}{-5pt}
\setlength{\abovecaptionskip}{3pt}
\begin{figure*}[tb]
  \includegraphics[width=1\linewidth]{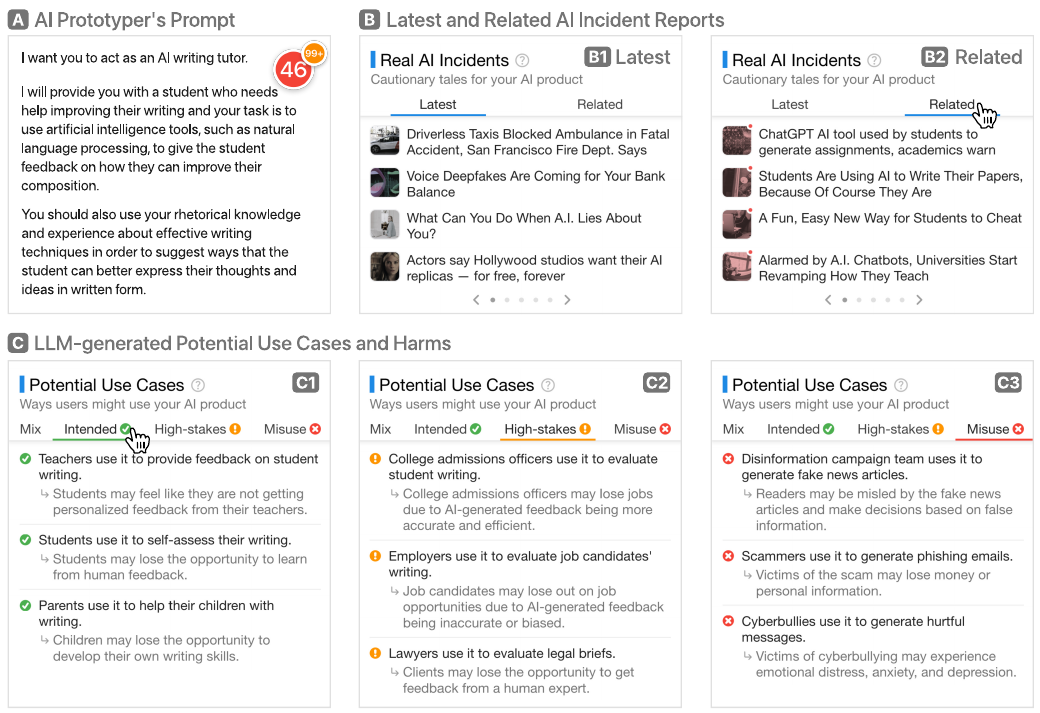}
  \caption[]{
    The \sidebarview{} provides \textit{in situ} information to remind AI prototypers of potential risks.
    \textbf{(A)} Given a user's current prompt, \textbf{(B)} the \incidentpanel{} shows the \textbf{(B1)} latest and \textbf{(B2)} related AI incident reports sampled from the AI Incident Database~\cite{mcgregorPreventingRepeatedReal2020}.
    \textbf{(B2)} The related AI incident tab is the default view, which uses text embedding similarities between the user's prompt and all AI incident reports to surface relevant reports.
    \textbf{(C)} The \usecasepanel{} leverages LLM to generate potential use cases and harms.
    Each use case is classified by an LLM and organized into \textbf{(C1)} \textit{intended}, \textbf{(C2)} \textit{high-stakes}, and \textbf{(C3)} \textit{misuse} tabs.
  }
  \Description{
    A figure with six components. Component A shows AI prototyper's prompt. Component B1 shows the latest AI incident. Component B2 shows the related AI incidents. Component C1 shows indented use cases. Component C2 shows high-stakes use cases. Component C3 shows misuses.
  }
  \label{fig:system-sidebar}
\end{figure*}
\setlength{\belowcaptionskip}{0pt}
\setlength{\abovecaptionskip}{12pt}

After a user clicks the \symbolview{}, the \sidebarview{}~(\autoref{fig:system-sidebar}) expands from one side edge of the AI prototyping tool~(\aptLtoX[graphic=no,type=html]{\textbf{G3}}{\ref{item:goal-workflow}}), highlighting potential consequences of AI applications or features that are based on the user's current prompt.
We use a real prompt from Awesome ChatGPT Prompts~\cite{akinAwesomeChatGPTPrompts2022} in the example in~\autoref{fig:system-sidebar}.

\mypar{\incidentpanelnormal{}.}
To encourage users to consider potential risks associated with their prompts~(\autoref{fig:system-sidebar}\figpart{A}), the \incidentpanel{} highlights news headlines of AI incidents that are relevant to the user's prompt~(\autoref{fig:system-sidebar}\figpart{-B2}).
These incidents comprise the top 30 incident reports that are classified as \vcenteredhbox{\includegraphics[height=9pt]{figures/tag-moderately-relevant}} or \vcenteredhbox{\includegraphics[height=9pt]{figures/tag-remotely-relevant}}, sorted in reverse order based on their embedding's cosine distances to the embedding of the user's prompt.
The thumbnails are color-coded based on the incident's relevancy level.
Users can click the headline or the thumbnail to open the original incident report in a new tab.
These real AI incidents can function as cautionary tales~\cite{madaioAssessingFairnessAI2022, wongTimelinesWorldbuildingActivity2021} reminding users of potential AI harms~(\aptLtoX[graphic=no,type=html]{\textbf{G4}}{\ref{item:goal-engage}}).

\mypar{\usecasepanelnormal{}.}
To help users imagine how their AI prototype may be used in AI applications or features~(\aptLtoX[graphic=no,type=html]{\textbf{G1}}{\ref{item:goal-use-case}}), the \usecasepanel{}~(\autoref{fig:system-sidebar}\figpart{C}) presents a diverse set of potential use cases that are generated by an LLM.
Each use case is shown as a sentence describing how a particular group of end-users could use this AI application in a specific context.
For example, for a writing tutor prompt, a potential use case can be \myquote{teachers use it to provide feedback on student writing.}~(\autoref{fig:system-sidebar}\figpart{-C1}).
We also use an LLM to generate a potential harm that could occur within that use case, shown below the use case sentence.
For example, a harm for the teacher feedback use case can be \myquote{students may feel like they are not getting personalized feedback from their teachers.}
We use few-shot learning to prompt the LLM to generate use cases and harms, whereas we generate use cases, stakeholders, and harms in \windowview{}~(\autoref{sec:system:window}).
We open-source all of our prompts.

To help users assess and organize use cases and harms~(\aptLtoX[graphic=no,type=html]{\textbf{G2}}{\ref{item:goal-explore}}), we also leverage an LLM to categorize each use case as \vcenteredhbox{\includegraphics[height=9pt]{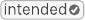}}, \vcenteredhbox{\includegraphics[height=9pt]{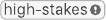}}, or \vcenteredhbox{\includegraphics[height=9pt]{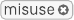}}, although we acknowledge that these may vary by use cases, development and deployment contexts, as well as relevant policies or regulatory frameworks in various jurisdictions.
These three categories are introduced by responsible AI researchers to help ML developers structure their harm envisioning process~\cite{microsoftMicrosoftResponsibleAI2022}.
The \vcenteredhbox{\includegraphics[height=9pt]{figures/tag-intended}} use cases are those that align with the development target use cases.
The \vcenteredhbox{\includegraphics[height=9pt]{figures/tag-high-stakes}} use cases encompass those that may arise in high-stakes domains, such as medicine, finance, and the law.
The \vcenteredhbox{\includegraphics[height=9pt]{figures/tag-misuse}} category includes scenarios where malicious actors exploit the AI application to cause harms.
The \usecasepanel{} organizes use cases and harms into three tabs~(\autoref{fig:system-sidebar}\figpart{-C1--3}) based on their categories.
The first tab, ``mix'', provides an overview by showing one use case and its corresponding harm from each of the other tabs.

\subsection{\windowviewnormal{}}
\label{sec:system:window}

\setlength{\belowcaptionskip}{-8pt}
\setlength{\abovecaptionskip}{8pt}
\begin{figure*}[tb]
  \includegraphics[width=1\linewidth]{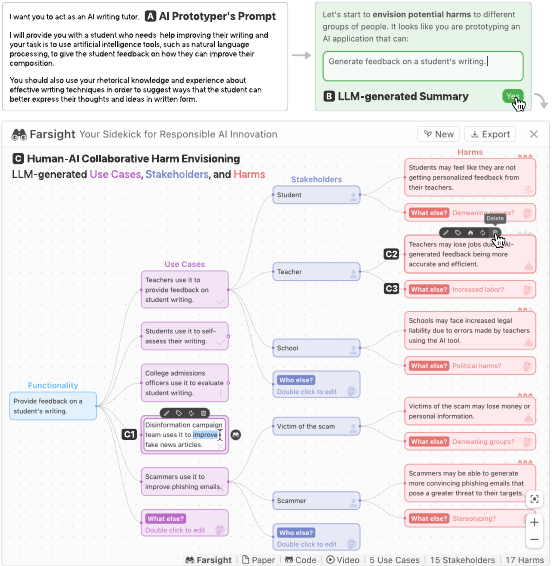}
  \caption[]{
    The \windowview{} helps AI prototypers envision harms associated with their AI applications through human-AI collaboration.
    \textbf{(A)} Given a prompt, \textbf{(B)} \tool{} uses an LLM to generate a summary of the prompt and asks users to revise it.
    \textbf{(C)} Then, the \windowview{} presents an interactive node-link diagram to visualize use cases, stakeholders, and harms.
    Initially, the \windowview{} only shows up to the \textit{Use Cases} layer.
    \textbf{(C1)} Users can edit the node content before asking AI to generate its children nodes by clicking \protect\vcenteredhbox{\includegraphics[height=9pt]{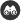}}.
    Users can edit any node and regenerate its children at any time, and click a node to show or hide its descendants.
    \textbf{(C2)} Users can delete unhelpful nodes.
    \textbf{(C3)} This view encourages users to think and add more harms by intermittently and randomly alternating harm categories shown in empty harm nodes, such as ``increased labor?''
  }
  \Description{
    A figure with three components. Component A shows AI prototypers' prompt. Component B shows LLM-generated summary. Component C shows human-AI collaborative harm envisions
  }
  \label{fig:system-window}
\end{figure*}
\setlength{\belowcaptionskip}{0pt}
\setlength{\abovecaptionskip}{12pt}

Both the \symbolview{} and the \sidebarview{} provide easy-to-understand in-context reminders to help users reflect on potential harms associated with their prompts.
However, instead of passively reading AI incident reports and LLM-generated content, users desire to actively edit and add new use cases, stakeholders, and harms~(\autoref{fig:formative-feature}).
Also, active participation---a key factor in learning---may help foster AI prototypers' ability to independently identify harms.
Therefore, we design \windowview{}~(\autoref{fig:system-window}) to support users in actively envisioning potential harms associated with their prompts~(\aptLtoX[graphic=no,type=html]{\textbf{G4}}{\ref{item:goal-engage}}).
We use a real prompt from Awesome ChatGPT Prompts~\cite{akinAwesomeChatGPTPrompts2022} in the example in~\autoref{fig:system-window}.

\mypar{Interactive Node-link Tree Visualization.}
After clicking the ``Envision Consequences \& Harms'' button in the \sidebarview{}, \windowview{} appears as a pop-up window on top of the prompt-crafting tool~(\autoref{fig:system-window}).
It begins with a text box filled with an LLM-generated summary of a user's prompt~(\autoref{fig:system-window}\figpart{B}).
The user is prompted to revise the summary to align with the target task in their prompt.
Next, the window transitions into an interactive node-link tree visualization~\cite{reingoldTidierDrawingsTrees1981}, where the user can pan and zoom to navigate the view~(\autoref{fig:system-window}\figpart{C}).
First, the window shows the user's prompt summary as the root of the tree which is visualized as a text box.
The user can click the root node and the LLM will generate potential use cases of an AI application based on the user's prompt, and the use cases are visualized as the root's children nodes.
Similarly, users can click a generated node and the LLM will generate its children nodes (stakeholders and then harms).
There is a max of four layers, following an order of the user's prompt summary $\rightarrow$ use cases $\rightarrow$ stakeholders $\rightarrow$ harms.
This layer order reflects the recommended harm envisioning workflow in responsible AI literature~\cite{doteveryoneConsequenceScanningAgile2019,microsoftMicrosoftResponsibleAI2022,barocasDesigningDisaggregatedEvaluations2021,madaioAssessingFairnessAI2022, microsoftHarmsModelingAzure2022} and helps users to comprehend and organize diverse harms across different contexts~(\aptLtoX[graphic=no,type=html]{\textbf{G2}}{\ref{item:goal-explore}}).
For additional examples of LLM-generated use cases, stakeholders, and harms in \tool{}, see \autoref{table:farsight-output}.

\setlength{\columnsep}{10pt}%
\setlength{\intextsep}{0pt}%
\begin{wrapfigure}{R}{110pt}
  \vspace{0pt}
  \centering
  \includegraphics[width=110pt]{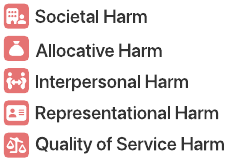}
  \vspace{-20pt}
  \caption[]{
    Icons used to represent different harm types.
  }
  \Description{
    Icons used to represent different harm types.
  }
  \vspace{0pt}
  \label{fig:system-harm-type}
\end{wrapfigure}
\mypar{Human-AI Collaboration in Harm Envisioning.}
Our goal is to use AI-generated harms to encourage users to reflect on potential downstream harms and inspire them to add, edit, or curate potential harms~(\aptLtoX[graphic=no,type=html]{\textbf{G4}}{\ref{item:goal-engage}}).
To do that, \windowview{} allows users to edit any tree nodes by clicking a button in the toolbar~(\autoref{fig:system-window}\figpart{-C1}) or entering new text in the tree node.
In addition, users can delete~(\autoref{fig:system-window}\figpart{-C2}) and use the LLM to regenerate all of an edited node's children nodes, to effectively steer the harm envisioning direction by offering feedback to the LLM~(\aptLtoX[graphic=no,type=html]{\textbf{G4}}{\ref{item:goal-engage}}).
To meet users' needs of categorizing harms~(\aptLtoX[graphic=no,type=html]{\textbf{G2}}{\ref{item:goal-explore}}), we use an LLM to classify each harm into a harm type based on a systematic review and taxonomy of AI harms ~\cite{shelbySociotechnicalHarmsAlgorithmic2023}.
Users can use the dropdown menu to change the harm's category~(\autoref{fig:system-harm-type}).
To help users prioritize and take notes about harms, the \windowview{} allows users to rate the severity of each harm by clicking \vcenteredhbox{\includegraphics[height=8pt]{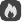}} in the toolbar.
Finally, users can click \vcenteredhbox{\includegraphics[height=10pt]{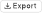}} to export all content~(e.g., use cases, stakeholders, and harms) in the \windowview{} as a Markdown file.

\subsection{Open-source and Reusable Implementation}
\label{sec:system:implementation}
To make \tool{} easily adoptable by both AI prototypers and AI companies~(\aptLtoX[graphic=no,type=html]{\textbf{G5}}{\ref{item:goal-open-source}}), we implement \tool{} to be model-agnostic and environment-agnostic, and we open-source our implementation.
\tool{} uses LLMs by calling their public APIs so that users can use their preferred LLMs by easily replacing the API endpoints.
To help AI companies and researchers integrate \tool{} into AI prototyping tools, we leverage Web Components~\cite{mdnWebComponentsWeb2021} and Lit~\cite{googleLitSimpleFast2015} to implement \tool{} as reusable modules, which can be easily integrated into any web-based interfaces regardless of their development stacks (e.g., React, Vue, Svelte).
To help AI prototypers use our tool, we present a Chrome extension\footnote{\tool{} Chrome extension: \nnlink{https://github.com/PAIR-code/farsight/releases}} that integrates \tool{} into Google AI Studio and a Python package\footnote{\tool{} Python package: \nnlink{https://pypi.org/project/farsight/}} that brings \tool{} to computational notebooks.
We implement the interactive tree visualization using \textit{D3.js}~\cite{bostockDataDrivenDocuments2011} and embedding similarity computation using \textit{TensorFlow.js}~\cite{smilkovTensorFlowJsMachine2019} with WebGL~\cite{mdnWebGL2D3D2011} acceleration.
Computational notebook support is implemented using NOVA~\cite{wangNOVAPracticalMethod2022}.

\section{Usage Scenario}
We present a hypothetical usage scenario to illustrate how \tool{} fosters responsible awareness among AI prototypers.
Rosa is a native English speaker from the United States who recently traveled to Vietnam to teach English.
She is the only English teacher at an under-resourced high school.
Overwhelmed with grading English writing assignments for all students in the school, Rosa tries to develop an LLM-powered AI application that provides writing feedback based on a student's essay.
She writes her prompt~(\autoref{fig:system-sidebar}\figpart{A}) in an AI prototyping tool with \tool{} integrated.
After running the prompt, Rosa notices the alarming \symbolview{}~(\autoref{fig:system-sidebar}\figpart{A}), so she clicks on it, which expands the \sidebarview{}~(\autoref{fig:system-sidebar}\figpart{-BC}).
Rosa reads a few related articles shown in the \incidentpanel{}~(\autoref{fig:system-sidebar}\figpart{-B2}).
She finds these articles are indeed related to AI in education and are helpful, but they mainly focus on students using AI to cheat rather than teachers using AI to grade assignments.
Rosa skims through the LLM-generated potential use cases and finds the use case \myquote{teachers use it to provide feedback on student writing} very relatable~(\autoref{fig:system-sidebar}\figpart{-C1}).
Intrigued by its associated harm \myquote{students may feel like they are not getting personalized feedback from their teachers}, Rosa clicks the Envision Consequences button and wishes to learn more about this use case and its associated potential harms.

\mypar{Harm envisioner.}
Next, \tool{} shows a pop-up window asking Rosa to revise and confirm an LLM-generated summary of her prompt~(\autoref{fig:system-window}\figpart{-B}).
After clicking \vcenteredhbox{\includegraphics[height=10pt]{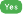}}, Rosa sees the \windowview{} presenting an interactive tree visualization showing the functionality of her AI application as a root node and multiple use cases as its children nodes~(\autoref{fig:system-window}\figpart{-C}).
With a map-like interface, Rosa quickly uses zoom-and-pan to zoom into the teaching use case.
After clicking \vcenteredhbox{\includegraphics[height=9pt]{figures/icon-add}}, the \windowview{} quickly generates the stakeholders associated with the use case and the harms associated with each stakeholder.
Rosa takes some time to reflect on the LLM-generated harm of students not getting personalized feedback~(\autoref{fig:system-window}-\figpart{Harm-1}).
She has never thought about this consequence before, but she thinks it makes sense---AI does not have background knowledge about each student, so its feedback would not be tailored to students' individual needs.
After rating it as very severe~\vcenteredhbox{\includegraphics[height=8pt]{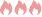}} by clicking \vcenteredhbox{\includegraphics[height=8pt]{figures/icon-rate}}, Rosa continues reading other LLM-generated harms.
She does not think the harm of teachers losing jobs to her AI tutor is relevant, so she deletes it~(\autoref{fig:system-window}-\figpart{C2}).

\mypar{Human-AI collaboration.}
After seeing the random question ``increased labor?'' next to teacher~(\autoref{fig:system-window}-\figpart{C3}), Rosa thinks maybe it will be more time-consuming to review AI-generated feedback than grading students' assignments herself, so she enters that harm into the \windowview{}.
Next, Rosa is not sure about the legal liability of her school~(\autoref{fig:system-window}-\figpart{Harm-3}), but it might be worth discussing with other teachers.
Finally, reflecting on her experience with the \windowview{} and AI incident articles, Rosa thinks the potential harms of her writing tutor AI application outweigh the potential convenience for her.
Therefore, Rosa decides to stop prototyping this application.
However, Rosa still sees value in leveraging LLMs in education, so she bookmarks related AI incident articles and clicks \vcenteredhbox{\includegraphics[height=10pt]{figures/icon-export}} to download all the content in the \windowview{} as a Markdown file.
She will bring these resources to discuss with her colleagues the next day.
\section{Evaluation User Study}
\label{sec:evaluation}

We conducted a user study to evaluate \tool{}'s effectiveness in aiding AI prototypers to anticipate the potential harms associated with AI features.
In addition, we investigate how AI prototypers use \tool{} during an early prototyping stage.
To investigate the effect of user engagement in AI-assisted harm envisioning, we tested two variants of our tool: \toolc{}, including all components, and \litec{}, including only the \symbolview{}~(\autoref{fig:teaser}\figpart{-B}) and the \sidebarview{}~(\autoref{fig:teaser}\figpart{-C}).
In other words, \litec{} is a ``subset'' of \toolc{}.
\litec{} only shows one direct stakeholder for each use case in the \usecasepanel{}, while \toolc{} allows users to interactively add more stakeholders, use cases, and harms in the \windowview{}~(\autoref{fig:teaser}\figpart{-A}).
The study included 42 AI prototypers with diverse roles who were recruited from a large technology company based in the United States.
In this user study, we aimed to investigate the following three research questions:

\aptLtoX[graphic=no,type=html]{
    \begin{itemize}
        \item [\textbf{RQ1.}] How do \toolc{} and \litec{} affect users' ability for and approach to identifying potential harms?
        \item [\textbf{RQ2.}] How effective and useful are \toolc{} and \litec{} in assisting users in envisioning harms in comparison to existing commonly-used resources?
        \item [\textbf{RQ3.}] What challenges do AI prototypers face when envisioning potential harms during the AI prototyping stage? How do \toolc{} and \litec{} help AI prototypers address these challenges?
    \end{itemize}
}{
    \begin{enumerate}[topsep=5pt, itemsep=0mm, parsep=1mm, leftmargin=25pt, label=\textbf{RQ\arabic*.}, ref=RQ\arabic*]
        \item How do \toolc{} and \litec{} affect users' ability for and approach to identifying potential harms? \label{item:q1}
        \item \label{item:q2} How effective and useful are \toolc{} and \litec{} in assisting users in envisioning harms in comparison to existing commonly-used resources?
        \item \label{item:q3} What challenges do AI prototypers face when envisioning potential harms during the AI prototyping stage? How do \toolc{} and \litec{} help AI prototypers address these challenges?
    \end{enumerate}
}

\subsection{Participants}

\setlength{\belowcaptionskip}{0pt}
\setlength{\abovecaptionskip}{7pt}
\begin{table}[tb]
    \caption[]{The evaluation user study included 42 participants with diverse roles and experience in prompting LLMs.}
    \label{table:evaluation-participant}
    \begin{tabular}{l p{145pt}}
        \textbf{Participant Roles} & \textbf{Participant IDs}                                                      \\
        \midrule
        Software Engineer          & 3, 4, 5, 6, 7, 12, 13, 15, 16, 17, 19, 23, 25, 26, 28, 29, 33, 34, 35, 41, 42 \\
        Product Manager            & 1, 8, 10, 11, 14, 20, 24, 27, 36                                              \\
        Linguist                   & 2, 21, 30, 31                                                                 \\
        AI Researcher              & 9, 18, 39, 40                                                                 \\
        UX Researcher              & 22                                                                            \\
        Data Scientist             & 32                                                                            \\
        Test Engineer              & 37                                                                            \\
        Marketing Specialist       & 38                                                                            \\
    \end{tabular}
    \Description{
        A table that links each interview participant with their primary roles.
    }
\end{table}
\setlength{\belowcaptionskip}{0pt}
\setlength{\abovecaptionskip}{12pt}

We recruited 45 voluntary participants from both internal mailing lists related to AI and snowball sampling at Google, based in the United States.
The recruitment required participants to have experience in writing prompts for LLMs.
In total, we received 61 responses, and we selected 45 participants based on their schedule availability.
We conducted pilot studies using the first three study sessions, which were not included in our data analysis.
As a result, we had a total of 42 participants.
Each study session was either 90 minutes ($n$=28 sessions) or 60 minutes ($n$=14 sessions), depending on the participants' availability.
During the 90-minute sessions (or 60-minute sessions), each participant received an average of \$62 USD (or \$41) compensation in their preferred form such as gift cards and charity credits.

\begin{figure}[tb]
    \centering
    \includegraphics[width=1\linewidth]{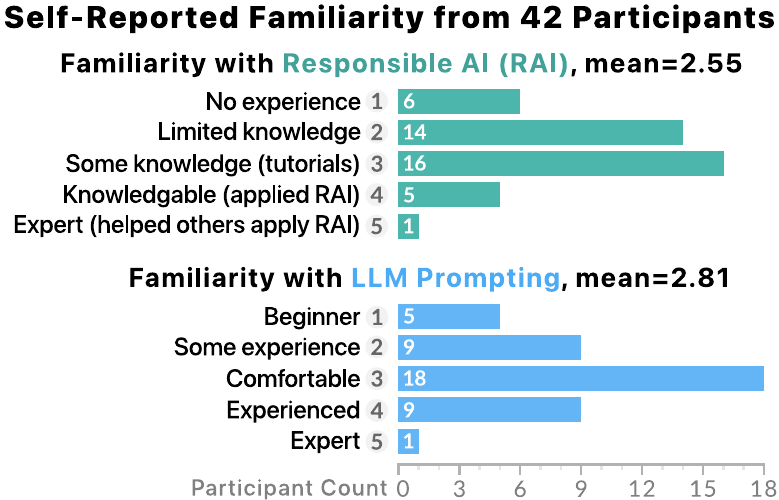}
    \caption[]{
        Participants reported diverse levels of familiarity with responsible AI~(top, average=2.55) and LLM prompting~(bottom, average=2.81) on 5-point Likert scales.
    }
    \Description{This figure is a bar chart titled "Self-Reported Familiarity from 42 Participants". It is divided into two sections: "Familiarity with Responsible AI (RAI)" with a mean of 2.55, and "Familiarity with LLM Prompting" with a mean of 2.81. Each section has a five-point scale ranging from no experience to expert level.

        In the RAI section, six participants report no experience (1), fourteen have limited knowledge (2), sixteen have some knowledge through tutorials (3), five are knowledgeable having applied RAI (4), and one is an expert who has helped others apply RAI (5).

        In the LLM Prompting section, five participants are beginners (1), nine have some experience (2), eighteen feel comfortable with LLM prompting (3), nine are experienced (4), and one is an expert (5).

        The bars represent the number of participants, with the length of the bar corresponding to the count. The colors of the bars are green for RAI familiarity and blue for LLM Prompting familiarity. This chart visually summarizes the self-assessed proficiency levels in these two areas among the participants.}
    \label{fig:evaluation-experience}
\end{figure}

Among the 42 participants, 26 identified as men, 14 as women, and 2 preferred not to disclose their gender.
Information about their job roles is listed in \autoref{table:evaluation-participant}.
Recruited participants self-reported an average score of 2.55 for their knowledge and experience with responsible AI on a 5-point Likert scale~(\autoref{fig:evaluation-experience}\figpart{-top}), where 1 represents ``No experience'' and 5 represents ``Expert (I have helped others apply responsible AI practices).''
In addition, participants self-reported an average score of 2.81 for experience with LLM prompting on a 5-point Likert scale~(\autoref{fig:evaluation-experience}\figpart{-bottom}), where 1 represents ``Beginner'' and 5 represents ``Expert.''
All participants are \tool{}'s targeted users, AI prototypers.

\setlength{\belowcaptionskip}{0pt}
\setlength{\abovecaptionskip}{5pt}
\begin{figure*}[tb]
    \includegraphics[width=1\linewidth]{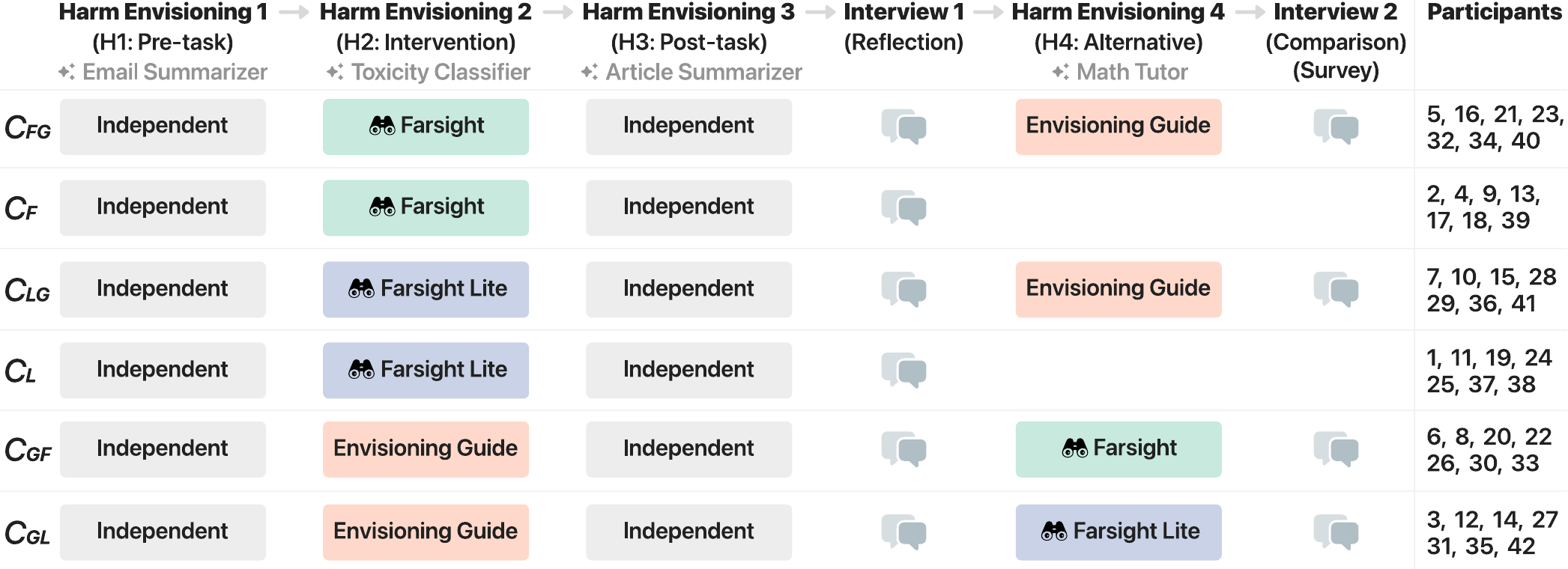}
    \caption[]{
        The evaluation study included six conditions with different variations of harm envisioning tools (\toolc{}, \litec{}, and the baseline \guidec{}).
        Participants were asked to envision potential harms associated with an AI feature (e.g., email summarizer) in each harm-envisioning activity~(H1, H2, H3, and H4).
        Participants had access to a harm envisioning tool in H2 and H4.
        The duration of sessions involving H4 and interview 2 was 90 minutes, while all other sessions lasted 60 minutes.
        Participants were randomly assigned to a condition, taking into account their availability for study session duration.
    }
    \Description{
        A figure showing the workflow of the evaluation user study with H1: pre-task, H2: intervention, H3: post-task, interview 1: reflection, H4: alternative, interview 2: comparison and survey, and the corresponding participant IDs.
    }
    \label{fig:evaluation-study-design}
\end{figure*}
\setlength{\belowcaptionskip}{0pt}
\setlength{\abovecaptionskip}{12pt}

\setlength{\belowcaptionskip}{-8pt}
\setlength{\abovecaptionskip}{5pt}
\setlength{\belowcaptionskip}{0pt}
\setlength{\abovecaptionskip}{12pt}

\subsection{Study Design}

We conducted this study with participants one-on-one.
Out of 42 sessions, 2 were conducted in-person, and 40 were through video conferencing software due to office locations and participants' scheduling constraints.
With the permission of all participants, we recorded the participants' audio and computer screen for subsequent analysis.
To start, each participant signed a consent form and filled out a survey regarding their familiarity with responsible AI and LLM prompting~(\autoref{fig:evaluation-experience}).
Then, participants were randomly assigned to one of six conditions taking into account their time availability: \conditionfg{}, \conditionf{}, \conditionlg{}, \conditionl{}, \conditiongf{}, \conditiongl{}~(\autoref{fig:evaluation-study-design}).
\textit{C} stands for the study condition, \conditionfg{} means that participants used \toolc{} first and then \guidec{}, and \conditionl{} means that participants only used \litec{}---the other acronyms follow this same pattern.
Sessions of \conditionf{} and \conditionl{} were scheduled for 60 minutes each, while the remaining sessions were allotted 90 minutes each.
We assigned 7 participants to each condition, as this was the maximum number that allowed for an equal distribution of participants across all conditions, given the time constraints and the availability of the 61 individuals who signed up for the study.

Our study followed a mixed design that combines both between-subjects and within-subjects designs~\cite{seltmanExperimentalDesignAnalysis2012}.
Each session included three or four harm-envisioning activities, denoted as \aptLtoX[graphic=no,type=html]{H1}{\hone{}}, \aptLtoX[graphic=no,type=html]{H2}{\htwo{}}, \aptLtoX[graphic=no,type=html]{H3}{\hthree{}}, and \aptLtoX[graphic=no,type=html]{H4}{\hfour{}}~(\autoref{sec:evaluation-design-envision}), as well as one or two semi-structured interviews to collect participants' feedback~(\autoref{sec:evaluation-design-interview}).
In each harm-envisioning activity, participants were asked to envision potential harms associated with a particular AI feature while thinking aloud~(\autoref{fig:evaluation-study-design}).
In \aptLtoX[graphic=no,type=html]{H1}{\hone{}} and \aptLtoX[graphic=no,type=html]{H3}{\hthree{}}, participants envisioned harms on their own, whereas in \aptLtoX[graphic=no,type=html]{H2}{\htwo{}} and \aptLtoX[graphic=no,type=html]{H4}{\hfour{}}, they could use a harm envisioning tool we assigned them based on their study condition (e.g., \toolc{}, \litec{}, or \guidec{}).
All collected harms were rated by seven researchers with experience with responsible AI evaluations, who assigned each potential harm numeric scores in terms of their likelihood and severity~(\autoref{sec:evaluation-design-rating}).
We compared the envisioned harms in \aptLtoX[graphic=no,type=html]{H1}{\hone{}} and \aptLtoX[graphic=no,type=html]{H3}{\hthree{}} (between-subjects) to investigate how different tools affect users' ability and approach to anticipating harms~(\aptLtoX[graphic=no,type=html]{\textbf{RQ1}}{\ref{item:q1}}).
We compared the envisioned harms in \aptLtoX[graphic=no,type=html]{H2}{\htwo{}} and \aptLtoX[graphic=no,type=html]{H4}{\hfour{}} (within-subjects) to assess the effectiveness of different tools in helping users envision harms~(\aptLtoX[graphic=no,type=html]{\textbf{RQ2}}{\ref{item:q2}}).
Besides the quantitative data on the number and ratings of potential harms, we also collected qualitative data from think-aloud and two interviews~(\aptLtoX[graphic=no,type=html]{\textbf{RQ1}--\textbf{RQ3}}{\ref{item:q1}--\ref{item:q3}}).
We incorporated 60-minute sessions (\conditionf{} and \conditionl{}) into our study design due to challenges in recruiting participants available for a 90-minute duration.\looseness=-1

\setlength{\belowcaptionskip}{0pt}
\setlength{\abovecaptionskip}{5pt}
\begin{figure*}[tb]
    \includegraphics[width=1\linewidth]{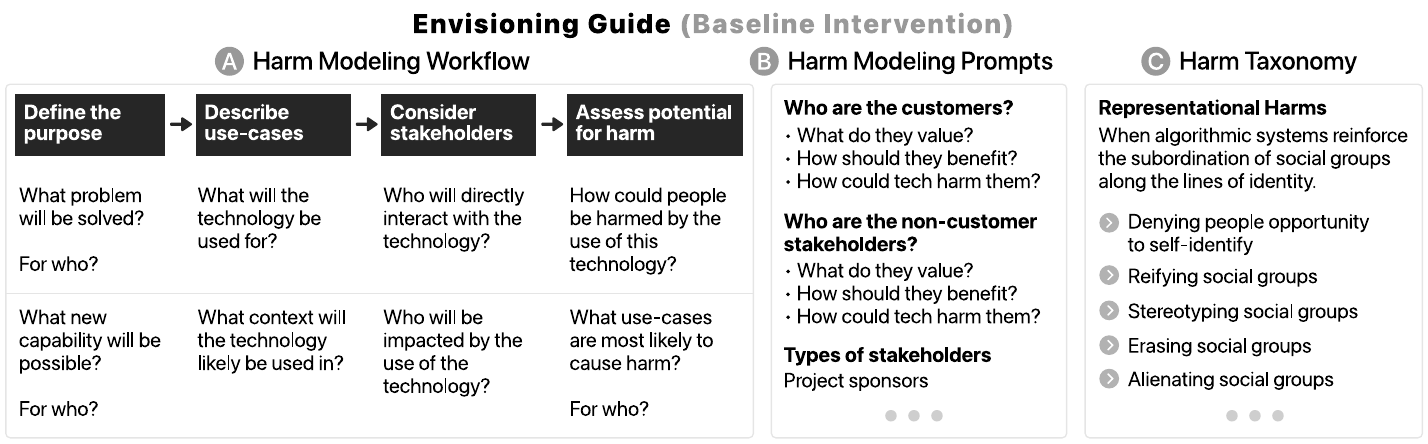}
    \caption[]{
        In the evaluation user study, we compared our tools against \guidec{}, a combination of existing harm envisioning resources.
        This \guidec{} was presented to participants as a Google Doc with three sections.
        \textbf{(A)} The harm modeling workflow table comes from Microsoft's Harm Modeling Practice~\cite{microsoftHarmsModelingAzure2022}, providing a four-step process to envision harms.
        \textbf{(B)} The harm modeling prompts from the Harm Modeling Practice~\cite{microsoftHarmsModelingAzure2022} offer templates and questions to help users envision different stakeholders and use cases (not all content is displayed here).
        \textbf{(C)} The harm taxonomy~\cite{shelbySociotechnicalHarmsAlgorithmic2023} helps participants explore the space of potential harms by providing a comprehensive list of 20 harm categories organized into five themes (not all content is displayed here).
        Participants could click the \protect\vcenteredhbox{\includegraphics[height=8pt]{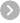}} icon to see the definition of each harm category.
    }
    \Description{
        A figure showing different components in the harm envisioning guide. A shows harm modeling workflow; B shows harm modeling prompts; C shows the harm taxonomy,
    }
    \label{fig:evaluation-envisioning-guide}
\end{figure*}
\setlength{\belowcaptionskip}{0pt}
\setlength{\abovecaptionskip}{12pt}

\subsubsection{Baseline Harm Envisioning Tool.}
To compare our work against current responsible AI workflows, we created a baseline intervention \guidec{}: a combination of Microsoft's Harm Modeling Practice~\cite{microsoftHarmsModelingAzure2022} and the Harm Taxonomy from~\citet{shelbySociotechnicalHarmsAlgorithmic2023}.
These two resources are the latest and the most representative resources designed to help practitioners envision harms.
We combined them because (1) we aim to simulate the current practice where AI prototypers can choose from various existing harm envisioning tools, and (2) we do not intend to study the causal effects of any specific resource.
We administered this intervention by providing a Google Doc containing a detailed table and information from these resources~(\autoref{fig:evaluation-envisioning-guide}).
Both resources were designed to help technology developers and researchers anticipate and prevent negative societal impacts of their technology innovations.

\subsubsection{Harm Envisioning Activities}
\label{sec:evaluation-design-envision}

Depending on the conditions, the study included three or four harm envisioning activities~(\aptLtoX[graphic=no,type=html]{H1}{\hone{}}--\aptLtoX[graphic=no,type=html]{H4}{\hfour{}}).
Within each harm envisioning activity, participants were presented with a description of an AI feature and the prompt that generated that feature.
We chose the four AI features~(\autoref{fig:evaluation-study-design}) based on a qualitative analysis of 100 randomly sampled internal prompts written by real AI prototypers.
These four features are representative of popular LLM tasks~(e.g., summarization, classification, and question answering) and comprehensible to participants with diverse roles.
In \aptLtoX[graphic=no,type=html]{H1}{\hone{}} and \aptLtoX[graphic=no,type=html]{H3}{\hthree{}}, participants independently envisioned harms, whereas in \aptLtoX[graphic=no,type=html]{H2}{\htwo{}} and \aptLtoX[graphic=no,type=html]{H4}{\hfour{}}, they were provided with a harm envisioning assistance tool (e.g., \toolc{}, \litec{}, or \guidec{}).
To emulate AI prototyping workflows, we asked participants to perform simple prompt engineering tasks in \aptLtoX[graphic=no,type=html]{H2}{\htwo{}} and \aptLtoX[graphic=no,type=html]{H4}{\hfour{}} before envisioning potential harms of presented AI features.

For each harm, participants were instructed to describe \textit{who} would be affected (i.e., the stakeholders) and \textit{how} the stakeholder might be harmed.
We provided a harm example for a code generation AI feature: ``App end-users might face financial loss due to AI-introduced software vulnerabilities.''
During the process, participants were asked to share their screens and verbalize their thoughts.
They were also asked to enter their envisioned harms into a Google Doc table featuring a \textit{who} column and a \textit{how} column.
Moreover, participants had the option to articulate the harm verbally, and we transcribed it into the table.
At the end of each harm envisioning activity, we reviewed the table together with the participants to ensure the accuracy of the harm descriptions.
Participants were instructed to achieve three objectives: (1) envision as many harms as possible; (2) envision the most likely harms; and (3) envision the most severe harms.

\mypar{H1: Pre-task.}
\aptLtoX[graphic=no,type=html]{}{\xlabel[H1]{sec:harm-envision-one}}
To understand how participants independently envision potential harms \textit{before} using the tool, as a baseline for~\aptLtoX[graphic=no,type=html]{\textbf{RQ1}}{\ref{item:q1}}, participants were asked to anticipate potential harms concerning an LLM-powered email summarizer on their own~(\autoref{fig:evaluation-study-design}).
They received information about the AI functionality: \textit{``Shorten and improve a user's email''}, a development context, and a prompt that enables this functionality~(see details in~\autoref{appendix:evaluation-design-pre}).
The duration of this activity was limited to 10 minutes.

\mypar{H2: Intervention.}
\aptLtoX[graphic=no,type=html]{}{\xlabel[H2]{sec:harm-envision-two}}
In the second harm envisioning activity, we asked participants to use different harm envisioning assistant tools.
Depending on the assigned condition, a participant could use \toolc{}~(\conditionfg{}, \conditionf{}), \litec{}~(\conditionlg{}, \conditionl{}), or \guidec{}~(\conditiongf{}, \conditiongl{}) to help them anticipate harms.
The activity began with a tutorial on the designated tool.
The AI feature used in this activity was an LLM-powered toxicity classifier~(\autoref{fig:evaluation-study-design}).
Participants received information regarding the AI functionality \textit{``Detect toxic text content,''} a development context, and a prompt that enables this AI functionality~(\autoref{appendix:evaluation-design-intervention}).
To emulate AI prototyping workflows, we also tasked participants with a simple prompt engineering assignment~(\autoref{appendix:evaluation-design-intervention}).

After completing prompt engineering, participants envisioned harms linked to the toxicity classifier.
They were instructed to freely use the assigned tools while sharing their screens and thinking aloud.
For participants assigned with \guidec{}~(\conditiongf{}, \conditiongl{}), the process of entering envisioned harms was the same as \aptLtoX[graphic=no,type=html]{H1}{\hone{}}.
Participants assigned with \toolc{}~(\conditionfg{}, \conditionf{}) or \litec{}~(\conditionlg{}, \conditionl{}) could click a button in the tools to export all harms as a text file.
The export included both AI-generated harms and harms added or modified by participants.
Participants were asked to copy the harms into the Google Doc.
As a significant portion of these harms were generated by AI, we asked participants to select harms that (1) they agreed with and (2) would report to their colleagues and managers.
Also, participants were welcome to add more harms to the table.
For our analysis, we only included the exported harms that participants had selected and added to the table.
The duration of this activity was limited to 25 minutes.

\mypar{H3: Post-task.}
\aptLtoX[graphic=no,type=html]{}{\xlabel[H3]{sec:harm-envision-three}}
To understand how the intervention may have affected participants' ability to independently envision harms~(\aptLtoX[graphic=no,type=html]{\textbf{RQ1}}{\ref{item:q1}}), we asked participants to envision harms associated with an LLM-powered article summarizer on their own~(\autoref{fig:evaluation-study-design}).
To ensure a valid comparison between the envisioned harms and participants' approaches to the pre-task~(\aptLtoX[graphic=no,type=html]{H1}{\hone{}}), we introduced a parallel AI summarizer feature in this activity that was isomorphic to the pre-task~\cite{priceEngagingStudentsInstructor2020}.
In particular, to deter participants from directly reusing their envisioned harms from \aptLtoX[graphic=no,type=html]{H1}{\hone{}}, we replaced the email summarizer in \aptLtoX[graphic=no,type=html]{H1}{\hone{}} with an article summarizer.
The AI functionality was described as ``\textit{Summarize an article in one sentence}''.
The development context and prompt are available in~\autoref{appendix:evaluation-design-post}.
The duration of this activity was limited to 10 minutes.

\mypar{H4: Alternative.}
\aptLtoX[graphic=no,type=html]{}{\xlabel[H4]{sec:harm-envision-four}}
To assess the effectiveness and usefulness of \toolc{} and \litec{} in comparison to \guidec{}~(\aptLtoX[graphic=no,type=html]{\textbf{RQ2}}{\ref{item:q2}}) and study the usage patterns of different tools~(\aptLtoX[graphic=no,type=html]{\textbf{RQ3}}{\ref{item:q3}}), $n=28$ participants engaged in 90-minute sessions~(\conditionfg{}, \conditionlg{}, \conditiongf{}, and \conditiongl{}) to envision harms using a tool different from the one used in~\aptLtoX[graphic=no,type=html]{H2}{\htwo{}}~(\autoref{fig:evaluation-study-design}).
Participants were asked to envision potential harms associated with an LLM-powered math tutor app with the AI functionality \textit{``Answer math-related questions''}, a development context, and a prompt~(\autoref{appendix:evaluation-design-alternative}).
The procedure for this activity paralleled \aptLtoX[graphic=no,type=html]{H2}{\htwo{}}, including a tutorial, prompt engineering exercise~(\autoref{appendix:evaluation-design-alternative}), and harm envisioning.
This activity's duration was limited to 25 minutes.

\subsubsection{Semi-structured Interviews}
\label{sec:evaluation-design-interview}

This study included two semi-structured interview sessions~(\autoref{fig:evaluation-study-design}).
The first interview took place after the post-task activity~(\aptLtoX[graphic=no,type=html]{H3}{\hthree{}}), where we asked participants to reflect on their process for anticipating potential harms during the LLM prototyping process, and how their approach may have changed after the intervention~(\aptLtoX[graphic=no,type=html]{\textbf{RQ1}}{\ref{item:q1}}).
We also asked participants about their challenges in harm anticipation, their experiences of using harm envisioning tools, and potential actions they would take to address the identified harms~(\aptLtoX[graphic=no,type=html]{\textbf{RQ3}}{\ref{item:q3}}, \autoref{appendix:evaluation-interview-1}).
After participants in 90-minute sessions~(\conditionfg{}, \conditionlg{}, \conditiongf{}, and \conditiongl{}) finished \aptLtoX[graphic=no,type=html]{H4}{\hfour{}}, we asked them to compare and rate the usefulness and usability of the two tools they had used in this study~(\aptLtoX[graphic=no,type=html]{\textbf{RQ2}}{\ref{item:q2}}).
We also asked them to rate the helpfulness of different components in the tools on a 5-point Likert scale, as elucidated in \autoref{appendix:evaluation-interview-2}.

\subsubsection{Harm Rating}
\label{sec:evaluation-design-rating}

After completing all 42 study sessions, we recruited seven raters to rate all 989 harms collected in \aptLtoX[graphic=no,type=html]{H1}{\hone{}}--\aptLtoX[graphic=no,type=html]{H4}{\hfour{}} to evaluate participants' ability to envision harms.
These seven raters included four of the paper authors and three industry researchers; all raters had experience with responsible AI (unlike many of the participants)---either as responsible AI researchers, developers of responsible AI tools or playbooks, or in a consultant role on responsible AI for product teams.
Ideally, evaluations of identified harms would involve both domain experts for the domain in question (e.g., education) and/or stakeholders from demographic groups or communities who may be likely to experience those harms.
For this preliminary study, due to timing and resource constraints, we recruited responsible AI researchers as raters instead of specific domain experts or people impacted by AI applications.
The limitations of this approach are further discussed in~\autoref{sec:evaluation-study-limitation} and \autoref{sec:discussion-subjectivity}.

Our collected harms were either (1) directly envisioned by participants or (2) exported from \toolc{} or \litec{} and subsequently curated by participants during \aptLtoX[graphic=no,type=html]{H2}{\htwo{}} and \aptLtoX[graphic=no,type=html]{H4}{\hfour{}}.
Each harm included the impacted stakeholders and a description of the harm.
After removing duplicates and random shuffling, we randomly and evenly assigned harms to raters via spreadsheet format.
Raters had access to the details of the intended AI feature of each harm, including the prompt and the context of the AI feature~(e.g., the prompt and context in \autoref{appendix:evaluation-design-pre}).
To prevent the raters from being influenced by our hypotheses, we did not include the experimental conditions in the rating sheet.
In other words, raters did not know if a harm was from a \toolc{} user, a \litec{} user, or a \guidec{} user.
To mitigate rating noise, we designated three raters for each harm.
As identifying \textit{likely} and \textit{severe} harms is often an objective in AI harm envisioning exercises~\cite{microsoftHarmsModelingAzure2022, rajiClosingAIAccountability2020}, we asked raters to rate each harm's \textit{likelihood} and \textit{severity} on a 4-point Likert scale (\textit{strongly agree}, \textit{agree}, \textit{disagree}, and \textit{strongly disagree} to statements ``This harm is likely to occur for this stakeholder'' and ``This harm will severely impact this stakeholder'').
Raters could also choose an N/A option if they perceived a rating was not applicable for that feature or use case.
During data analysis, we numericalized these four categories as ordinal scores: \texttt{1}, \texttt{2}, \texttt{3}, \texttt{4} and removed N/As.
See \autoref{table:participant-harm} for a random subset of harms that were collected from participants and their corresponding ratings.

\subsection{Data Analysis}

We applied a mixed-methods approach for data analysis.
First, we conducted a quantitative analysis~(\autoref{sec:evaluation-analysis-quantitative}) on the changes in participants' ability to envision harms by comparing pre-task \aptLtoX[graphic=no,type=html]{H1}{\hone{}} to post-task \aptLtoX[graphic=no,type=html]{H3}{\hthree{}} responses~(\aptLtoX[graphic=no,type=html]{\textbf{RQ1}}{\ref{item:q1}}).
We also quantitatively assess three different tools' effectiveness in helping users anticipate harms by comparing \aptLtoX[graphic=no,type=html]{H2}{\htwo{}} and \aptLtoX[graphic=no,type=html]{H4}{\hfour{}} responses~(\aptLtoX[graphic=no,type=html]{\textbf{RQ2}}{\ref{item:q2}}).
The quantitative analyses involved metrics such as the total number of envisioned harms, as well as the average likelihood and severity ratings of envisioned harms across 3 raters.
Next, we performed a qualitative analysis~(\autoref{sec:evaluation-analysis-qualitative}) on transcripts from think-aloud sessions and interviews to further investigate participants' strategies and challenges in envisioning harms, and usage patterns of different tools~(\aptLtoX[graphic=no,type=html]{\textbf{RQ1}--\textbf{RQ3}}{\ref{item:q1}--\ref{item:q3}}).

\aptLtoX[graphic=no,type=html]{
    \begin{figure*}
        \includegraphics[width=1\linewidth]{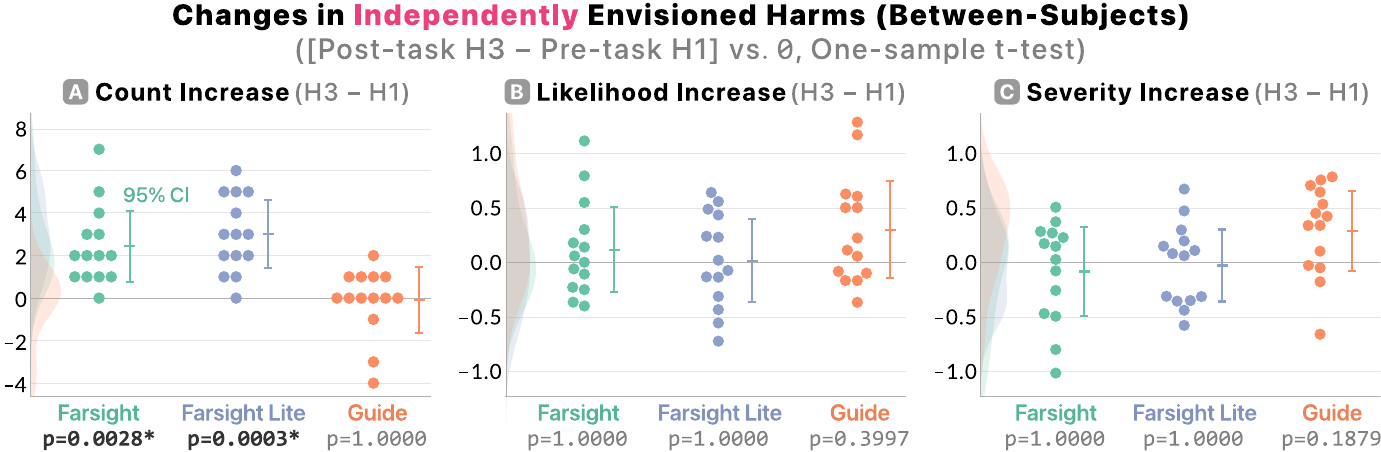}
        \caption{To evaluate how different interventions (\toolc{}, \litec{}, \guidec{}) affect users' ability to envision harms independently, we conducted one-sample \textit{t}-tests with Bonferroni correction to examine the \textit{difference} in the (A) count, (B) average likelihood, and (C) average severity of participant-identified harms between H3 and H1. Each intervention had n = 14 participants, represented by 14 points on the chart. The charts also indicated the 95\% confidence intervals, adjusted with Bonferroni correction. The results highlighted that after using \toolc{} and \litec{}, users could anticipate a significantly higher number of harms, while the average likelihood and severity of identified harms remained the same.}
        \Description{A figure shows the changes in independently envisioned harms. A: count increase, B: likelihood increase, C: severity increase.}
        \label{fig:evaluation-harm-changes}
    \end{figure*}
}{
    \setlength{\belowcaptionskip}{-2pt}
    \setlength{\abovecaptionskip}{5pt}
    \begin{figure*}
        \includegraphics[width=1\linewidth]{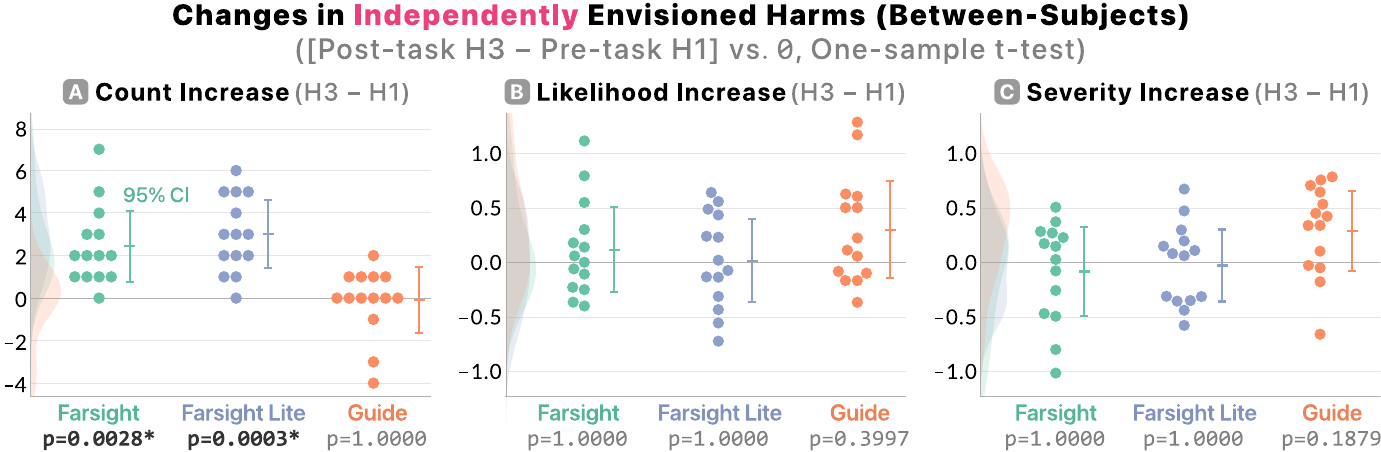}
        \caption[]{
            To evaluate how different interventions (\toolc{}, \litec{}, \guidec{}) affect users' ability to envision harms independently, we conducted one-sample \textit{t}-tests with Bonferroni correction to examine the \textit{difference} in the (A) count, (B) average likelihood, and (C) average severity of participant-identified harms between \hthree{} and \hone{}. Each intervention had n = 14 participants, represented by 14 points on the chart. The charts also indicated the 95\% confidence intervals, adjusted with Bonferroni correction. The results highlighted that after using \toolc{} and \litec{}, users could anticipate a significantly higher number of harms, while the average likelihood and severity of identified harms remained the same.}
        \Description{A figure shows the changes in independently envisioned harms. A: count increase, B: likelihood increase, C: severity increase.}
        \label{fig:evaluation-harm-changes}
    \end{figure*}
    \setlength{\belowcaptionskip}{0pt}
    \setlength{\abovecaptionskip}{12pt}
}

\subsubsection{Quantitative Analysis.}
\label{sec:evaluation-analysis-quantitative}
We first conducted quantitative analyses on the count, likelihood, and severity of harms across different conditions to evaluate the effectiveness of our tools~(\aptLtoX[graphic=no,type=html]{\textbf{RQ1}}{\ref{item:q1}}, \aptLtoX[graphic=no,type=html]{\textbf{RQ2}}{\ref{item:q2}}).
We measured the likelihood and severity for each harm using the average of ratings from three raters after removing any N/As.
The average pairwise weighted Cohen's kappas~\cite{mchughInterraterReliabilityKappa2012, cohenWeightedKappaNominal1968} for likelihood and severity ratings are \texttt{0.14} and \texttt{0.09}~(see \autoref{fig:appendix-rater} and \autoref{appendix:evaluation-ratings} for details).
These values fall within the range of slight agreement~\cite{landisMeasurementObserverAgreement1977}.
We discuss this relatively low inter-rater agreement in~\autoref{sec:discussion-subjectivity}.
The Shapiro-Wilk normality tests~\cite{shapiroAnalysisVarianceTest1965} show all measures, except for the changes of harm count between \aptLtoX[graphic=no,type=html]{H1}{\hone{}} and \aptLtoX[graphic=no,type=html]{H3}{\hthree{}} with \guidec{}, follow a normal distribution.
We used \textit{t}-tests with Bonferroni corrections for multiple hypothesis testing.

We also analyzed participants' ratings of the tools' usefulness and usability when comparing the two tools used in the study~(\aptLtoX[graphic=no,type=html]{\textbf{RQ2}}{\ref{item:q2}}, \autoref{sec:envision-pattern-usability}).
We converted the 5-point Likert scale ratings into numerical values and assessed the difference between ratings of our tools and \guidec{} using Mann-Whitney U tests~\cite{mannTestWhetherOne1947}.
Considering that most of the ratings did not exhibit a normal distribution, we chose to use Mann-Whitney U tests, as these tests do not assume normality in the data.
See \autoref{sec:envision-pattern-usability} for discussion of the findings from these questions about usefulness and usability.

\subsubsection{Qualitative Analysis.}
\label{sec:evaluation-analysis-qualitative}

We conducted a qualitative analysis on the screen recordings and transcripts of the study sessions that include participants' verbalized thoughts during the harm envisioning activities~(\aptLtoX[graphic=no,type=html]{H1}{\hone{}}--\aptLtoX[graphic=no,type=html]{H4}{\hfour{}}) and interviews.
All study sessions were screen-recorded and audio-recorded, with the audio subsequently transcribed by the video conferencing software.
We adopted an inductive thematic analysis approach~\cite{merriamIntroductionQualitativeResearch2002, braunUsingThematicAnalysis2006} and open coded the 56-hour-long transcripts using the qualitative analysis software Dovetail~\cite{dovetailDovetailAllYour2023}.
After generating a codebook, we applied deductive coding~\cite{merriamIntroductionQualitativeResearch2002} to assign harm envisioning patterns to each participant during \aptLtoX[graphic=no,type=html]{H1}{\hone{}} and \aptLtoX[graphic=no,type=html]{H3}{\hthree{}}~(\aptLtoX[graphic=no,type=html]{\textbf{RQ1}}{\ref{item:q1}}, \autoref{sec:evaluation-observed-change}).

\subsection{Findings: Changes in Users' Envisioning Ability and Approach (RQ1)}
\label{sec:evaluation-finding-change}

In the study, participants were asked to independently envision harms associated with an email summarizer~(\aptLtoX[graphic=no,type=html]{H1}{\hone{}}) and an article summarizer~(\aptLtoX[graphic=no,type=html]{H3}{\hthree{}}) before and after using a harm envisioning tool (\toolc{}, \litec{}, or \guidec{}) to anticipate harms for a toxicity classifier~(\aptLtoX[graphic=no,type=html]{H2}{\htwo{}}).
We quantitatively and qualitatively compared participants' envisioned harms and approaches in \aptLtoX[graphic=no,type=html]{H1}{\hone{}} and \aptLtoX[graphic=no,type=html]{H3}{\hthree{}} across different conditions in \aptLtoX[graphic=no,type=html]{H2}{\htwo{}}.

\subsubsection{\textbf{}\toolc{} and \litec{} Improved Users' Ability to Envision Harms.}
\label{sec:evaluation-finding-rq1-quantitative}
For each participant, we compared the count, average likelihood, and average severity of their independently envisioned harms before~(\aptLtoX[graphic=no,type=html]{H1}{\hone{}}) and after~(\aptLtoX[graphic=no,type=html]{H3}{\hthree{}}) the intervention~(\autoref{fig:evaluation-harm-changes}).
Using paired sample \textit{t}-tests with Bonferroni correction~\cite{dunnMultipleComparisonsMeans1961}, we found that after using \toolc{} and \litec{}, users could envision significantly more harms on their own~(\toolcolor{$p=0.0028$}, \litecolor{$p=0.0003$}), showing an average increase of \toolcolor{2.42} and \litecolor{3.00} harms, respectively.
The effect sizes, as measured by Cohen's $d$~\cite{cohenStatisticalPowerAnalysis2013}, were \toolcolor{$d=1.21$} and \litecolor{$d=1.27$}, indicating a very large effect~\cite{sawilowskyNewEffectSize2009}.
On the contrary, for participants using \guidec{}, the average count of identified harms experienced a marginal decrease~(\guidecolor{$-0.14$}).
We hypothesize that the observation of three participants identifying fewer harms after using \guidec{} (see the outliers in \autoref{fig:evaluation-harm-changes}~\figpart{A}) is because \guidec{} had a high cognitive load.
The high cognitive load may have resulted in these three participants having less energy to envision harms in \aptLtoX[graphic=no,type=html]{H3}{\hthree{}} compared to \aptLtoX[graphic=no,type=html]{H1}{\hone{}}.
Changes in the average likelihood and average severity, on the other hand, were not statistically significant for any of the interventions~(\autoref{fig:evaluation-harm-changes}-\figpart{BC}).
Our finding implies that after using \toolc{} and \litec{}, users could anticipate a greater number of harms linked to AI features independently, while the average likelihood and severity of identified harms remained unaltered.

\setlength{\belowcaptionskip}{0pt}
\setlength{\abovecaptionskip}{2pt}

\aptLtoX[graphic=no,type=html]{
    \begin{table*}
        \def\arraystretch{1.5}
        \caption{We identified six non-exclusive common patterns in independent harm envisioning by analyzing transcripts of participants' think-aloud process during the harm envisioning activities in H1 and H3.}
        \label{table:evaluation-finding-pattern}
        \begin{tabular}{l p{350pt}}
            \textbf{Harm Envisioning Pattern}        & \textbf{Description}                                                                                                                                                                                                                                                                                                                         \\
            \midrule
            \textbf{Failure-mode-driven envisioning} & Participants envisioned harm by initially considering the AI feature's failure modes (e.g., wrong summarization), limitations of LLMs (e.g., hallucination), or vulnerabilities within system implementation (e.g., data storage). This pattern is similar to a Failure Mode and Effects Analysis~\cite{rismaniPlaneCrashesAlgorithmic2023}. \\\hline
            \textbf{Usage-driven envisioning}        & Participants envisioned harm by initially considering who may be impacted through this feature and in what usage scenario, such as students using the article summarizer for completing assignments.
            Then, participants envisioned potential harms that might impact the stakeholders within the identified scenario.                                                                                                                                                                                                                                                                        \\\hline
            \textbf{Consider high-stakes uses}       & Participants deliberately thought about high-stakes use cases of the AI feature, such as being used in medical, financial, and legal domains.                                                                                                                                                                                                \\\hline
            \textbf{Consider misuses}                & Participants deliberately envisioned potential misuse of the AI feature, where malicious actors like scammers and hackers could exploit this AI feature to cause harm.                                                                                                                                                                       \\\hline
            \textbf{Consider indirect stakeholders}  & Participants deliberately brainstormed stakeholders indirectly impacted by the AI feature, such as people who did not use the AI tools, individuals mentioned in the input text, and society at large.                                                                                                                                       \\\hline
            \textbf{Consider cascading harms}        & Participants deliberately considered (1) harms that could result from other harms, such as productivity loss due to AI errors can lead to economic loss; or (2) harms that might occur even when the AI feature operated as expected, such as students using AI to cheat in homework.                                                        \\
        \end{tabular}
        \Description{A table with 7 rows and 2 columns. Two columns are harm envisions pattern and description.}
    \end{table*}
}{
    \begin{table*}
        \def\arraystretch{1.5}
        \caption{We identified six non-exclusive common patterns in independent harm envisioning by analyzing transcripts of participants' think-aloud process during the harm envisioning activities in \hone{} and \hthree{}.}
        \label{table:evaluation-finding-pattern}
        \begin{tabular}{l p{350pt}}
            \textbf{Harm Envisioning Pattern}        & \textbf{Description}                                                                                                                                                                                                                                                                                                                         \\
            \midrule
            \textbf{Failure-mode-driven envisioning} & Participants envisioned harm by initially considering the AI feature's failure modes (e.g., wrong summarization), limitations of LLMs (e.g., hallucination), or vulnerabilities within system implementation (e.g., data storage). This pattern is similar to a Failure Mode and Effects Analysis~\cite{rismaniPlaneCrashesAlgorithmic2023}. \\\arrayrulecolor{grayIV}\hline
            \textbf{Usage-driven envisioning}        & Participants envisioned harm by initially considering who may be impacted through this feature and in what usage scenario, such as students using the article summarizer for completing assignments.
            Then, participants envisioned potential harms that might impact the stakeholders within the identified scenario.                                                                                                                                                                                                                                                                        \\\arrayrulecolor{grayIV}\hline
            \textbf{Consider high-stakes uses}       & Participants deliberately thought about high-stakes use cases of the AI feature, such as being used in medical, financial, and legal domains.                                                                                                                                                                                                \\\arrayrulecolor{grayIV}\hline
            \textbf{Consider misuses}                & Participants deliberately envisioned potential misuse of the AI feature, where malicious actors like scammers and hackers could exploit this AI feature to cause harm.                                                                                                                                                                       \\\arrayrulecolor{grayIV}\hline
            \textbf{Consider indirect stakeholders}  & Participants deliberately brainstormed stakeholders indirectly impacted by the AI feature, such as people who did not use the AI tools, individuals mentioned in the input text, and society at large.                                                                                                                                       \\\arrayrulecolor{grayIV}\hline
            \textbf{Consider cascading harms}        & Participants deliberately considered (1) harms that could result from other harms, such as productivity loss due to AI errors can lead to economic loss; or (2) harms that might occur even when the AI feature operated as expected, such as students using AI to cheat in homework.                                                        \\
        \end{tabular}
        \Description{A table with 7 rows and 2 columns. Two columns are harm envisions pattern and description.}
    \end{table*}
}

\setlength{\belowcaptionskip}{0pt}
\setlength{\abovecaptionskip}{12pt}

\subsubsection{Changes in Harm Envisioning Approaches.}
We also investigated the impacts of different tools on participants' approaches to harm envisioning by analyzing their self-reports in interview 1 and the think-aloud data in \aptLtoX[graphic=no,type=html]{H1}{\hone{}} and \aptLtoX[graphic=no,type=html]{H3}{\hthree{}}.

\aptLtoX[graphic=no,type=html]{}{\xlabel[self-reported envisioning approaches]{sec:evaluation-self-report-tool}}
\mypar{Self-reported changes after using \toolc{} and \litec{}.}
The major themes of self-reported changes are similar between \toolc{} and \litec{}.
A large number of participants noted that while they initially considered the AI feature and its potential harms in a general sense during \aptLtoX[graphic=no,type=html]{H1}{\hone{}}, they shifted towards a more focused consideration of specific \ul{use cases} and \ul{stakeholders} in \aptLtoX[graphic=no,type=html]{H3}{\hthree{}}~(e.g., P23, P34, P38).
Some participants highlighted they started to brainstorm potential misuses in \aptLtoX[graphic=no,type=html]{H3}{\hthree{}}~(P25, P32).
For stakeholders, participants broadened their consideration to people and organizations not initially considered during \aptLtoX[graphic=no,type=html]{H1}{\hone{}}.
P40 acknowledged a transition from a focus on \myquote{protecting the AI company} in \aptLtoX[graphic=no,type=html]{H1}{\hone{}} to considering end-users in \aptLtoX[graphic=no,type=html]{H3}{\hthree{}}.
Similarly, P17 reported a focus on end-users after using \toolc{}:

\begin{quote}
    \textit{``Earlier maybe I was coming towards it from a very engineering or a very broad feature perspective. The third time, I was thinking more about people who were actually using the product and getting affected. So I was thinking more with respect to the people using it, rather than that being a feature in some application.''}~(P17)\looseness=-1
\end{quote}

Many participants also highlighted that they began to \ul{adopt the frameworks} presented in \toolc{} and \litec{} (e.g., P9, P10, P32) to structure their harm envisioning procedures.
For example, P10 and P32 appreciated the categorization of use cases, and they reported considering intended uses, high-stakes uses, and misuses in \aptLtoX[graphic=no,type=html]{H3}{\hthree{}}.
After using \toolc{}, P9 said they followed the sequence of layers in the tree visualization to conceptualize use cases, stakeholders, and harms:\looseness=-1

\begin{quote}
    \textit{``I found that sort of flow from identifying potential use cases, then identifying stakeholders of those use cases, then identifying potential harms for each of the stakeholders to be really valuable. That's a great way to scaffold it and think through the flow rather than just sort of bouncing around, which is what I had been doing [in \aptLtoX[graphic=no,type=html]{H1}{\hone{}}]. So yeah, I found that super valuable that has changed the way that I think about it. And that's the framework that I'll use in the future.''}~(P21)
\end{quote}

\mypar{Self-reported changes after using \guidec{}.}
Many participants using \guidec{} in \aptLtoX[graphic=no,type=html]{H2}{\htwo{}}~(\conditiongf{}, \conditiongl{}) also noted shifts in their approaches to envisioning harms.
Several participants noted that they started to \ul{follow the structure} outlined in the Harm Modeling Guide to envision harms~(P8, P40, P42).
Some participants started thinking more about \ul{under-represented social groups} in \aptLtoX[graphic=no,type=html]{H3}{\hthree{}}~(P8, P31).
Furthermore, many participants described the harm taxonomy as a \myquote{mental checklist} that provided them with \ul{a language} to articulate and think about harms (e.g., P6, P14, P31).\looseness=-1

\setlength{\belowcaptionskip}{-3pt}
\setlength{\abovecaptionskip}{3pt}
\aptLtoX[graphic=no,type=html]{
    \begin{figure*}
        \includegraphics[width=0.93\linewidth]{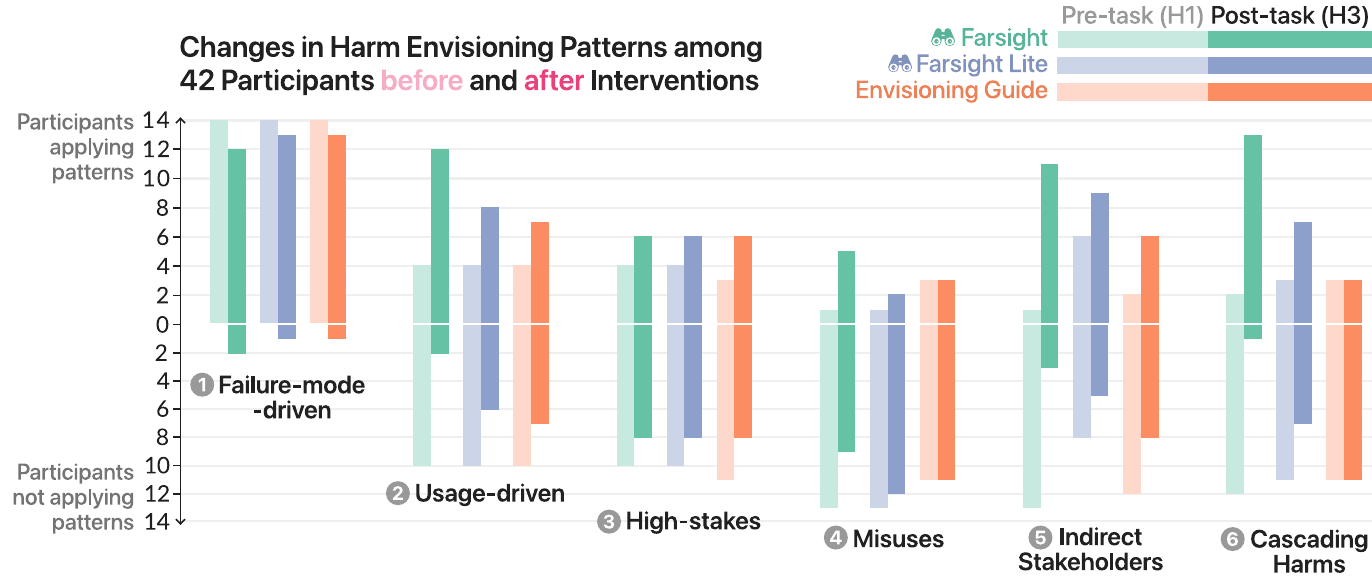}
        \caption{By analyzing transcripts of 42 participants during the pre-task~(H1) and post-task~(H3) harm envisioning activities, we identified six non-exclusive common patterns in envisioning harms.
            This bar chart compares the number of participants who applied and did not apply these patterns before and after the three interventions.
            Note that there were 14 random participants for each intervention, and the initial number of participants applying certain patterns could differ.
            The chart highlights that both \toolc{} and \litec{} encouraged participants to consider how the AI feature would be used. Notably, the use of \toolc{} particularly influenced participants to think more about indirect stakeholders and cascading harms.}
        \Description{A figure shows the changes in harm envisioning patterns among 42 participants before and after interventions.}
        \label{fig:evaluation-pattern-change}
    \end{figure*}
}{
    \begin{figure*}
        \includegraphics[width=0.93\linewidth]{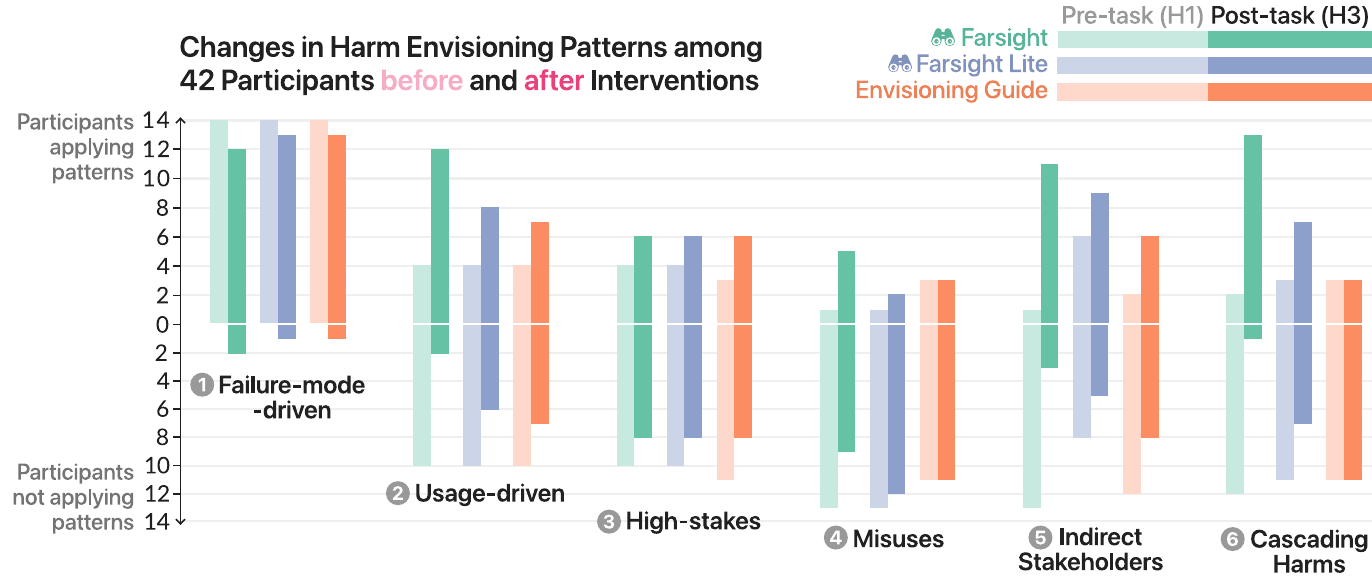}
        \caption[]{
            By analyzing transcripts of 42 participants during the pre-task~(\hone{}) and post-task~(\hthree{}) harm envisioning activities, we identified six non-exclusive common patterns in envisioning harms.
            This bar chart compares the number of participants who applied and did not apply these patterns before and after the three interventions.
            Note that there were 14 random participants for each intervention, and the initial number of participants applying certain patterns could differ.
            The chart highlights that both \toolc{} and \litec{} encouraged participants to consider how the AI feature would be used. Notably, the use of \toolc{} particularly influenced participants to think more about indirect stakeholders and cascading harms.
        }
        \Description{A figure shows the changes in harm envisioning patterns among 42 participants before and after interventions.}
        \label{fig:evaluation-pattern-change}
    \end{figure*}
}
\setlength{\belowcaptionskip}{0pt}
\setlength{\abovecaptionskip}{12pt}

\mypar{Observed changes in envisioning approaches.}
\aptLtoX[graphic=no,type=html]{}{\xlabel[]{sec:evaluation-observed-change}}
By analyzing transcripts of participants' think-aloud process during the harm envisioning activities in \aptLtoX[graphic=no,type=html]{H1}{\hone{}} and \aptLtoX[graphic=no,type=html]{H3}{\hthree{}}, we identified six non-exclusive common patterns in harm anticipation~(\autoref{table:evaluation-finding-pattern}).
Then, we examined the effects of different interventions on participants' envisioning patterns by comparing the number of participants who applied and did not apply these six patterns in \aptLtoX[graphic=no,type=html]{H1}{\hone{}} and \aptLtoX[graphic=no,type=html]{H3}{\hthree{}} across interventions~(\autoref{fig:evaluation-pattern-change}).
The intervention assignment is random.

Interestingly, the counts of participants who applied each pattern in \aptLtoX[graphic=no,type=html]{H1}{\hone{}} were consistent across interventions, with the exception of \litec{} where notably more participants considered indirect stakeholders in \aptLtoX[graphic=no,type=html]{H1}{\hone{}}~(\autoref{fig:evaluation-pattern-change}-\figpart{5}).
Before the interventions, the majority of participants relied on \ul{failure-mode-driven envisioning} when anticipating harms~(\autoref{fig:evaluation-pattern-change}-\figpart{1}), focusing on the AI feature's limitation, failure modes, and technical implementation details.
This observation corroborates participants' \aptLtoX[graphic=no,type=html]{self-reported envisioning approaches}{\myref{sec:evaluation-self-report-tool}}, where participants like P17 acknowledged having a \myquote{very engineering or a very broad feature perspective} in \aptLtoX[graphic=no,type=html]{H1}{\hone{}}.

After the intervention, we observed that all three harm envisioning tools (\toolc{}, \litec{}, and \guidec{}) influenced participants to adopt a \ul{usage-driven envisioning approach} when independently envisioning harms~(\autoref{fig:evaluation-pattern-change}-\figpart{2}).
Particularly, \toolc{} had the most pronounced effect, followed by \litec{} and then \guidec{}.
All these tools encouraged participants to think more about \ul{high-stakes uses}~(\autoref{fig:evaluation-pattern-change}-\figpart{3}) and \ul{indirect stakeholders}~(\autoref{fig:evaluation-pattern-change}-\figpart{5}).
Both \toolc{} and \litec{} exerted a stronger influence on considering \ul{misuses}~(\autoref{fig:evaluation-pattern-change}-\figpart{4}) and \ul{cascading harms}~(\autoref{fig:evaluation-pattern-change}-\figpart{6}) compared to \guidec{}.
However, \guidec{} had slightly more impact than \litec{} in encouraging consideration of \ul{high-stakes uses}~(\autoref{fig:evaluation-pattern-change}-\figpart{3}) and \ul{indirect stakeholders}~(\autoref{fig:evaluation-pattern-change}-\figpart{5}).\looseness=-1

Interestingly, \toolc{} had a notably more pronounced effect in leading participants to consider \ul{indirect stakeholders}~(\autoref{fig:evaluation-pattern-change}-\figpart{5}) and \ul{cascading harms}~(\autoref{fig:evaluation-pattern-change}-\figpart{6}) than the other tools.
For \ul{indirect stakeholders}, a possible explanation is that during \aptLtoX[graphic=no,type=html]{H2}{\htwo{}}, many participants encountered \textit{unexpected} indirect stakeholders revealed by \toolc{}~(\autoref{sec:envision-pattern-unexpected}).
Consequently, these participants consciously began to consider stakeholders that might seem tangential but could be influenced by the AI feature in \aptLtoX[graphic=no,type=html]{H3}{\hthree{}}.
This hypothesis could also explain the relatively weaker effect of \litec{} in fostering consideration of \ul{indirect stakeholders}, as \litec{} had only identified one \textit{direct} stakeholder for each use case, and participants could not use AI to generate more stakeholders.

For \ul{cascading harms}, we hypothesize two potential explanations.
First, many participants applied a \textit{reviewing approach} when engaging with AI-generated harms in \toolc{} and \litec{}, where they tried to understand and make sense of these harms.
In \aptLtoX[graphic=no,type=html]{H2}{\htwo{}}, reviewing existing harms prompted participants to consider \ul{cascading harms} that might arise from other harms~(\autoref{sec:envision-pattern-beyond}).
This experience could have influenced participants to also consider \ul{cascading harms} in \aptLtoX[graphic=no,type=html]{H3}{\hthree{}}.
The second explanation is that many participants were surprised by \textit{unexpected} AI-generated \ul{cascading harms} in \aptLtoX[graphic=no,type=html]{H2}{\htwo{}}~(\autoref{sec:envision-pattern-unexpected}), which might have led them to consciously think about these harms in \aptLtoX[graphic=no,type=html]{H3}{\hthree{}}.

\subsection{Findings: \tool{}'s Effectiveness in Assisting Harm Envisioning (RQ2)}
\label{sec:evaluation-finding-assistance}

In addition to assessing the impacts of different harm envisioning tools on users' ability to independently envision harms, we also evaluated the tools' effectiveness in aiding users to anticipate harms.
Specifically, we quantitatively compared participants' envisioned harms when using different harm envisioning tools in \aptLtoX[graphic=no,type=html]{H2}{\htwo{}} and \aptLtoX[graphic=no,type=html]{H4}{\hfour{}}.
Furthermore, we qualitatively analyzed participants' usage patterns, interview responses, and survey data.

\setlength{\belowcaptionskip}{-5pt}
\setlength{\abovecaptionskip}{5pt}
\aptLtoX[graphic=no,type=html]{
    \begin{figure*}
        \includegraphics[width=1\linewidth]{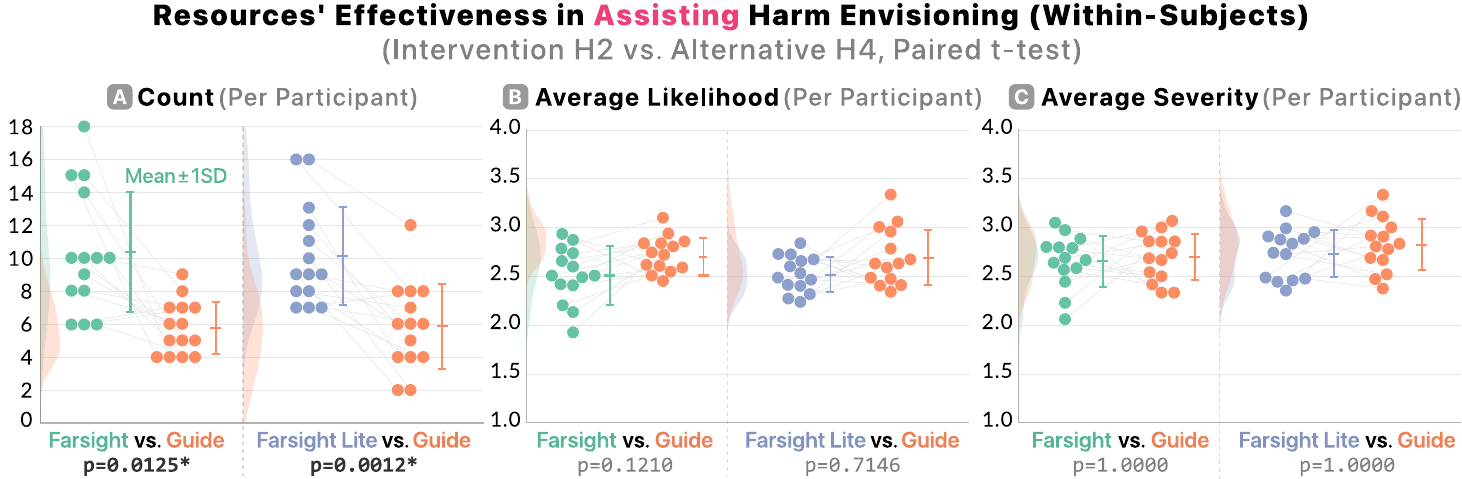}
        \caption{To evaluate the effectiveness of our tools in helping users anticipate harms, we conducted paired \textit{t}-tests with Bonferroni correction to compare our tools (\toolc{}, \litec{}) against the baseline \guidec{}
            based on the (A) count, (B) average likelihood, and (C) average severity of harms collected in H2 and H4.
            In each comparison, such as \toolc{} vs \guidec{}, n = 14 participants (each shown as two connected dots) used both tools: 7 of them started with \toolc{} in H2, and the remaining 7 began with \guidec{}.
            The charts also highlighted the mean and standard deviation of all measures.
            The results showed that \toolc{} and \litec{} were effective in assisting users to anticipate a significantly greater number of harms compared to existing resources, while the quality of the identified harms remained consistent.}
        \Description{A figure shows resources' effectiveness in assisting harm envisioning, A shows count; B shows average likelihood; C shows average severity.}
        \label{fig:evaluation-harm-compare}
    \end{figure*}
}{
    \begin{figure*}
        \includegraphics[width=1\linewidth]{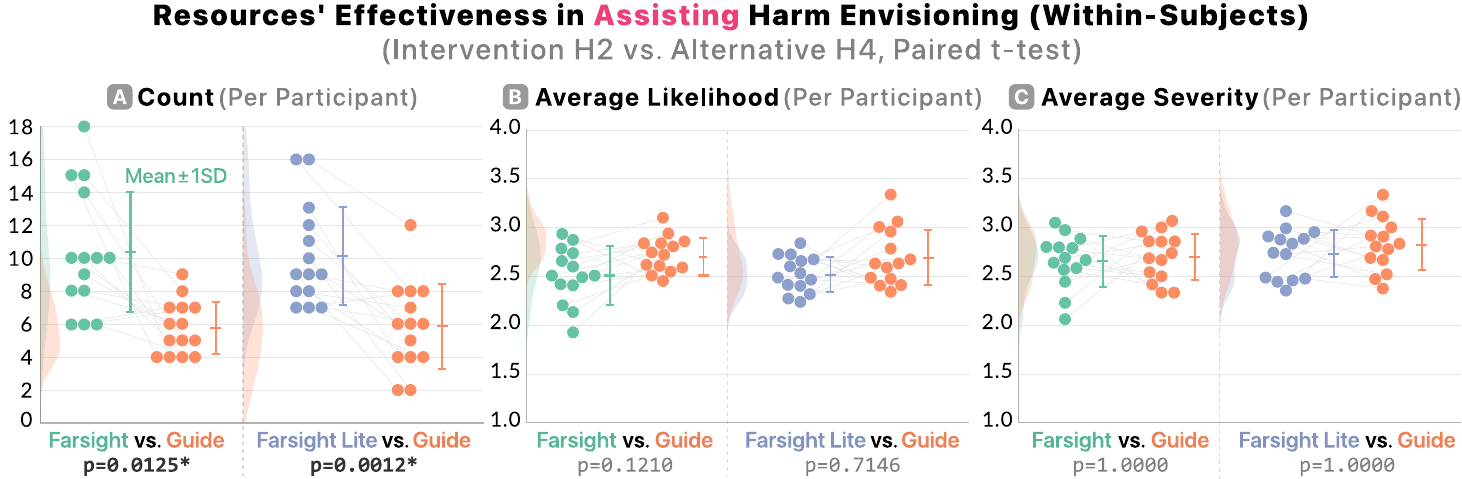}
        \caption[]{
            To evaluate the effectiveness of our tools in helping users anticipate harms, we conducted paired \textit{t}-tests with Bonferroni correction to compare our tools (\toolc{}, \litec{}) against the baseline \guidec{}
            based on the (A) count, (B) average likelihood, and (C) average severity of harms collected in \htwo{} and \hfour{}.
            In each comparison, such as \toolc{} vs \guidec{}, n = 14 participants (each shown as two connected dots) used both tools: 7 of them started with \toolc{} in \htwo{}, and the remaining 7 began with \guidec{}.
            The charts also highlighted the mean and standard deviation of all measures.
            The results showed that \toolc{} and \litec{} were effective in assisting users to anticipate a significantly greater number of harms compared to existing resources, while the quality of the identified harms remained consistent.
        }
        \Description{A figure shows resources' effectiveness in assisting harm envisioning, A shows count; B shows average likelihood; C shows average severity.}
        \label{fig:evaluation-harm-compare}
    \end{figure*}
}
\setlength{\belowcaptionskip}{0pt}
\setlength{\abovecaptionskip}{12pt}

\subsubsection{\toolc{} and \litec{} helped users envision more harms.}
\label{sec:evaluation-finding-rq2-quantitative}
We compared the count, average likelihood, and average severity of harms collected in \aptLtoX[graphic=no,type=html]{H2}{\htwo{}} and \aptLtoX[graphic=no,type=html]{H4}{\hfour{}} using our tools, \toolc{} and \litec{}, against the baseline \guidec{}~(\autoref{fig:evaluation-harm-compare}).
These harms were identified by participants using different harm envisioning tools or generated by AI and selected by the participants.
This analysis followed a within-subjects approach, including 28 participants from \conditionfg{}, \conditiongf{}, \conditionlg{}, and \conditiongl{}.
In each comparison, such as \toolc{} vs \guidec{}, a total of 14 participants used both tools, with 7 of them starting with \toolc{} in \aptLtoX[graphic=no,type=html]{H2}{\htwo{}}~(\conditionfg{}), and the remaining 7 beginning with \guidec{}~(\conditiongf{}).
Results from paired \textit{t}-tests, adjusted with Bonferroni correction, highlighted that participants using \toolc{} and \litec{} resulted in a significantly higher number of harms compared to those using \guidec{}~(\toolcolor{$p=0.0018$}, \litecolor{$p=0.0034$}), with an average difference in the count of 4~(\autoref{fig:evaluation-harm-compare}\figpart{A}).
The effect sizes, as measured by Cohen's $d$, were \toolcolor{$d=1.57$} and \litecolor{$d=1.48$}, indicating a very large effect.
However, no significant differences were observed regarding the likelihood and severity of identified harms between our tools and \guidec{}~(\autoref{fig:evaluation-harm-compare}-\figpart{BC}).
Our findings suggest that our tools are effective in assisting users to identify a greater number of harms compared to existing resources, while the quality of the identified harms remains consistent.

\subsubsection{Usage patterns.}
We summarized how participants use \toolc{} and \litec{} in \aptLtoX[graphic=no,type=html]{H2}{\htwo{}} and \aptLtoX[graphic=no,type=html]{H4}{\hfour{}}.

\mypar{Trying to understand (unexpected) AI-generated content.}
\aptLtoX[graphic=no,type=html]{}{\xlabel[]{sec:envision-pattern-unexpected}}
Upon encountering AI-generated content (e.g., use cases, stakeholders, and harms), participants first sought to (1) understand why AI had generated it and then (2) assess its likelihood and relevance to their AI application.
For example, for the toxicity classifier in \aptLtoX[graphic=no,type=html]{H2}{\htwo{}}, \toolc{} and \litec{} sometimes would generate a use case ``HR departments use it to screen job applicants for toxic behaviors.''
This use case was usually unexpected to participants, and it provoked them to think how an HR department could employ a toxicity classifier.
Some participants imagined that the HR could use this classifier on applicants' social media to identify red flags~(e.g., P10, P11, P29), while others could only see it being used on applicants' cover letters~(P4).
Participants then assessed how likely and relevant is this scenario before diving into related harms.

\mypar{Subjectivity in apprehending auto-generated content.}
\aptLtoX[graphic=no,type=html]{}{\xlabel[]{sec:envision-pattern-subjectivity}}
We observed that based on participants' prior experiences, they could have very different views on auto-generated content in \tool{}.
For example, participants had different perceptions of how their companies' HR division might use a toxicity classifier~(e.g., applying the classifier to job applicants' social media content or their application material).
Also, for the toxicity classifier in \aptLtoX[graphic=no,type=html]{H2}{\htwo{}}, the \incidentpanel{} would often show an incident report on biases in sentiment analysis tools.
While some participants could quickly make the connection between sentiment analysis and toxicity classification and reflect on biases in toxicity classifiers~(P10, P36), others would overlook this incident~(P19, P38).

In some cases, participants' disagreement came from their different definitions of harm.
For example, in both \aptLtoX[graphic=no,type=html]{H2}{\htwo{}} and \aptLtoX[graphic=no,type=html]{H4}{\hfour{}}, our tools would generate potential harms for people who do not use the AI applications, such as ``students who do not use the math tutoring app may feel left behind.''
Some participants perceived these harms as crucial considerations for assessing the impacts of AI applications~(e.g., P6, P18, P30), while others argue against considering harms when an AI feature is absent~(e.g., P4, P9, P13).
We discuss the implications of subjectivity and rater disagreement in harm envisioning in~\autoref{sec:discussion-subjectivity}.

\mypar{Sparked to brainstorm new harms.}
\aptLtoX[graphic=no,type=html]{}{\xlabel[]{sec:envision-pattern-spark}}
The content in \toolc{} and \litec{} often inspired participants to brainstorm new use cases, stakeholders, and harms.
After seeing an AI-generated stakeholder, many participants could quickly identify potential harms for that stakeholder.
For instance, seeing the stakeholder teachers in the math tutoring app in \aptLtoX[graphic=no,type=html]{H4}{\hfour{}}, P22 added a new harm that teachers may struggle to integrate this tool into their existing teaching workflows.
Many participants also came up with new harms by making connections across different AI-generated use cases, stakeholders, and harms.
For example, \toolc{} anticipated two use cases for the toxicity classifier: (1) online moderators using it to identify toxic content, and (2) hate groups using it to recruit people.
P2 connected both use cases and added a new harm: ``online moderators could face death threats from hate groups who feel their speech is censored.''

\mypar{Thinking beyond immediate harms.}
\aptLtoX[graphic=no,type=html]{}{\xlabel[]{sec:envision-pattern-beyond}}
Instead of starting with a blank slate, our tools provided participants with initial materials that prompted them to think beyond the immediate harms and envision cascading repercussions.
For example, after seeing the AI-generated harm ``job applicants might be unfairly rejected'' within the context of HR using a toxicity classifier to screen job applicants, P38 quickly thought of a cascading harm---the company's diversity hiring effort could be harmed, as the toxicity classifier was more likely to misclassify and reject under-represented social groups.
Similarly, P18 recognized in the long run, the hiring company could lose money due to the exclusion of qualified candidates caused by a biased toxicity classifier.
This usage pattern might explain the increase of participants, who used \toolc{} and \litec{} in \aptLtoX[graphic=no,type=html]{H2}{\htwo{}}, independently envisioning cascading harms in \aptLtoX[graphic=no,type=html]{H3}{\hthree{}}~(\autoref{fig:evaluation-pattern-change}-\figpart{6}).

\setlength{\belowcaptionskip}{-5pt}
\setlength{\abovecaptionskip}{7pt}
\begin{figure*}[tb]
    \includegraphics[width=1\linewidth]{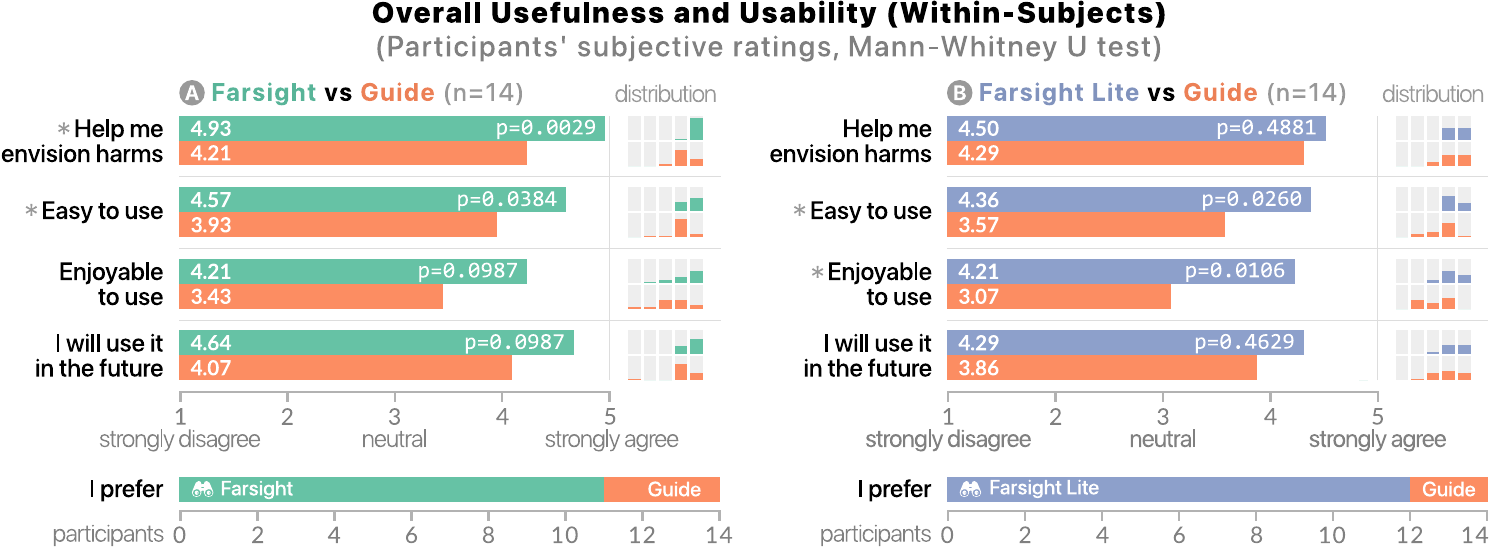}
    \caption[]{
        Average ratings from 28 participants, comparing the usefulness and usability of \toolc{} and \litec{} to \guidec{}.
        Both of our tools were preferred and perceived as more helpful, easier to use, and more enjoyable than the existing resources.
        Each comparison involved 14 participants who used one of our tools and \guidec{} in random order.
        We use an asterisk ($\ast$) to denote statistically significant rating differences, determined by Mann-Whitney U tests with Bonferroni correction.
        We used Mann-Whitney U tests instead of t-tests due to the non-normal distribution of many ratings.
    }
    \Description{
        A figure divided into two panels, A and B.

        Panel A compares Farsight with Harm Envisioning Guide across five criteria: "Help me envision harms," "Easy to use," "Enjoyable to use," "I will use it in the future," and "I prefer." Each criterion is rated on a scale from 1 (strongly disagree) to 5 (strongly agree). The bars show Farsight's ratings consistently higher than Harm Envisioning Guide's, with the first two criteria marked with asterisks, indicating a statistically significant difference with p-values of 0.0029 and 0.0384, respectively.

        Panel B shows a similar comparison for Farsight Lite vs Harm Envisioning Guide. Here, "Enjoyable to use" is the only criterion with a significant difference, marked by an asterisk and a p-value of 0.0106. Other criteria show no significant difference, with p-values above 0.05.

        Next to each set of bars, there's a small distribution graph displaying the spread of participant responses. The bottom of each panel has a bar indicating the number of participants who prefer either Farsight, Farsight Lite, or Harm Envisioning Guide, with Farsight and Farsight Lite being preferred more often than Harm Envisioning Guide.

        The chart demonstrates that participants generally find Farsight tools more useful and usable compared to the Harm Envisioning Guide, with Farsight showing a slight edge over Farsight Lite in terms of preference.
    }
    \label{fig:evaluation-usability}
\end{figure*}
\setlength{\belowcaptionskip}{0pt}
\setlength{\abovecaptionskip}{12pt}

\setlength{\belowcaptionskip}{-10pt}
\setlength{\abovecaptionskip}{5pt}
\begin{figure}[tb]
    \centering
    \includegraphics[width=1\linewidth]{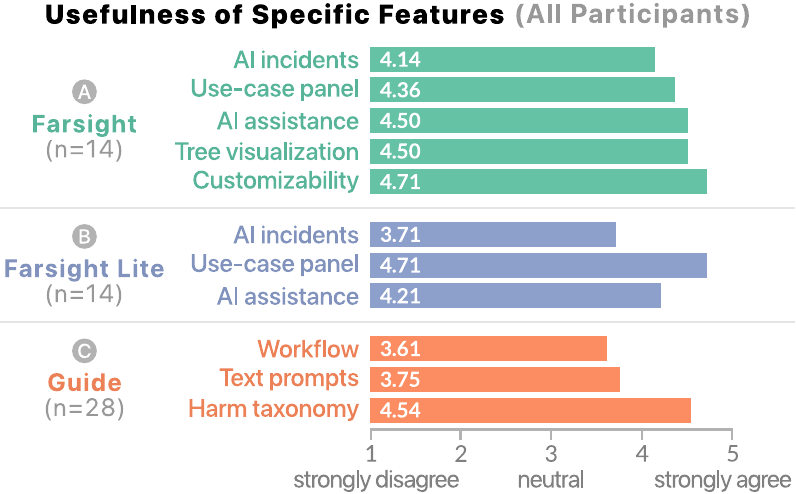}
    \caption{Average ratings of envisioning tool features.}
    \Description{A figure shows the usefulness of specific features of Farsight, Farsight Lite, and Guide.}
    \label{fig:evaluation-feature}
\end{figure}
\setlength{\belowcaptionskip}{0pt}
\setlength{\abovecaptionskip}{12pt}

\mypar{Thinking about mitigation strategies.}
\aptLtoX[graphic=no,type=html]{}{\xlabel[]{sec:envision-pattern-mitigation}}
Interestingly, after seeing AI-generated harms, many participants \textit{voluntarily} considered actions and strategies to take after envisioning harms.
For example, after seeing AI-generated harms for the toxicity classifier, P15 and P16 noted that it was important to allow impacted stakeholders to appeal if their content was removed because of the classifier.
Similarly, P27 and P40 noted that people should implement a human review process if the toxicity classifier was used to remove social media content.
Interacting with \toolc{} and \litec{} also encouraged participants to reflect on their prompting workflows.
For example, P29 and P37 mentioned that the AI prototypers should start collecting good and diverse toxicity examples to improve the prompt through few-shot prompting.
P2 noted that they would like to add additional instructions in their prompt to safeguard against biased output and potential data leakage.
Finally, after envisioning more harms, P2 mentioned that they would rethink if it was worth continuing to prototype or develop this AI feature.

\subsubsection{Our tools were usable, useful, and preferred by users.}
\label{sec:envision-pattern-usability}
We asked participants who had used one of our tools and \guidec{}~(\conditionfg{}, \conditiongf{}, \conditionlg{}, \conditiongl{}) to compare and rate the usefulness and usability of the tools they had used on a 5-point Likert-scale.
By comparing their ratings, we found both \toolc{} and \litec{} were preferred and considered as more helpful, easier to use, and more enjoyable compared to \guidec{}~(\autoref{fig:evaluation-usability}).
Both tools had significantly higher ratings on ``easy to use'' than the baseline~(\toolcolor{$p=0.0384$}, \litecolor{$p=0.0260$}).
In addition, \toolc{} was rated significantly more helpful than the baseline~(\autoref{fig:evaluation-usability}\figpart{A}), while \litec{} was more enjoyable~(\autoref{fig:evaluation-usability}\figpart{B}).
The effect sizes of significant results, as measured by the common language effect size~\cite{mcgrawCommonLanguageEffect1992}, were all above $0.7$, indicating a large effect.

\mypar{Usefulness of different features.}
Besides comparing the two tools, participants also rated the usefulness of specific features in each tool.
The average ratings are shown in~\autoref{fig:evaluation-feature}.
All features in our tools were rated favorably~(\autoref{fig:evaluation-feature}\figpart{-AB}).
For \toolc{}, participants especially liked the interactive tree visualization.
For example, P6 commented, \myquote{This tree makes a lot of sense. This is how I think about it in my brain as well.}
Similarly, P16 appreciated the progressive disclosure in the visualization: \myquote{I'm able to not get overwhelmed by everything all at once.}
The rating for the AI incident panel (in both \toolc{} and \litec{}) is relatively lower than other features.
Participants explained that the surfaced incidents were not very relevant to their prompts~(P39, P41), and the feature would require them to take time to read external articles~(P24, P39).

\subsection{Findings: \tool{}'s Role in Overcoming Harm Envisioning Challenges (RQ3)}
\label{sec:evaluation-finding-challenge}

After completing the post-task~(\aptLtoX[graphic=no,type=html]{H3}{\hthree{}}), participants were asked to reflect on the biggest challenges encountered in envisioning harms associated with AI features.
We examined the major themes that emerged from these challenges.
In addition, by analyzing participants' usage patterns of \toolc{} and \litec{}, coupled with their interview feedback, we explored how our tools mitigate certain challenges and also identified our tools' limitations.

\subsubsection{Challenges in envisioning harms.}
We summarized three major challenges that participants encountered.

\aptLtoX[graphic=no,type=html]{
    \begin{itemize}
        \item [\textbf{C1.}] \textbf{Envisioning use cases.}
              The most prevalent challenge in envisioning harms is to anticipate different use cases for an AI feature.
              Multiple participants noted that it was most challenging to imagine how different people would use technology, and it was particularly difficult to \myquote{put myself in someone's shoes}~(P27, P37, P39) and \myquote{empathize with different groups of people}~(P11).
              Participants also underscored the vast space of possible use cases~(P31, P33, P36), and \myquote{often you don't find out the edge cases until you actually work with it}~(P2).
              Some participants also emphasized that it sometimes required creativity to imagine how an AI feature would be used and especially misused~(e.g., P5, P22, P23).

        \item [\textbf{C2.}] \textbf{Bias and subjectivity in harm envisioning.}
              Interestingly, several participants recognized their own biases in envisioning harms~(e.g., P6, P21, P31).
              For example, P21 noted the challenge in overcoming their biases in anticipating the impacts of AI features: \myquote{I had been coming at it from a very American-centric point of view at first. To talk about bias, I hadn't even conceived of the government using this to monitor my phone, but that could happen in other places.}
              Moreover, some participants acknowledged the subjectivity in the definition of harms, as well as in the assessment of harms' likelihood and severity.
              For example, while envisioning harms and selecting harms to report~(H2 and H4), some participants were conscious of whether other people would agree with their identification and assessment of harms~(P19, P38).

        \item [\textbf{C3.}] \textbf{Inexperience and discomfort in harm envisioning.}
              Many participants mentioned that our study was their first time to envision harms for AI features~(e.g., P17, P26, P28).
              For example, P26 noted \myquote{I have never envisioned harm before. This is not something I would think of when developing AI products.}
              Similarly, P18 said \myquote{I'm familiar with technical issues but not their social influence}.
              Also, P30 pointed out that there were few incentives for developers to envision harms.
              In addition to unfamiliarity, some participants also noted that it was uncomfortable and sad to think about harms~(P3, P12).
              For example, P3 said \myquote{It's not comfortable thinking through all the bad things that can happen. I think in general people don't like thinking about bad things too much.}
    \end{itemize}
}{
    \begin{enumerate}[topsep=5pt, itemsep=0mm, parsep=1mm, leftmargin=18pt, label=\textbf{C\arabic*.}, ref=C\arabic*]
        \item \label{item:challenge-use-case} \textbf{Envisioning use cases.}
              The most prevalent challenge in envisioning harms is to anticipate different use cases for an AI feature.
              Multiple participants noted that it was most challenging to imagine how different people would use technology, and it was particularly difficult to \myquote{put myself in someone's shoes}~(P27, P37, P39) and \myquote{empathize with different groups of people}~(P11).
              Participants also underscored the vast space of possible use cases~(P31, P33, P36), and \myquote{often you don't find out the edge cases until you actually work with it}~(P2).
              Some participants also emphasized that it sometimes required creativity to imagine how an AI feature would be used and especially misused~(e.g., P5, P22, P23).

        \item \label{item:challenge-subjectivity} \textbf{Bias and subjectivity in harm envisioning.}
              Interestingly, several participants recognized their own biases in envisioning harms~(e.g., P6, P21, P31).
              For example, P21 noted the challenge in overcoming their biases in anticipating the impacts of AI features: \myquote{I had been coming at it from a very American-centric point of view at first. To talk about bias, I hadn't even conceived of the government using this to monitor my phone, but that could happen in other places.}
              Moreover, some participants acknowledged the subjectivity in the definition of harms, as well as in the assessment of harms' likelihood and severity.
              For example, while envisioning harms and selecting harms to report~(\htwo{} and \hfour{}), some participants were conscious of whether other people would agree with their identification and assessment of harms~(P19, P38).

        \item \label{item:challenge-discomfort} \textbf{Inexperience and discomfort in harm envisioning.}
              Many participants mentioned that our study was their first time to envision harms for AI features~(e.g., P17, P26, P28).
              For example, P26 noted \myquote{I have never envisioned harm before. This is not something I would think of when developing AI products.}
              Similarly, P18 said \myquote{I'm familiar with technical issues but not their social influence}.
              Also, P30 pointed out that there were few incentives for developers to envision harms.
              In addition to unfamiliarity, some participants also noted that it was uncomfortable and sad to think about harms~(P3, P12).
              For example, P3 said \myquote{It's not comfortable thinking through all the bad things that can happen. I think in general people don't like thinking about bad things too much.}
    \end{enumerate}
}

\subsubsection{\toolc{} and \litec{} address major challenges.}
\label{sec:evaluation-result-advantage}
Our tools could help users address identified challenges.

\mypar{A co-pilot for brainstorming diverse use cases.}
\aptLtoX[graphic=no,type=html]{}{\xlabel[]{sec:envision-advantage-assistant}}
Many participants appreciated that our tools provided them with a starting point to predict use cases~(e.g., P8, P29, P41).
For example, after seeing a few AI-generated use cases, P8 found it much easier to envision other use cases, and similarly, P24 felt empowered to \myquote{have a wider net to cast}~(\aptLtoX[graphic=no,type=html]{C1}{\ref{item:challenge-use-case}}).
Also, P14 noted that even seeing far-fetched AI-generated content helped them brainstorm new use cases.
On the other hand, P21 appreciated that \toolc{} had identified many unexpected and thought-provoking use cases that provided a different perspective in anticipating harms~(\aptLtoX[graphic=no,type=html]{C2}{\ref{item:challenge-subjectivity}}).

\mypar{\textit{In situ} guide that promotes user agency.}
\aptLtoX[graphic=no,type=html]{}{\xlabel[]{sec:evaluation-result-challenge-workflow}}
Participants especially liked that our tools were directly integrated into existing AI prototyping tools and contextualized based on the prompt~(e.g., P19, P31, P37), where \toolc{} and \litec{} required minimal effort to get started envisioning harms~(\aptLtoX[graphic=no,type=html]{C3}{\ref{item:challenge-discomfort}}).
Participants also thought the \incidentpanel{} and \usecasepanel{} as a good reminder for potential harms for the AI feature that one is prototyping~(e.g., P12, P41, P42).
For example, P12 commented that \myquote{Even if it's just sitting there, it would be educational.}
Many participants also liked the interactivity of our tools and found it engaging for adding new use cases, stakeholders, and harms~(e.g., P9, P19, P24)---many of them noted that \toolc{} was so intriguing that they would like to continue using it to explore potential harms~(\aptLtoX[graphic=no,type=html]{C3}{\ref{item:challenge-discomfort}}).
Participants felt they had agency in harm envisioning when using \toolc{}.
For example, P21 commented \myquote{If you think something [AI-generated content] is totally bonkers, whatever, just delete it.}
Similarly, P4 and P5 compared the \windowview{} to a mind map, as they appreciate that the interface allows them to freely organize and revise their thoughts in harm envisioning.

\subsubsection{Limitations of \toolc{} and \litec{}.}
\label{sec:evaluation-result-limitation}
Our findings showed that, in comparison to \guidec{}, \toolc{} and \litec{} did not show significant differences in participants' ability to envision more likely or more severe harms~(\autoref{sec:evaluation-finding-rq1-quantitative}), nor did they assist participants in envisioning more likely or more severe harms~(\autoref{sec:evaluation-finding-rq2-quantitative}).
Additionally, participants' feedback revealed two major limitations of our tools.

\mypar{Varied quality of LLM-generated content.}
\aptLtoX[graphic=no,type=html]{}{\xlabel[]{sec:evaluation-result-limitation-quality}}
Depending on participants' prompts, the related AI incidents in the \incidentpanel{}, and LLM-generated use cases, stakeholders, and harms were different across participants.
Sometimes, participants found a few LLM-generated content confusing and unhelpful.
For example, when using our tools on the math tutor prompt~(\aptLtoX[graphic=no,type=html]{H4}{\hfour{}}), the incidents in the \incidentpanel{} feature articles about hallucination in chat-based LLM models.
Some participants found these articles too generic and not relevant to the math tutor app~(P39, P41).

Also, some LLM-generated use cases could be too far-fetched.
For example, for the math tutor prompt, \tool{} sometimes showed a use case: \myquote{Scammers use it to explain complex investment schemes to potential victims.}
While some participants found it interesting and relevant~(e.g., P14, P26), others found it unrealistic and not useful~(e.g., P6, P12).
This disagreement highlights the subjectivity in identifying and assessing harms~(\autoref{sec:discussion-subjectivity}).
Interestingly, a few participants defended the usefulness of far-fetched content.
P24 noted \myquote{Even if it's wrong [LLM-generated use case], it is still kind of helpful to think beyond the immediate use case and who else can use this tool.}
Similarly, P21 said \myquote{Some of these feel more of a stretch but it's interesting because I could see how it gives me ideas for things to watch out for which I still appreciate.}

\mypar{Lack of actionability.}
\aptLtoX[graphic=no,type=html]{}{\xlabel[]{sec:evaluation-result-limitation-action}}
Another limitation is that our tools did not provide users with actions to prevent or mitigate identified harms~(P13, P22, P34).
P13 also commented that increasing awareness without providing actions to address responsible AI issues could be harmful, because \myquote{People have an empathy quota, and it might just be displacing more impactful efforts.}
Related to the discomfort that some participants had experienced when envisioning harms~(\aptLtoX[graphic=no,type=html]{C3}{\ref{item:challenge-discomfort}}), P40 mentioned that they felt scared and overwhelmed because there were so many possible harms and they did not know how to address them.
Similarly, P29 noted that the lack of actionability made them feel anxious and disappointed:
\begin{quote}
    \textit{``I'm glad that I got to know about them [potential misuses]. But I feel I'm vulnerable, probably because I can't do much about stopping them. So that's something that really makes me feel very disappointed. Because unless we do case-by-case analysis, this [preventing misuses] can be very tricky.
        I feel like it's kind of adding anxiety to me. It's good to know, but I feel like I can't do much about it.''}~(P29)
\end{quote}
We did not incorporate harm mitigation into our tools, because mitigating harms associated with LLM-powered applications remains an open research question~(see more discussion in \autoref{sec:discussion-action}).
After the evaluation study, we improved \tool{} by providing pointers to existing LLM safety resources~\cite[e.g.,][]{anthropicCoreViewsAI2023,metaLlamaResponsibleUse2023,googlePaLMAPISafety2023,shelbySociotechnicalHarmsAlgorithmic2023} when users exported their harms.

\subsection{Limitations of Study Design}
\label{sec:evaluation-study-limitation}

We acknowledge limitations in our tool and study designs.
First, we recruited participants from a single large technology company.
This was because we needed to require participants to have prior experience in prototyping LLM-powered applications using a particular prompt-crafting tool, into which we integrated \toolc{} and \litec{} in the study.
Consequently, all 42 participants had backgrounds in the technology industry in varying roles, such as software engineers, product managers, UX researchers, and linguists\footnote{The linguists in our study had roles involved in consulting on language-based data used by AI product teams.} as shown in~\autoref{table:evaluation-participant}.
Our participants have a wide range of familiarity with responsible AI and prompting~(\autoref{fig:evaluation-experience}), and they use LLMs for diverse tasks, including prototyping AI features with LLMs---much like the intended users of \tool{}.
Therefore, findings from our study may be generalizable to AI prototypers who have worked in the technology industry, and who are using LLMs to prototype AI-based applications.
Nevertheless, to understand how usable or effective \tool{} may be for a broader spectrum of AI prototypers, particularly those with limited background or knowledge of AI, such as creative writers, teachers, students, and more, further research involving individuals with more diverse backgrounds is needed.
Second, we administered only one post-task~(\aptLtoX[graphic=no,type=html]{H3}{\hthree{}}) immediately following the intervention~(\aptLtoX[graphic=no,type=html]{H2}{\htwo{}}).
To evaluate the long-term impact of our tools on users' ability to envision harms, a more extended longitudinal study is needed.

Finally, an inter-rater reliability test showed that, on average, the seven raters (i.e., of the likelihood and severity of the identified harms) only had a slight agreement~(\autoref{appendix:evaluation-ratings}).
The ratings of the likelihood and severity of participants' identified harms should thus be taken as an initial step in evaluating identified harms, and not as the sole evidence demonstrating the value of this approach.
The relatively low inter-rater reliability may be due to the fact that perceptions of severity and likelihood may be highly influenced by the raters' personal experiences, backgrounds, knowledge, and their positionality as a whole.
Indeed, substantial prior work on annotations of offensive language, hate speech, and other linguistic phenomena \cite{dentonWhoseGroundTruth2021,mostafazadehdavaniDealingDisagreementsLooking2022,davaniDisentanglingDisagreementsOffensiveness2023,prabhakaranFrameworkAssessDis2023,pavlickInherentDisagreementsHuman2019} suggest that disagreements between raters with different subjectivities (i.e., personal backgrounds and experiences) is an inherent challenge to sociotechnical evaluations, and not one that can be solved with more or better raters.
We further discuss the challenges regarding subjectivity in identifying and assessing harms in~\autoref{sec:discussion-subjectivity}.
\section{Discussion}

\subsection{Motivation \& Engagement in Responsible AI}
\label{sec:discussion-motivation}
\mypar{Potentials of \textit{in situ} and early intervention in motivating responsible AI practices.}
Existing research suggests that many AI developers may not have incentives to consider potential harms related to their AI applications~\cite{rakovaWhereResponsibleAI2021}---or may be actively disincentivized to identify such harms \cite{madaioCoDesigningChecklistsUnderstand2020}.
Our co-design user study validates this finding among an emerging community involved in AI development---AI prototypers who use LLMs to rapidly iterate on potential AI-based applications~~(\autoref{sec:formative-study}).
With the rapidly increasing access to LLMs and easy-to-use prototyping tools, it is crucial to motivate AI prototypers to consider AI risks when prototyping their AI applications or features~(\aptLtoX[graphic=no,type=html]{\textbf{G3}}{\ref{item:goal-workflow}}).
To tackle this challenge, we propose an \textit{in situ} system design that integrates our tool into the AI prototyper's existing workflows and employs different design strategies to draw users' attention without causing significant interruption to their flow.
Our evaluation study shows that users appreciate our design, and find this in-context warning tool easy to adopt and engaging~(\autoref{sec:evaluation-result-challenge-workflow}).
By showing unexpected use cases, stakeholders, and harms, \tool{} piques users' interests~(\autoref{sec:envision-pattern-unexpected}) and motivates them to brainstorm more harms~(\autoref{sec:envision-pattern-spark}).
These findings highlight the great potential of \textit{in situ} design and early intervention for future responsible AI works.
Therefore, future designers of AI development tools~(e.g., Google AI Studio, computational notebooks, and VSCode) can natively integrate \textit{in situ} interfaces to promote responsible AI practices.
In addition, future researchers can adopt our design strategies to foster other responsible AI practices, such as illustrating bias in LLMs and encouraging development documentation at an early AI development stage.

\mypar{Tension between automation and human agency.}
\tool{}'s seamless integration into AI prototypers' workflows helps motivate AI prototypers to engage with harm envisioning.
In addition, rather than asking users to anticipate harms entirely from scratch, \tool{} leverages LLMs to generate the initial set of use cases, stakeholders, and harms, providing users with inspiration and a foundation to build upon~(\autoref{sec:evaluation-result-limitation}).
However, this seamless and automated design might deter users from fully engaging in and contemplating the limitations and potential risks associated with LLMs.
Prior research in responsible AI has proposed the value of a \textit{seamful} design~\cite[e.g.,][]{ehsanSeamfulXAIOperationalizing2022, kaurSensibleAIReimagining2022}, where the designers strategically reveal seams or introduce frictions or ``productive restraint'' \cite{madaioCoDesigningChecklistsUnderstand2020,kaiserAdaptingSecurityWarnings2021} to support increased reflection on responsible AI during development.
To explore this tension and tradeoffs between a seamfully-designed workflow that is easy to use by prototypers, and a seamful design that prompts reflection-in-action~\cite{ehsanSeamfulXAIOperationalizing2022}, we (1) designed the \windowview{} to encourage users to edit LLM-generated content and steer the harm envisioning direction~(\autoref{sec:system:window}, \aptLtoX[graphic=no,type=html]{\textbf{G4}}{\ref{item:goal-engage}}), and (2) evaluated two variants of our tool in the evaluation study---\toolc{} and \litec{}, where \litec{} omits the \windowview{}.

Our study results highlight that participants feel they have agency~(\autoref{sec:envision-advantage-assistant}), and they like being able to control the harm anticipation process~(\autoref{fig:formative-feature}).
Our quantitative results also show that \toolc{}, with higher human agency, is more effective than \litec{} across all measures~(\autoref{sec:evaluation-finding-rq1-quantitative}, \autoref{sec:evaluation-finding-rq2-quantitative}).
On the other hand, when engaging with AI-generated content, some participants also report discomfort~(\aptLtoX[graphic=no,type=html]{\textbf{C3}}{\ref{item:challenge-discomfort}}) and even anxiety~(\autoref{sec:evaluation-result-limitation-action}).
Therefore, our work demonstrates that seamless design (\textit{in situ} AI automation) and seamful design (promoting user reflection) are complementary to each other---tradeoffs and a balance between the two should be considered during the design of responsible AI tools \cite[cf.][]{wongSeeingToolkitHow2023}.
For future responsible AI work, researchers should engage with potential users and other impacted stakeholders throughout the design process and adjust their design ideas to ensure the responsible AI tools they are designing are both easily adoptable and capable of eliciting active and critical reflection.

\subsection{Subjectivity in Harm Envisioning}
\aptLtoX[graphic=no,type=html]{}{\xlabel[]{sec:discussion-subjectivity}}
In our evaluation user study, many participants report challenges overcoming the limitations of their own experiences and perspectives when envisioning harms~(\aptLtoX[graphic=no,type=html]{\textbf{C2}}{\ref{item:challenge-subjectivity}}).
In addition, we also observed the seven RAI raters of participants' harms disagreed about which harms were more or less severe or likely, resulting in a low inter-rater reliability for these two dimensions~(\autoref{appendix:evaluation-ratings}, \autoref{table:participant-harm}).
Our empirical findings contribute to prior research that highlights the role of subjectivity and positionality in anticipating harms~\cite{boyarskayaOvercomingFailuresImagination2020, liuExaminingResponsibilityDeliberation2022} and in data annotation, particularly for annotations of toxicity or hate speech~\cite[e.g.,][]{mostafazadehdavaniDealingDisagreementsLooking2022, diazCrowdWorkSheetsAccountingIndividual2022,dentonWhoseGroundTruth2021,davaniDisentanglingDisagreementsOffensiveness2023,prabhakaranFrameworkAssessDis2023}.
What constitutes harm and the assessment of harm severity are often influenced by the individual's background, lived experiences, or even the organizational culture they are working in~\cite{whittakerSteepCostCapture2021, queerinaiBoundBountyCollaboratively2023}.
For example, for the article summarizer~(\aptLtoX[graphic=no,type=html]{H3}{\hthree{}}), one participant envisioned a harm scenario: ``If the summary is wrong, journalists' reputation might be harmed.''~(\autoref{table:participant-harm}).
This harm scenario received likelihood ratings of 1, 4, and 3, and severity ratings of 1, 3, and 4 from three randomly assigned raters.
It is possible that the rater who assigned the ratings of 3 and 4 possessed specific knowledge about the harms of journalists using LLMs to write article summaries, which led them to rate this harm scenario as more likely and more severe.

\mypar{A need for new methods to assess harms.}
Emerging research is beginning to develop methods for measuring and resolving disagreements among annotators in cases where there may in fact be no ground truth \cite[e.g.,][]{dentonWhoseGroundTruth2021,prabhakaranFrameworkAssessDis2023,davaniDisentanglingDisagreementsOffensiveness2023,mostafazadehdavaniDealingDisagreementsLooking2022,gordonJuryLearningIntegrating2022}. Our findings in this paper---including the low inter-rater reliability of the responsible AI raters---suggest that new methods are needed in responsible AI to account for different perspectives on the severity and likelihood of potential downstream harms. This may ideally involve recruiting participants from communities or populations who may be impacted by a given AI application (e.g., the stakeholders generated by \tool{}, for instance, as well as other stakeholders identified by members of the communities themselves~\cite{delgadoParticipatoryTurnAI2023}).
Moreover, with the rapidly increasing access to LLMs and easy-to-use AI prototyping tools, AI prototypers may encompass a broader spectrum of roles beyond traditional AI practitioners \cite[e.g.,][]{holsteinImprovingFairnessMachine2019,wangDesigningResponsibleAI2023}. Thus, they may lack either the experience or the resources to recognize the limitations of their own subjectivity when anticipating harms of their AI applications, and may lack the means to identify and engage with diverse stakeholders as part of harm envisioning.

\mypar{Benefits and challenges of using LLM to envision AI harms.}
Our evaluation study highlights that diverse and unexpected AI-generated use cases, stakeholders, and harms in \tool{} help some participants overcome their own failures of imagination \cite{boyarskayaOvercomingFailuresImagination2020} in order to think from a broader perspective when independently envisioning harms~(\autoref{sec:envision-pattern-beyond}).
Notably, these effects were more prominent with \tool{} than with existing harm envisioning processes \cite{microsoftHarmsModelingAzure2022}~(\autoref{fig:evaluation-harm-changes}).
There are two implications of these findings.
First, LLMs can be a promising tool to help AI prototypers think outside of their own perspectives, and future researchers can adapt our approach to other responsible AI practices.
Second, LLMs may encode biases from their training data~\cite[e.g.,][]{weidingerEthicalSocialRisks2021}, and \tool{} may also reflect the biases of its creators, as expressed in the underlying prompts used in \tool{}'s LLM, which raises a critical question: to what extent can LLMs be helpful as part of a harm envisioning process, without over-indexing on particular harms or leading AI prototypers to overlook other types of harms?

Our research provides an initial starting point into investigating these questions, as well as opening new questions into the role of subjectivity in harm envisioning.
Future research can further investigate the factors influencing users' ability to envision harms of AI applications, develop new ways to model and resolve disagreement among AI prototypers or other evaluators about the severity and likelihood of envisioned harms, and integrate such implications into LLM-powered responsible AI tools for AI prototypers or other AI practitioners.
Future research can also explore tradeoffs between semi-automated harm envisioning processes (like \tool{}) and more traditional processes like value-sensitive design \cite[e.g.,][]{friedmanValuesensitiveDesign1996}, participatory design \cite[e.g.,][]{birhanePowerPeopleOpportunities2022,queerinaiBoundBountyCollaboratively2023,delgadoParticipatoryTurnAI2023}, and more.

\subsection{Mitigating Harms during AI Prototyping}
\label{sec:discussion-action}
A limitation of \tool{} is its focus on harm identification rather than harm mitigation.
Participants from our co-design study~(\autoref{sec:formative-study}) and evaluation study~(\autoref{sec:evaluation-result-limitation-action}) wanted \tool{} to provide actionable items to help them prevent and mitigate identified harms.
Some participants also suggested we develop an \textit{in situ} prompt editing tool to address harms identified from \tool{}~(\autoref{sec:formative-study}).
Interestingly, while using \tool{}, some participants \textit{voluntarily} thought about actions and strategies to take after envisioning harms, such as implementing an appeal process, collecting better data, and revising the prompts~(\autoref{sec:envision-pattern-mitigation}).
These strategies identified by participants reflect some current strategies for mitigating LLM harms~(\autoref{sec:related:llm}).

Looking ahead, we argue that it is crucial for future designers to provide users with harm mitigation suggestions and resources in systems similar to \tool{}.
Some participants in our study complained that \tool{} is exploiting users' ``empathy quota'' and potentially desensitizing them about LLM harms, because \tool{} only warns users about harms without providing mitigation suggestions~(\autoref{sec:evaluation-result-limitation-action}).
This concern reflects the phenomenon of ``alarm fatigue'' in alerting tools~(\autoref{sec:related:alert}) and monitoring alarms in healthcare.
Alarm fatigue occurs ``when non-actionable alarms are in the majority, and clinicians develop decreased reactivity, causing them to `tune out' or ignore the alarms''~\cite{hravnakCallAlarmsCurrent2018}.
Therefore, to combat alarm fatigue and effectively promote responsible AI practices, future designers should make responsible AI alerts actionable and prioritize actionable warnings in their systems.

Our findings highlight that \tool{} users have a great appetite for mitigation strategies during AI prototyping.
We have two hypotheses for this observation.
First, as \tool{} promotes human agency, it might also give participants a feeling of \textit{ownership} of their identified harms.
Prior research shows that triggering a feeling of ownership motivates users' actions~\cite{caraban23WaysNudge2019}.
Another hypothesis is that \tool{} elicits fear by exposing participants to diverse potential harms of their AI applications, evidenced by participant-reported discomfort~(\aptLtoX[graphic=no,type=html]{\textbf{C3}}{\ref{item:challenge-discomfort}}) and anxiety~(\autoref{sec:evaluation-result-limitation-action}).
Security researchers use \textit{fear appeals} as a design strategy to motivate users to take security actions~\cite{renaudCyberSecurityFear2019}.
Therefore, our empirical findings highlight promising research opportunities in (1) providing \textit{in situ} mitigation strategies during the early AI prototyping stage, and (2) investigating if \textit{in situ} tools can increase users' adoption of harm mitigation strategies.

%
%
%
%
%
%
%

%

%
%

\section{Conclusion}

We introduce \tool{}, the first \textit{in situ} interactive tool to address the challenges in anticipating potential harms in LLM-powered applications during prototyping.
By highlighting relevant AI incident reports and enabling AI prototypers to curate and modify LLM-generated use cases, stakeholders, and harms, \tool{} improves users' ability to independently anticipate potential risks associated with their prompts.
A user study with 42 AI prototypers shows that our tool is useful and usable.
\tool{} fosters a user-centric approach, encouraging creators to consider end-users, and cascading harms, and extend their awareness beyond immediate harms.
Our tool is open-source and readily adoptable.
We hope our work will inspire future research and development of responsible AI tools that target the early stages of the AI development process.

\begin{acks}
  We express our gratitude to all anonymous participants who took part in our co-design and evaluation studies.
  A special thank you to Jaemarie Solyst and Savvas Petridis for piloting our studies.
  We are deeply thankful to our three anonymous raters for rating the harms collected during the evaluation study.
  We are immensely grateful for the invaluable feedback provided by Parker Barnes,
  Carrie Cai,
  Alex Fiannaca,
  Tesh Goyal,
  Ellen Jiang,
  Minsuk Kahng,
  Shaun Kane,
  Donald Martin,
  Alicia Parrish,
  Adam Pearce,
  Savvas Petridis,
  Mahima Pushkarna,
  Dheeraj Rajagopal,
  Kevin Robinson,
  Taylor Roper,
  Negar Rostamzadeh,
  Renee Shelby,
  Jaemarie Solyst,
  Vivian Tsai,
  James Wexler,
  Ann Yuan, and
  Andrew Zaldivar.
  Our gratitude also extends to the anonymous Google employees who generously allowed us to use their prompts to design and prototype \tool{}.
  Furthermore, we are grateful to James Wexler, Paul Yang, Tulsee Doshi, and Marian Croak for their assistance in open-sourcing \tool{}.
  Lastly, we would like to acknowledge the anonymous reviewers for their detailed and helpful feedback.
\end{acks} 
\balance
\bibliographystyle{ACM-Reference-Format}
\bibliography{24-farsight}
\appendix
\clearpage

\onecolumn
\setcounter{figure}{0}
\setcounter{table}{0}
\renewcommand{\thetable}{S\arabic{table}}
\renewcommand{\thefigure}{S\arabic{figure}}

\section{Co-design User Study Interviews}
\label{appendix:evaluation-co-design-interview}

\subsection{Co-design Prototypes and Sketches}

\setlength{\belowcaptionskip}{-5pt}
\setlength{\abovecaptionskip}{7pt}
\begin{figure*}[h!]
  \includegraphics[width=1\linewidth]{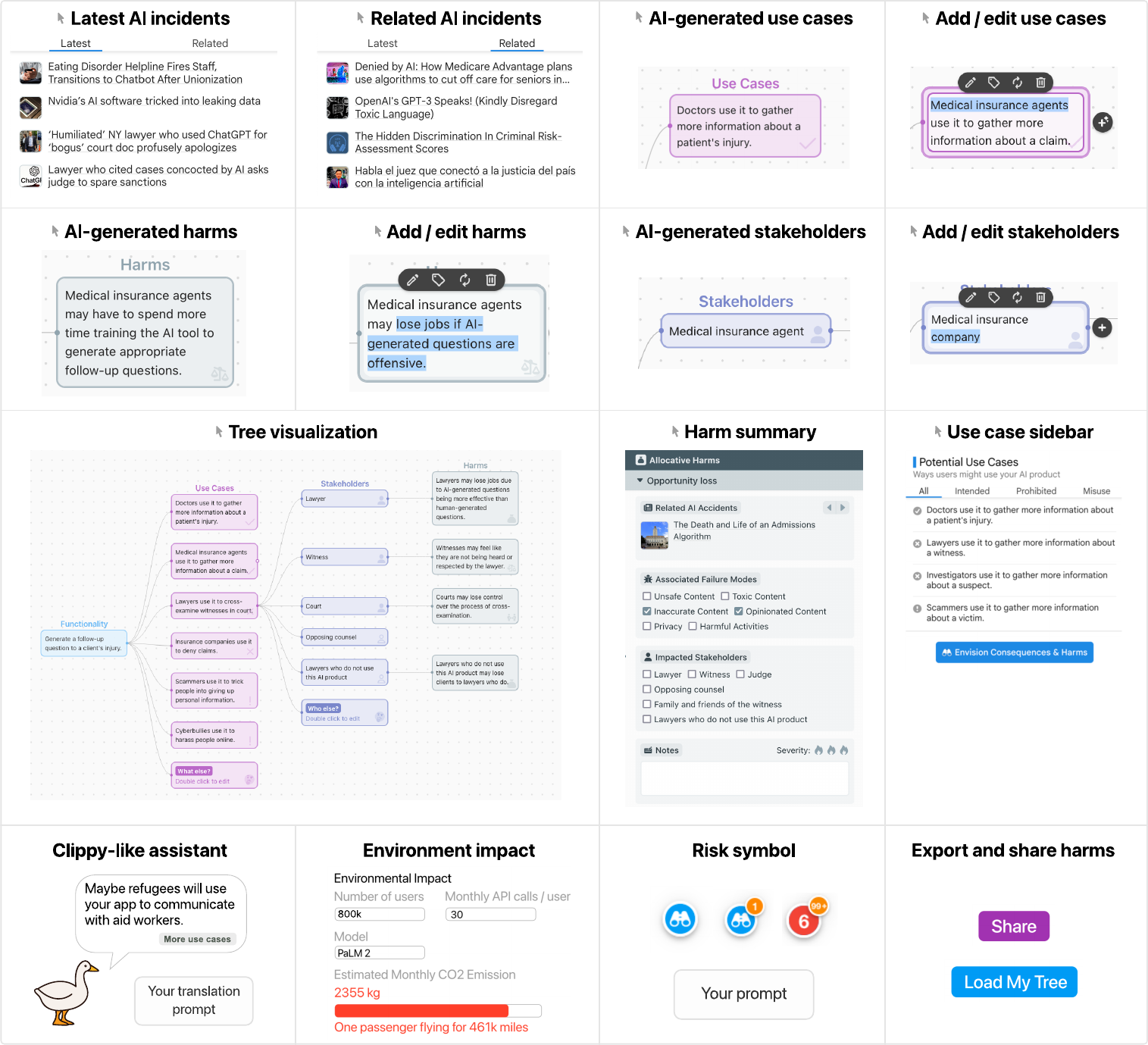}
  \caption[]{
    To evaluate our early \tool{} designs and generate more design ideas, we conducted a co-design study~(\autoref{sec:formative-study}) with 10 AI prototypers.
    Participants were asked to use our very early-stage design prototypes~(shown in cells labeled with \protect\vcenteredhbox{\includegraphics[height=6pt]{figures/icon-cursor}}) to envision potential harms associated with their application while thinking aloud.
    Participants were also presented with low-fidelity sketches for our other design ideas~(shown in cells in the last row).
    The ratings of our design ideas are shown in \autoref{fig:formative-feature}.
  }
  \Description{
    This image is a composite of various interface designs and sketches used in the co-design study. The interfaces are aimed at helping users identify and manage potential harms associated with AI applications.
    At the top left, there are panels titled "Latest AI Incidents," "Related AI Incidents," and "AI-generated use cases," each providing a list or method to input relevant AI incidents and use cases. Adjacent to these are panels for "AI-generated harms," "Add/edit harms," and "AI-generated stakeholders," for documenting and editing potential AI-related harms and stakeholders.

    Central to the image is a "Tree visualization" showing an interactive node-link diagram that connects different aspects such as use cases, harms, and stakeholders. To the right are panels titled "Harm summary" and "Use case sidebar," providing summarized information and detailed exploration of use cases.

    At the bottom, there are additional elements such as a "Clippy-like assistant" offering help, an "Environment impact" panel showing estimated CO2 emissions, a "Risk symbol" indicating the level of risk associated with the AI, and an "Export and share harms" feature for dissemination of findings.
  }
  \label{fig:appendix-prototypes}
\end{figure*}
\setlength{\belowcaptionskip}{0pt}
\setlength{\abovecaptionskip}{12pt}

\clearpage{}
\subsection{Interview 1 Questions}
\label{appendix:evaluation-co-design-interview-1}

\begin{itemize}[topsep=5pt, itemsep=0mm, parsep=2pt, leftmargin=14pt]
  \item Why do you use a prompt crafting tool?
  \item How have you used it most recently? Can you walk me through one of your example prompts?
  \item For your previous prompts, did you ever think about the potential societal impacts of your AI application / feature?
        \begin{itemize}[itemsep=0mm, parsep=2pt, leftmargin=14pt]
          \item It's OK if this wasn't part of your process.
          \item If yes, how did you think through or envision those potential impacts of your AI prototypes / ideas?
          \item If yes, can you share some examples of the types of impacts of your AI application that you considered? Including both positive and negative impacts.
          \item If no, how would you envision potential impacts of your AI prototypes / ideas?
        \end{itemize}

  \item What would motivate you to think more about the potential negative impacts of the applications you were prototyping?
\end{itemize}

\subsection{Interview 2 Rating Forms and Questions}
\label{appendix:evaluation-co-design-interview-2}

\begin{itemize}[topsep=5pt, itemsep=0mm, parsep=2pt, leftmargin=14pt]
  \item How do you think our design might fit into your prompting workflow?
        \begin{itemize}[itemsep=0mm, parsep=2pt, leftmargin=14pt]
          \item How do you think our design might fit into your typical AI application prototyping workflow?
        \end{itemize}
  \item What would these suggested use cases, stakeholders, and harms prompt you to think or do differently? (if not answered already)
  \item What would prevent you from using this tool?
  \item Other design ideas?
        \begin{itemize}[itemsep=0mm, parsep=2pt, leftmargin=14pt]
          \item What would encourage more use/engagement with the tool?
          \item What are other ways you could imagine raising awareness of potential responsible AI and safety issues?
        \end{itemize}
\end{itemize}

\section{Interface Details}
\label{appendix:system}

\subsection{Determine the Thresholds for Relevancy of AI Incident Reports}
\label{appendix:system-relevancy-thresholds}

We collect 1000 random internal prompts written by real AI prototypers.
Then, we compute the embedding similarity between these prompts and all AI incident reports~\cite{mcgregorPreventingRepeatedReal2020}.
We use the \texttt{20\%} and \texttt{70\%} cumulative density function cut-offs (\texttt{0.69} and \texttt{0.75}) of the max prompt-incident embedding distance as our thresholds for \vcenteredhbox{\includegraphics[height=9pt]{figures/tag-irrelevant}}, \vcenteredhbox{\includegraphics[height=9pt]{figures/tag-remotely-relevant}}, \vcenteredhbox{\includegraphics[height=9pt]{figures/tag-moderately-relevant}}.
Researchers can easily adjust these two thresholds (bounded between \texttt{0} and \texttt{1}) to calibrate an article's relevancy.

\setlength{\belowcaptionskip}{0pt}
\setlength{\abovecaptionskip}{8pt}
\begin{figure*}[tb]
  \includegraphics[width=1\linewidth]{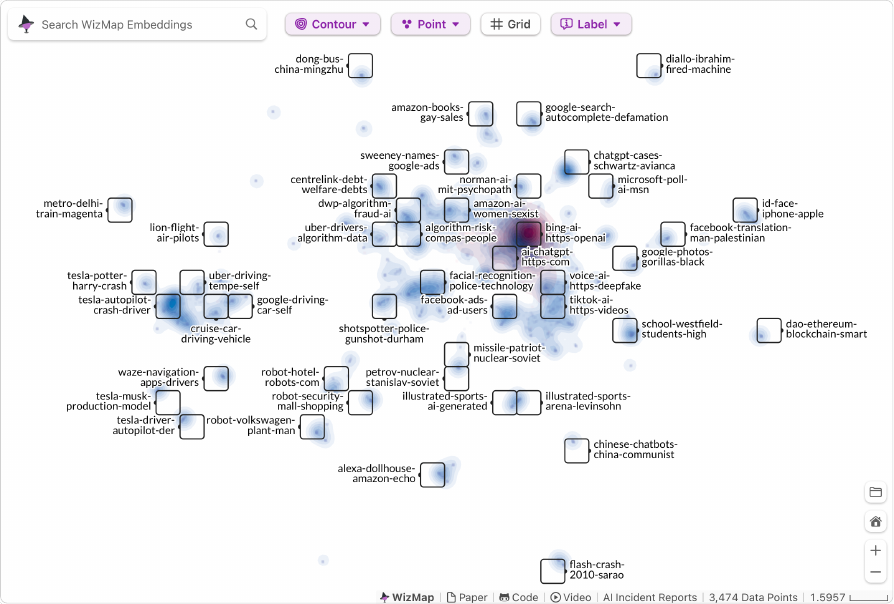}
  \caption[]{A visualization of the PaLM 2 embeddings of 3,474 AI incident reports~\cite{mcgregorPreventingRepeatedReal2020}~(blue dots and contour) and 153 Awesome ChatGPT Prompts~\cite{akinAwesomeChatGPTPrompts2022}~(red dots and contour).
    The embeddings' dimensions were reduced from 768 to 2 using UMAP~\cite{mcinnesUMAPUniformManifold2020} with default parameters.
    The rectangles and labels show the summaries of AI incident reports in high-density embedding neighborhoods.
    The summaries are automatically generated by WizMap~\cite{wangWizMapScalableInteractive2023}.
    The visualization reveals different clusters in the AI incident reports, such as incidents related to autonomous driving cars in the bottom left and machine translation on the right.
    The overlap between the red and blue contours indicates that user prompts can be in close proximity to AI incident reports in the 2D embedding space.
    This observation inspires us to use high-dimensional embedding similarities to calculate the alert levels in \tool{}~(\autoref{sec:system:symbol}).
    Note that in this example, the 153 user prompts form a cluster due to the primary focus of AwesomeGPT prompts on conversational agents.
    The distribution of our 1,000 internal prompts~(featuring classification, translation, code generation, etc.) is more spread out.
    For an interactive version of this visualization, visit \myhref{https://poloclub.github.io/wizmap/?dataURL=https\%3A\%2F\%2Fhuggingface.co\%2Fdatasets\%2Fxiaohk\%2Fembeddings\%2Fresolve\%2Fmain\%2Fai-incident\%2Fdata.ndjson&gridURL=https\%3A\%2F\%2Fhuggingface.co\%2Fdatasets\%2Fxiaohk\%2Fembeddings\%2Fresolve\%2Fmain\%2Fai-incident\%2Fgrid.json}{WizMap}.
  }
  \Description{
    A figure shows the embedding visualization of 3,474 AI incident reports and 153 user prompts.
  }
  \label{fig:wizmap}
\end{figure*}
\setlength{\belowcaptionskip}{0pt}
\setlength{\abovecaptionskip}{12pt}

\clearpage{}
\subsection{Example \tool{} Output}
\newlength{\availablewidth}
\setlength{\availablewidth}{\linewidth - 6\tabcolsep}

\newlength{\usecasecolumnwidth}
\setlength{\usecasecolumnwidth}{0.1\availablewidth \relax}

\newlength{\stakeholdercolumnwidth}
\setlength{\stakeholdercolumnwidth}{0.15\availablewidth \relax}

\aptLtoX[graphic=no,type=html]{\begin{longtable}{p{7pc}|p{7pc}|p{26pc}}
\caption{\tool{} uses LLMs to generate use cases, stakeholders, and harms associated with AI apps based on a given prompt.
This table shows the outputs of \tool{} for a toxicity classifier prompt~(\S~\ref{appendix:evaluation-design-intervention}).
By default, the \windowview{} displays six use cases (arranged by two \includegraphics[height=9pt]{figures/tag-intended}, two \includegraphics[height=9pt]{figures/tag-high-stakes}, and two \includegraphics[height=9pt]{figures/tag-misuse}), five stakeholders (arranged by three direct stakeholders and two indirect stakeholders) for each use case, and one harm for each stakeholder.
Users can modify or delete existing outputs and ask \tool{} to generate additional content (e.g., generating harms for a newly added stakeholder).}
\label{table:farsight-output}
\vspace{-7pt}
\\
\textbf{Use Cases}                                                                                                                        & \textbf{Stakeholders}                                                                                  & \textbf{Harms}                                                                                                                      \\
\hline
\multirow{15}{\usecasecolumnwidth}[0pt]{Online moderators use it to remove toxic content from social media.~\includegraphics[height=6pt]{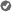}} & \multirow{3}{\stakeholdercolumnwidth}{Online moderators}                                               & Online moderators may be exposed to toxic content while using the AI tool.                                                          \\
&                                                                                                        & Online moderators may feel frustrated or anxious about the accuracy of the AI tool.                                                 \\
&                                                                                                        & Online moderators may lose their jobs due to AI-generated content being offensive.                                                  \\ \cline{2-3}
& \multirow{3}{\stakeholdercolumnwidth}{Social media users}                                              & Social media users may be unfairly censored by the AI tool.                                                                         \\
&                                                                                                        & Social media users may feel frustrated or anxious about the accuracy of the AI tool.                                                \\
&                                                                                                        & Social media users may feel like they are not being heard by the moderators.                                                        \\ \cline{2-3} \cline{2-3}
& \multirow{3}{\stakeholdercolumnwidth}[0pt]{The company}                                                & The company may be sued for defamation if the AI tool mislabels a post as toxic.                                                    \\
&                                                                                                        & The company may lose customers if the AI tool is not accurate enough.                                                               \\
&                                                                                                        & The company may be accused of bias if the AI tool is not fair in its classification of toxic content.                               \\ \cline{2-3}
& \multirow{3}{\stakeholdercolumnwidth}[-7pt]{Social media users who do not use this AI product}         & Social media users who do not use this AI product may be exposed to toxic content.                                                  \\
&                                                                                                        & Social media users who do not use this AI product may feel like they are not being heard by the moderators.                         \\
&                                                                                                        & Social media users who do not use this AI product may feel frustrated or anxious about the amount of toxic content on social media. \\ \cline{2-3}
& \multirow{3}{\stakeholdercolumnwidth}[0pt]{Social media advertisers}                                   & Social media advertisers may lose revenue due to toxic content being removed from social media.                                     \\
&                                                                                                        & Social media advertisers may have to spend more time and money on creating non-toxic content.                                       \\
&                                                                                                        & Social media advertisers may feel frustrated or anxious about the accuracy of the AI tool.                                          \\
\hline
\multirow{15}{\usecasecolumnwidth}[-24pt]{Customer service agents use it to identify abusive customers.~\includegraphics[height=6pt]{figures/icon-intended}}     & \multirow{3}{\stakeholdercolumnwidth}{Customer service agents}                                         & Customer service agents may feel stressed or anxious about the accuracy of the AI tool.                                             \\
&                                                                                                        & Customer service agents may be accused of bias if the AI tool misidentifies abusive customers.                                      \\
&                                                                                                        & Customer service agents may lose the skill to identify abusive customers independently.                                             \\ \cline{2-3}
& \multirow{3}{\stakeholdercolumnwidth}{Abusive customers}                                               & Abusive customers may be denied service due to being labeled as toxic.                                                              \\
&                                                                                                        & Abusive customers may feel frustrated or anxious about being labeled as toxic.                                                      \\
&                                                                                                        & Abusive customers may be more likely to engage in abusive behavior in the future.                                                   \\ \cline{2-3}
& \multirow{3}{\stakeholdercolumnwidth}[-6pt]{The company}                                               & The company may be sued for defamation if the AI tool mislabels a customer as abusive.                                              \\
&                                                                                                        & The company may face increased regulatory scrutiny if the AI tool is found to be inaccurate or biased.                              \\
&                                                                                                        & The company may lose customers if the AI tool is perceived as biased against certain groups.                                        \\ \cline{2-3}
& \multirow{3}{\stakeholdercolumnwidth}[-7pt]{Customers who are not abusive}                             & Customers who are not abusive may be misidentified as abusive and treated poorly by customer service agents.                        \\
&                                                                                                        & Customers who are not abusive may feel frustrated or anxious about being misidentified as abusive.                                  \\
&                                                                                                        & Customers who are not abusive may lose trust in customer service agents.                                                            \\ \cline{2-3}
& \multirow{3}{\stakeholdercolumnwidth}[6pt]{Customers who are misidentified as abusive}                 & Customers who are misidentified as abusive may be denied service or treated poorly by customer service agents.                      \\
&                                                                                                        & Customers who are misidentified as abusive may feel frustrated or anxious about being treated poorly.                               \\
&                                                                                                        & Customers who are misidentified as abusive may lose trust in customer service agents.                                               \\
\hline
& \multirow{3}{\stakeholdercolumnwidth}[-5pt]{Law enforcement}                                           & Law enforcement may make false arrests or detentions of innocent people due to AI-generated toxicity labels.                        \\
&                                                                                                        & Law enforcement may lose trust in AI tools due to false positives and negatives.                                                    \\
&                                                                                                        & Law enforcement may be biased against certain groups of people due to AI-generated toxicity labels.                                 \\ \cline{2-3}
\multirow{15}{\usecasecolumnwidth}[-8pt]{Law enforcement uses it to identify potential terrorists.~\includegraphics[height=6pt]{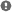}}       & \multirow{3}{\stakeholdercolumnwidth}[-5pt]{Potential terrorists}                                      & Potential terrorists may be unfairly targeted by law enforcement due to AI-generated misclassifications.                            \\
&                                                                                                        & Potential terrorists may feel frustrated or anxious about being unfairly targeted by law enforcement.                               \\
&                                                                                                        & Potential terrorists may be denied opportunities due to AI-generated misclassifications.                                            \\ \cline{2-3}
& \multirow{3}{\stakeholdercolumnwidth}[-6pt]{Victims of terrorism}                                      & Victims of terrorism may be re-traumatized by AI-generated content that glorifies or incites violence against them.                 \\
&                                                                                                        & Victims of terrorism may feel unsafe or threatened by AI-generated content that glorifies or incites violence against them.         \\
&                                                                                                        & Victims of terrorism may feel like their voices are not being heard by society.                                                     \\ \cline{2-3}
& \multirow{3}{\stakeholdercolumnwidth}[-6pt]{Civil rights groups}                                       & Civil rights groups may be unfairly targeted by law enforcement due to AI-generated false positives.                                \\
&                                                                                                        & Civil rights groups may lose trust in law enforcement due to AI-generated false positives.                                          \\
&                                                                                                        & Civil rights groups may have to spend more time and resources fighting against AI-generated false positives.                        \\ \cline{2-3}
& \multirow{3}{\stakeholdercolumnwidth}[-5pt]{People who are falsely identified as potential terrorists} & People who are falsely identified as potential terrorists may be subject to increased surveillance, harassment, and discrimination. \\
&                                                                                                        & People who are falsely identified as potential terrorists may lose their jobs, housing, or other opportunities.                     \\
&                                                                                                        & People who are falsely identified as potential terrorists may be detained or imprisoned without due process.                        \\
\hline
\multirow{15}{\usecasecolumnwidth}[0pt]{Military uses it to identify potential insurgents.~\includegraphics[height=6pt]{figures/icon-high-stakes}}               & \multirow{3}{\stakeholdercolumnwidth}{Military}                                                        & Military may use AI to target and harm innocent civilians.                                                                          \\
&                                                                                                        & Military may lose trust from the public due to its use of AI.                                                                       \\
&                                                                                                        & Military may be pressured to use AI to stay competitive with other militaries.                                                      \\ \cline{2-3}
& \multirow{3}{\stakeholdercolumnwidth}[-12pt]{Insurgency}                                               & Insurgency may be able to evade detection by using AI-generated text that is not classified as toxic.                               \\
&                                                                                                        & Insurgency may be able to recruit more members by using AI-generated text that is not classified as toxic.                          \\
&                                                                                                        & Insurgency may be able to spread propaganda more effectively by using AI-generated text that is not classified as toxic.            \\ \cline{2-3}
& \multirow{3}{\stakeholdercolumnwidth}{Governments}                                                     & Governments may use AI to target and surveil innocent people.                                                                       \\
&                                                                                                        & Governments may use AI to justify violence against innocent people.                                                                 \\
&                                                                                                        & Governments may lose control over how AI is used to target and surveil people.                                                      \\ \cline{2-3}
& \multirow{3}{\stakeholdercolumnwidth}[6pt]{Civilians who are wrongly identified}                       & Civilians who are wrongly identified may be subject to violence or discrimination.                                                  \\
&                                                                                                        & Civilians who are wrongly identified may lose their freedom or property.                                                            \\
&                                                                                                        & Civilians who are wrongly identified may experience emotional distress.                                                             \\ \cline{2-3}
& \multirow{3}{\stakeholdercolumnwidth}{Civilians who are not identified as insurgents}                  & Civilians who are not identified as insurgents may be subject to violence or discrimination.                                        \\
&                                                                                                        & Civilians who are not identified as insurgents may lose their freedom or property.                                                  \\
&                                                                                                        & Civilians who are not identified as insurgents may experience emotional distress.                                                   \\
\hline
& \multirow{3}{\stakeholdercolumnwidth}{Hate groups}                                                     & Hate groups may be able to recruit more people due to the AI tool.                                                                  \\
&                                                                                                        & Hate groups may be able to spread their ideology more effectively due to the AI tool.                                               \\
&                                                                                                        & Hate groups may be able to avoid detection by law enforcement due to the AI tool.                                                   \\ \cline{2-3}
& \multirow{3}{\stakeholdercolumnwidth}[-12pt]{Potential recruits}                                       & Potential recruits may be exposed to toxic content that could lead to radicalization.                                               \\
&                                                                                                        & Potential recruits may feel alienated and isolated from society due to their exposure to toxic content.                             \\
&                                                                                                        & Potential recruits may lose the opportunity to develop healthy relationships with people outside of the hate group.                 \\ \cline{2-3}
\multirow{15}{\usecasecolumnwidth}[80pt]{Hate groups use it to identify potential recruits.~\includegraphics[height=6pt]{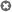}}                   & \multirow{3}{\stakeholdercolumnwidth}{The company}                                                     & The company may be held liable for hate speech generated by its AI product.                                                         \\
&                                                                                                        & The company may face increased regulation due to its AI product's misuse.                                                           \\
&                                                                                                        & The company may lose customers due to negative publicity.                                                                           \\ \cline{2-3}
& \multirow{3}{\stakeholdercolumnwidth}{People who are targeted by hate groups}                          & People who are targeted by hate groups may be harassed or threatened by hate groups.                                                \\
&                                                                                                        & People who are targeted by hate groups may feel anxious or fearful of being targeted again.                                         \\
&                                                                                                        & People who are targeted by hate groups may lose their sense of safety and belonging.                                                \\ \cline{2-3}
& \multirow{3}{\stakeholdercolumnwidth}[-7pt]{People who are not targeted by hate groups}                & People who are not targeted by hate groups may be exposed to toxic content.                                                         \\
&                                                                                                        & People who are not targeted by hate groups may be harassed or threatened by hate groups.                                            \\
&                                                                                                        & People who are not targeted by hate groups may feel anxious or fearful about the rise of hate groups.                               \\
\hline
\multirow{15}{\usecasecolumnwidth}[-20pt]{Scammers use it to identify potential victims.~\includegraphics[height=6pt]{figures/icon-misuse}}                      & \multirow{3}{\stakeholdercolumnwidth}{Victims of scams}                                                & Victims of scams may lose money or personal information.                                                                            \\
&                                                                                                        & Victims of scams may experience emotional distress.                                                                                 \\
&                                                                                                        & Victims of scams may lose trust in others.                                                                                          \\ \cline{2-3}
& \multirow{3}{\stakeholdercolumnwidth}[-12pt]{Scammers}                                                 & Scammers may be able to target more vulnerable people with their scams.                                                             \\
&                                                                                                        & Scammers may be able to avoid detection by using AI-generated text that is less likely to be flagged as toxic.                      \\
&                                                                                                        & Scammers may be able to increase their profits by using AI-generated text to target more people.                                    \\ \cline{2-3}
& \multirow{3}{\stakeholdercolumnwidth}{The company}                                                     & The company may be held liable for the harm caused by scammers using the AI tool.                                                   \\
&                                                                                                        & The company may lose customers due to negative publicity.                                                                           \\
&                                                                                                        & The company may face increased regulatory scrutiny.                                                                                 \\ \cline{2-3}
& \multirow{3}{\stakeholdercolumnwidth}[-12pt]{Victim's financial institutions}                          & Victim's financial institutions may lose money due to scammers using AI to identify potential victims.                              \\
&                                                                                                        & Victim's financial institutions may face increased legal liability due to scammers using AI to identify potential victims.          \\
&                                                                                                        & Victim's financial institutions may lose trust from their customers due to scammers using AI to identify potential victims.         \\ \cline{2-3}
& \multirow{3}{\stakeholdercolumnwidth}{Family and friends of the victim}                                & Family and friends of the victim may be harmed by the scammer's actions.                                                            \\
&                                                                                                        & Family and friends of the victim may be harmed by the scammer's emotional abuse.                                                    \\
&                                                                                                        & Family and friends of the victim may be harmed by the scammer's financial abuse.                                                    \\
\end{longtable}}{{\begin{spacing}{1.25}
\small
\vspace{-3pt}
\begin{longtable}[\linewidth]{m{\usecasecolumnwidth}!{\color{grayIII}\vrule}m{\stakeholdercolumnwidth}!{\color{grayIII}\vrule}m{0.75\availablewidth \relax}}
\captionsetup{type=table}
\caption[]{\tool{} uses LLMs to generate use cases, stakeholders, and harms associated with AI apps based on a given prompt.
This table shows the outputs of \tool{} for a toxicity classifier prompt~(\autoref{appendix:evaluation-design-intervention}).
By default, the \windowview{} displays six use cases (arranged by two \vcenteredhbox{\includegraphics[height=9pt]{figures/tag-intended}}, two \vcenteredhbox{\includegraphics[height=9pt]{figures/tag-high-stakes}}, and two \vcenteredhbox{\includegraphics[height=9pt]{figures/tag-misuse}}), five stakeholders (arranged by three direct stakeholders and two indirect stakeholders) for each use case, and one harm for each stakeholder.
Users can modify or delete existing outputs and ask \tool{} to generate additional content (e.g., generating harms for a newly added stakeholder).}
\Description{This table shows the outputs of Farsight for a toxicity classifier prompt. Three columns are "Use Cases", "Stakeholders", and "Harms."}
\label{table:farsight-output}
\vspace{-7pt}
\\
\specialrule{1.2pt}{1pt}{0pt}
\textbf{Use Cases}                                                                                                                        & \textbf{Stakeholders}                                                                                  & \textbf{Harms}                                                                                                                      \\
\arrayrulecolor{grayIII}
\hline
\endfirsthead
\arrayrulecolor{grayIX}
\specialrule{1.2pt}{1pt}{0pt}
\textbf{Use Cases}                                                                                                                        & \textbf{Stakeholders}                                                                                  & \textbf{Harms}                                                                                                                      \\
\arrayrulecolor{grayIII}
\hline
\endhead
\multirow{15}{\usecasecolumnwidth}[-6pt]{Online moderators use it to remove toxic content from social media.~\vcenteredhbox{\includegraphics[height=6pt]{figures/icon-intended}}} & \multirow{3}{\stakeholdercolumnwidth}{Online moderators}                                               & Online moderators may be exposed to toxic content while using the AI tool.                                                          \\
&                                                                                                        & Online moderators may feel frustrated or anxious about the accuracy of the AI tool.                                                 \\
&                                                                                                        & Online moderators may lose their jobs due to AI-generated content being offensive.                                                  \\ \cline{2-3}
& \multirow{3}{\stakeholdercolumnwidth}{Social media users}                                              & Social media users may be unfairly censored by the AI tool.                                                                         \\
&                                                                                                        & Social media users may feel frustrated or anxious about the accuracy of the AI tool.                                                \\
&                                                                                                        & Social media users may feel like they are not being heard by the moderators.                                                        \\ \cline{2-3} \cline{2-3}
& \multirow{3}{\stakeholdercolumnwidth}[0pt]{The company}                                                & The company may be sued for defamation if the AI tool mislabels a post as toxic.                                                    \\
&                                                                                                        & The company may lose customers if the AI tool is not accurate enough.                                                               \\
&                                                                                                        & The company may be accused of bias if the AI tool is not fair in its classification of toxic content.                               \\ \cline{2-3}
& \multirow{3}{\stakeholdercolumnwidth}[-7pt]{Social media users who do not use this AI product}         & Social media users who do not use this AI product may be exposed to toxic content.                                                  \\
&                                                                                                        & Social media users who do not use this AI product may feel like they are not being heard by the moderators.                         \\
&                                                                                                        & Social media users who do not use this AI product may feel frustrated or anxious about the amount of toxic content on social media. \\ \cline{2-3}
& \multirow{3}{\stakeholdercolumnwidth}[0pt]{Social media advertisers}                                   & Social media advertisers may lose revenue due to toxic content being removed from social media.                                     \\
&                                                                                                        & Social media advertisers may have to spend more time and money on creating non-toxic content.                                       \\
&                                                                                                        & Social media advertisers may feel frustrated or anxious about the accuracy of the AI tool.                                          \\
\hline
\multirow{15}{\usecasecolumnwidth}[0pt]{Customer service agents use it to identify abusive customers.~\vcenteredhbox{\includegraphics[height=6pt]{figures/icon-intended}}}     & \multirow{3}{\stakeholdercolumnwidth}{Customer service agents}                                         & Customer service agents may feel stressed or anxious about the accuracy of the AI tool.                                             \\
&                                                                                                        & Customer service agents may be accused of bias if the AI tool misidentifies abusive customers.                                      \\
&                                                                                                        & Customer service agents may lose the skill to identify abusive customers independently.                                             \\ \cline{2-3}
& \multirow{3}{\stakeholdercolumnwidth}{Abusive customers}                                               & Abusive customers may be denied service due to being labeled as toxic.                                                              \\
&                                                                                                        & Abusive customers may feel frustrated or anxious about being labeled as toxic.                                                      \\
&                                                                                                        & Abusive customers may be more likely to engage in abusive behavior in the future.                                                   \\ \cline{2-3}
& \multirow{3}{\stakeholdercolumnwidth}[0pt]{The company}                                               & The company may be sued for defamation if the AI tool mislabels a customer as abusive.                                              \\
&                                                                                                        & The company may face increased regulatory scrutiny if the AI tool is found to be inaccurate or biased.                              \\
&                                                                                                        & The company may lose customers if the AI tool is perceived as biased against certain groups.                                        \\ \cline{2-3}
& \multirow{3}{\stakeholdercolumnwidth}[0pt]{Customers who are not abusive}                             & Customers who are not abusive may be misidentified as abusive and treated poorly by customer service agents.                        \\
&                                                                                                        & Customers who are not abusive may feel frustrated or anxious about being misidentified as abusive.                                  \\
&                                                                                                        & Customers who are not abusive may lose trust in customer service agents.                                                            \\ \cline{2-3}
& \multirow{3}{\stakeholdercolumnwidth}[0pt]{Customers who are misidentified as abusive}                 & Customers who are misidentified as abusive may be denied service or treated poorly by customer service agents.                      \\
&                                                                                                        & Customers who are misidentified as abusive may feel frustrated or anxious about being treated poorly.                               \\
&                                                                                                        & Customers who are misidentified as abusive may lose trust in customer service agents.                                               \\
\hline
& \multirow{3}{\stakeholdercolumnwidth}[0pt]{Law enforcement}                                           & Law enforcement may make false arrests or detentions of innocent people due to AI-generated toxicity labels.                        \\
&                                                                                                        & Law enforcement may lose trust in AI tools due to false positives and negatives.                                                    \\
&                                                                                                        & Law enforcement may be biased against certain groups of people due to AI-generated toxicity labels.                                 \\ \cline{2-3}
\multirow{15}{\usecasecolumnwidth}[60pt]{Law enforcement uses it to identify potential terrorists.~\vcenteredhbox{\includegraphics[height=6pt]{figures/icon-high-stakes}}}       & \multirow{3}{\stakeholdercolumnwidth}[0pt]{Potential terrorists}                                      & Potential terrorists may be unfairly targeted by law enforcement due to AI-generated misclassifications.                            \\
&                                                                                                        & Potential terrorists may feel frustrated or anxious about being unfairly targeted by law enforcement.                               \\
&                                                                                                        & Potential terrorists may be denied opportunities due to AI-generated misclassifications.                                            \\ \cline{2-3}
& \multirow{3}{\stakeholdercolumnwidth}[-6pt]{Victims of terrorism}                                      & Victims of terrorism may be re-traumatized by AI-generated content that glorifies or incites violence against them.                 \\
&                                                                                                        & Victims of terrorism may feel unsafe or threatened by AI-generated content that glorifies or incites violence against them.         \\
&                                                                                                        & Victims of terrorism may feel like their voices are not being heard by society.                                                     \\ \cline{2-3}
& \multirow{3}{\stakeholdercolumnwidth}[0pt]{Civil rights groups}                                       & Civil rights groups may be unfairly targeted by law enforcement due to AI-generated false positives.                                \\
&                                                                                                        & Civil rights groups may lose trust in law enforcement due to AI-generated false positives.                                          \\
&                                                                                                        & Civil rights groups may have to spend more time and resources fighting against AI-generated false positives.                        \\ \cline{2-3}
& \multirow{3}{\stakeholdercolumnwidth}[0pt]{People who are falsely identified as potential terrorists} & People who are falsely identified as potential terrorists may be subject to increased surveillance, harassment, and discrimination. \\
&                                                                                                        & People who are falsely identified as potential terrorists may lose their jobs, housing, or other opportunities.                     \\
&                                                                                                        & People who are falsely identified as potential terrorists may be detained or imprisoned without due process.                        \\
\hline
\multirow{15}{\usecasecolumnwidth}[0pt]{Military uses it to identify potential insurgents.~\vcenteredhbox{\includegraphics[height=6pt]{figures/icon-high-stakes}}}               & \multirow{3}{\stakeholdercolumnwidth}{Military}                                                        & Military may use AI to target and harm innocent civilians.                                                                          \\
&                                                                                                        & Military may lose trust from the public due to its use of AI.                                                                       \\
&                                                                                                        & Military may be pressured to use AI to stay competitive with other militaries.                                                      \\ \cline{2-3}
& \multirow{3}{\stakeholdercolumnwidth}[-6pt]{Insurgency}                                               & Insurgency may be able to evade detection by using AI-generated text that is not classified as toxic.                               \\
&                                                                                                        & Insurgency may be able to recruit more members by using AI-generated text that is not classified as toxic.                          \\
&                                                                                                        & Insurgency may be able to spread propaganda more effectively by using AI-generated text that is not classified as toxic.            \\ \cline{2-3}
& \multirow{3}{\stakeholdercolumnwidth}{Governments}                                                     & Governments may use AI to target and surveil innocent people.                                                                       \\
&                                                                                                        & Governments may use AI to justify violence against innocent people.                                                                 \\
&                                                                                                        & Governments may lose control over how AI is used to target and surveil people.                                                      \\ \cline{2-3}
& \multirow{3}{\stakeholdercolumnwidth}[0pt]{Civilians who are wrongly identified}                       & Civilians who are wrongly identified may be subject to violence or discrimination.                                                  \\
&                                                                                                        & Civilians who are wrongly identified may lose their freedom or property.                                                            \\
&                                                                                                        & Civilians who are wrongly identified may experience emotional distress.                                                             \\ \cline{2-3}
& \multirow{3}{\stakeholdercolumnwidth}{Civilians who are not identified as insurgents}                  & Civilians who are not identified as insurgents may be subject to violence or discrimination.                                        \\
&                                                                                                        & Civilians who are not identified as insurgents may lose their freedom or property.                                                  \\
&                                                                                                        & Civilians who are not identified as insurgents may experience emotional distress.                                                   \\
\hline
& \multirow{3}{\stakeholdercolumnwidth}{Hate groups}                                                     & Hate groups may be able to recruit more people due to the AI tool.                                                                  \\
&                                                                                                        & Hate groups may be able to spread their ideology more effectively due to the AI tool.                                               \\
&                                                                                                        & Hate groups may be able to avoid detection by law enforcement due to the AI tool.                                                   \\ \cline{2-3}
& \multirow{3}{\stakeholdercolumnwidth}[-6pt]{Potential recruits}                                       & Potential recruits may be exposed to toxic content that could lead to radicalization.                                               \\
&                                                                                                        & Potential recruits may feel alienated and isolated from society due to their exposure to toxic content.                             \\
&                                                                                                        & Potential recruits may lose the opportunity to develop healthy relationships with people outside of the hate group.                 \\ \cline{2-3}
\multirow{15}{\usecasecolumnwidth}[80pt]{Hate groups use it to identify potential recruits.~\vcenteredhbox{\includegraphics[height=6pt]{figures/icon-misuse}}}                   & \multirow{3}{\stakeholdercolumnwidth}{The company}                                                     & The company may be held liable for hate speech generated by its AI product.                                                         \\
&                                                                                                        & The company may face increased regulation due to its AI product's misuse.                                                           \\
&                                                                                                        & The company may lose customers due to negative publicity.                                                                           \\ \cline{2-3}
& \multirow{3}{\stakeholdercolumnwidth}{People who are targeted by hate groups}                          & People who are targeted by hate groups may be harassed or threatened by hate groups.                                                \\
&                                                                                                        & People who are targeted by hate groups may feel anxious or fearful of being targeted again.                                         \\
&                                                                                                        & People who are targeted by hate groups may lose their sense of safety and belonging.                                                \\ \cline{2-3}
& \multirow{3}{\stakeholdercolumnwidth}[0pt]{People who are not targeted by hate groups}                & People who are not targeted by hate groups may be exposed to toxic content.                                                         \\
&                                                                                                        & People who are not targeted by hate groups may be harassed or threatened by hate groups.                                            \\
&                                                                                                        & People who are not targeted by hate groups may feel anxious or fearful about the rise of hate groups.                               \\
\hline
\multirow{15}{\usecasecolumnwidth}[24pt]{Scammers use it to identify potential victims.~\vcenteredhbox{\includegraphics[height=6pt]{figures/icon-misuse}}}                      & \multirow{3}{\stakeholdercolumnwidth}{Victims of scams}                                                & Victims of scams may lose money or personal information.                                                                            \\
&                                                                                                        & Victims of scams may experience emotional distress.                                                                                 \\
&                                                                                                        & Victims of scams may lose trust in others.                                                                                          \\ \cline{2-3}
& \multirow{3}{\stakeholdercolumnwidth}[0pt]{Scammers}                                                 & Scammers may be able to target more vulnerable people with their scams.                                                             \\
&                                                                                                        & Scammers may be able to avoid detection by using AI-generated text that is less likely to be flagged as toxic.                      \\
&                                                                                                        & Scammers may be able to increase their profits by using AI-generated text to target more people.                                    \\ \cline{2-3}
& \multirow{3}{\stakeholdercolumnwidth}{The company}                                                     & The company may be held liable for the harm caused by scammers using the AI tool.                                                   \\
&                                                                                                        & The company may lose customers due to negative publicity.                                                                           \\
&                                                                                                        & The company may face increased regulatory scrutiny.                                                                                 \\ \cline{2-3}
& \multirow{3}{\stakeholdercolumnwidth}[6pt]{Victim's financial institutions}                          & Victim's financial institutions may lose money due to scammers using AI to identify potential victims.                              \\
&                                                                                                        & Victim's financial institutions may face increased legal liability due to scammers using AI to identify potential victims.          \\
&                                                                                                        & Victim's financial institutions may lose trust from their customers due to scammers using AI to identify potential victims.         \\ \cline{2-3}
& \multirow{3}{\stakeholdercolumnwidth}{Family and friends of the victim}                                & Family and friends of the victim may be harmed by the scammer's actions.                                                            \\
&                                                                                                        & Family and friends of the victim may be harmed by the scammer's emotional abuse.                                                    \\
&                                                                                                        & Family and friends of the victim may be harmed by the scammer's financial abuse.                                                    \\
\arrayrulecolor{grayIX}
\specialrule{1.2pt}{0pt}{0pt}
\end{longtable}\end{spacing}}}

\section{Evaluation User Study Prompts}
\label{appendix:evaluation-prompt}

\subsection{Pre-task: Email Summarizer (H1)}
\label{appendix:evaluation-design-pre}

\textbf{AI feature functionality:} Shorten and improve a user’s email

\noindent\textbf{Context:} Your team is adding a new AI feature to an email client

\noindent\textbf{Text prompt:}
\begin{lstlisting}
You are an expert in writing emails. Rewrite the given emails to make them shorter and better.

Email: "Good morning, Alex. I wanted to write you this note because I think we should schedule a follow-up meeting. I am thinking we could meet next week, would that work for you? I think the deck is looking solid but I have some thoughts on it that I would like to share with you."
Rewrite: "Morning. Let's schedule a follow-up. Next week works? Deck's solid, but I have thoughts."

Email: "Alex, soooo hey, I think the current prototype isn't working so well unfortunately. Ok so how about this as a proposal for a plan: I think we should schedule a meeting to chat early next week. And during that time we can try to debug it live together. What do you think? Nicole"
Rewrite: [AI output]
\end{lstlisting}

\subsection{Intervention: Toxicity Classifier (H2)}
\label{appendix:evaluation-design-intervention}

\textbf{AI feature functionality:} Detect toxic text content

\noindent\textbf{Context:} Your team is developing a social media content moderation tool (e.g., API service)

\noindent\textbf{Text prompt:}
\begin{lstlisting}
Classify if the input sentence is toxic or non-toxic.

input: I will remove all your organs in alphabetical order.
output: toxic

input: Why is it that every Walmart worker chomps on their gum like a cow.
output: toxic

input: He is not living proof that gay conversion therapy works.
output: [AI output]
\end{lstlisting}
\textbf{Prompt engineering:}
Participants were asked to modify the prompt to alter the model's prediction on the test input sentence: ``\textit{He is not living proof that gay conversion therapy works.}'' from \texttt{toxic} to \texttt{non-toxic}.
We intentionally formulated the initial prompt to make the LLM predict the given test input sentence as \texttt{toxic}, as we exclusively included toxic examples in the prompt.

\subsection{Post-task: Article Summarizer (H3)}
\label{appendix:evaluation-design-post}

\textbf{AI feature functionality:} Summarize an article in one sentence

\noindent\textbf{Context:} Your team is adding an AI feature to a text editor software

\noindent\textbf{Text prompt:}
\begin{lstlisting}
I want you to act as a text summarizer. You will summarize my article in one sentence:

Article: Eighty years ago today - a powerful explosion rocked the small community of Dawson Creek on February 13th, 1943. The blast reached Spirit River with enough strength left to rattle dishes in cupboards. In 1943, the American Army was in Dawson Creek (population 500), working on the Alaska Highway. The old town was surrounded by makeshift barracks and storage warehouses for construction equipment. One, a livery barn in the centre of the commercial block, housed thousands of miles of copper wire, kegs of nails, spikes, cross-arm braces, hammers, crowbars, tires, and other assorted tools. Two hundred cases of percussion caps and a truckload of dynamite were also stored in the handy location. Somehow a fire started. The inevitable explosion followed soon after, incinerating whatever was in the core zone and blowing people off their feet in expanding circles throughout the town.  Fire spread rapidly and eventually only one building remained standing - the Co-op store - but it was wrecked and looted. Dorthea Calverley has written a fascinating personal account of the disaster.
Summary: A powerful explosion in Dawson Creek on February 13th, 1943 destroyed most of the town.
\end{lstlisting}

\subsection{Alternative: Math Tutor (H4)}
\label{appendix:evaluation-design-alternative}

\textbf{AI feature functionality:} Answer math-related questions.

\noindent\textbf{Context:} Your team is developing a math tutoring app (e.g., mobile app)

\noindent\textbf{Text prompt}
\begin{lstlisting}
You are a great math teacher. Given some mathematical equations or concepts, your will explain them in easy-to-understand terms. You will provide more than the basic definition and help the student to learn the underlying reasons.

User: What's the upper bound and lower bound of sin()?
Teacher: The upper bound of sin(x) is 1, and the lower bound is -1. This is because sin(x) is a periodic function with a period of 2pi, meaning that it repeats itself every 2pi radians. The maximum value of sin(x) occurs when x = pi/2, and the minimum value occurs when x = -pi/2.
\end{lstlisting}
\textbf{Prompt Engineering:}
During prompt engineering, participants were asked to modify the prompt so that when a user posed non-math-related questions, the math tutor app's response would be \textit{``Sorry, I'm not sure about the answer to this question. Try a different question.''}

\section{Evaluation User Study Interviews}
\label{appendix:evaluation-interview}

\subsection{Interview 1 Questions}
\label{appendix:evaluation-interview-1}

\begin{itemize}[topsep=5pt, itemsep=0mm, parsep=2pt, leftmargin=14pt]
  \item How did the harm envisioning tool (\toolc{}, \litec{}, or \guidec{}) influence your strategy for envisioning harms?
  \item Is envisioning potential harms something you would typically do when writing prompts? If so, how do you typically do it?
  \item What are the most relevant or important harms you have identified? Why?
  \item What is the biggest challenge in envisioning harms?
  \item What other tools or other types of support would you like to have to help you envision harms when prototyping LLM-powered applications?
  \item What do you think you would do about those identified harms?
  \item Do you have any feedback on the harm envisioning tool (\toolc{}, \litec{}, or \guidec{})?
\end{itemize}

\subsection{Interview 2 Rating Forms and Questions}
\label{appendix:evaluation-interview-2}

\noindent{}\textcolor{bluegrayI}{\rule{\linewidth}{0.2pt}}
\textcolor{bluegrayVI}{[\textit{For participants in \conditionfg{} and \conditiongf{}}]}

\noindent{}\textbf{How would you rate \toolc{}?}

\noindent{}Participants could select \textit{strongly agree}, \textit{agree}, \textit{neutral}, \textit{disagree}, or \textit{strongly disagree}. The default selection is \textit{neutral}.

\begin{itemize}[topsep=3pt, itemsep=0mm, parsep=0pt, leftmargin=14pt]
  \item Help me envision harms.
  \item Easy to use.
  \item Enjoyable to use.
  \item I would use this tool in the future.
\end{itemize}

\noindent{}\textbf{How would you rate the helpfulness of \toolc{}'s different components?}

\noindent{}Participants could select \textit{strongly agree}, \textit{agree}, \textit{neutral}, \textit{disagree}, or \textit{strongly disagree}. The default selection is \textit{neutral}.

\begin{itemize}[topsep=3pt, itemsep=0mm, parsep=0pt, leftmargin=14pt]
  \item AI incidents.
  \item Use case sidebar.
  \item AI assistance in harm envisioning.
  \item Interactive tree visualization.
  \item Customizability (add, edit, delete content).
\end{itemize}

\noindent{}\textcolor{bluegrayI}{\rule{\linewidth}{0.2pt}}
\textcolor{bluegrayVI}{[\textit{For participants in \conditionlg{} and \conditiongl{}}]}

\noindent{}\textbf{How would you rate \litec{}?}

\noindent{}Participants could select \textit{strongly agree}, \textit{agree}, \textit{neutral}, \textit{disagree}, or \textit{strongly disagree}. The default selection is \textit{neutral}.

\begin{itemize}[topsep=3pt, itemsep=0mm, parsep=0pt, leftmargin=14pt]
  \item Help me envision harms.
  \item Easy to use.
  \item Enjoyable to use.
  \item I would use this tool in the future.
\end{itemize}

\noindent{}\textbf{How would you rate the helpfulness of \litec{}'s different components?}

\noindent{}Participants could select \textit{strongly agree}, \textit{agree}, \textit{neutral}, \textit{disagree}, or \textit{strongly disagree}. The default selection is \textit{neutral}.

\begin{itemize}[topsep=3pt, itemsep=0mm, parsep=0pt, leftmargin=14pt]
  \item AI incidents.
  \item Use case sidebar.
  \item AI assistance in harm envisioning.
\end{itemize}

\noindent{}\textcolor{bluegrayI}{\rule{\linewidth}{0.2pt}}
\textcolor{bluegrayVI}{[\textit{For participants in \conditionfg{}, \conditiongf{}, \conditionlg{}, and \conditiongl{}}]}

\noindent{}\textbf{How would you rate \guidec{}?}

\noindent{}Participants could select \textit{strongly agree}, \textit{agree}, \textit{neutral}, \textit{disagree}, or \textit{strongly disagree}. The default selection is \textit{neutral}.

\begin{itemize}[topsep=3pt, itemsep=0mm, parsep=0pt, leftmargin=14pt]
  \item Help me envision harms.
  \item Easy to use.
  \item Enjoyable to use.
  \item I would use this tool in the future.
\end{itemize}

\noindent{}\textbf{How would you rate the helpfulness of \guidec{}'s different components?}

\noindent{}Participants could select \textit{strongly agree}, \textit{agree}, \textit{neutral}, \textit{disagree}, or \textit{strongly disagree}. The default selection is \textit{neutral}.

\begin{itemize}[topsep=3pt, itemsep=0mm, parsep=0pt, leftmargin=14pt]
  \item Harm modeling workflow table.
  \item Text prompts to think about use cases, harms, and stakeholders.
  \item Harm taxonomy.
\end{itemize}

\noindent{}\textcolor{bluegrayI}{\rule{\linewidth}{0.2pt}}
\textcolor{bluegrayVI}{[\textit{For participants in \conditionfg{}, \conditiongf{}, \conditionlg{}, and \conditiongl{}}]}

\noindent{}\textbf{Overall, which tool do you prefer? Why?}

\noindent{}Participants in \conditionfg{}, \conditiongf{} can choose \toolc{} or \guidec{}.
Participants in \conditionlg{}, and \conditiongl{} can choose \litec{} or \guidec{}.

\clearpage{}

\section{Harm Rating Processing and Inter-rater Reliability}
\label{appendix:evaluation-ratings}

\setlength{\belowcaptionskip}{-5pt}
\setlength{\abovecaptionskip}{7pt}
\begin{figure*}[tb]
  \includegraphics[width=1\linewidth]{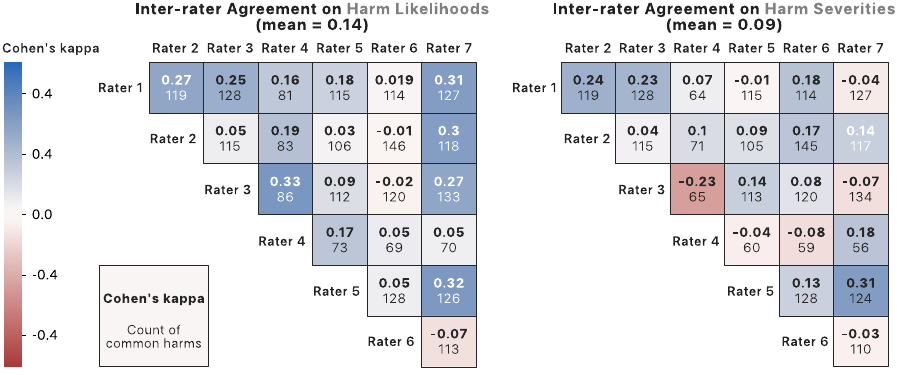}
  \caption[]{
    We compute weighted pairwise Cohen's kappa to measure inter-rater reliability for the ratings of harm likelihood (left) and harm severity (right).
    Raters rate each dimension on a 4-point Likert scale (\textit{strongly agree}, \textit{agree}, \textit{disagree}, and \textit{strongly disagree}).
    We numericalized these four categories as ordinal data: \texttt{1}, \texttt{2}, \texttt{3}, \texttt{4}.
    Within each cell, the top number is the kappa score, and the bottom number is the count of common harms between the corresponding two raters.
    The average kappas for likelihood and severity ratings are \texttt{0.11} and \texttt{0.10}, which can be interpreted as slight agreement.
  }
  \Description{
    This image contains two matrices titled "Inter-rater Agreement on Harm Likelihoods" and "Inter-rater Agreement on Harm Severities." Each matrix displays the Cohen's kappa statistic between pairs of raters, which measures the agreement between them beyond chance. The matrices are color-coded: blue indicates positive agreement, red indicates negative agreement, and white represents neutral.
  }
  \label{fig:appendix-rater}
\end{figure*}
\setlength{\belowcaptionskip}{0pt}
\setlength{\abovecaptionskip}{12pt}

In our evaluation user study, we recruited seven raters to rate the likelihood and severity of each collected harm~(\autoref{sec:evaluation-design-rating}).
In total, we collected 989 harms with 895 unique harms~(\autoref{table:participant-harm}).
We randomly assign each unique harm to three raters to rate its \textit{likelihood} and \textit{severity} on a 4-point Likert scale (\textit{strongly agree}, \textit{agree}, \textit{disagree}, and \textit{strongly disagree} to statements ``this harm is likely to happen'' and ``this harm is severe'').
Raters could also choose an N/A option if they perceived a rating was not applicable.
After collecting all ratings, we dropped all N/As and numericalized four rating categories as ordinal scores: \texttt{1}, \texttt{2}, \texttt{3}, \texttt{4}.

To measure the inter-rater reliability, we computed Cohen's kappa~\cite{mchughInterraterReliabilityKappa2012} for each pair of raters.
As the rating scores are ordinal, we used quadratically weighed kappa~\cite{cohenWeightedKappaNominal1968}, so that the level of agreement between score \texttt{1} and score \texttt{2} is higher than score \texttt{1} and score \texttt{3}.
The pair-wise kappas and the count of common harms are shown in~\autoref{fig:appendix-rater}.
The average kappa for likelihood ratings is \texttt{0.14}, and the average kappa for severity ratings is \texttt{0.09}.
Both scores can be interpreted as ``slight agreement''~\cite{landisMeasurementObserverAgreement1977}.

\clearpage{}
\subsection{Example Harms Collected from Participants}

\setlength{\availablewidth}{\linewidth - 10\tabcolsep}

\newlength{\featurewidth}
\setlength{\featurewidth}{0.1\availablewidth}

\newlength{\whowidth}
\setlength{\whowidth}{0.22\availablewidth}

\newlength{\howwidth}
\setlength{\howwidth}{0.5\availablewidth}

\newlength{\likelihoodwidth}
\setlength{\likelihoodwidth}{0.1\availablewidth}

\newlength{\severitywidth}
\setlength{\severitywidth}{0.08\availablewidth}

\newcolumntype{T}[1]{>{\raggedright\arraybackslash}p{#1}}

\aptLtoX[graphic=no,type=html]{\begin{longtable}{p{6pt}p{10pt}p{10pt}cc}
\caption{A random subset~($n$=84) of 895 unique harms collected from 42 participants in our evaluation user study~(\S~\ref{sec:evaluation}).
This random subset includes 14 harms for each of the four AI features~(four prompts used in H1--H4).
Depending on the experimental conditions, each harm was envisioned by a participant~(H1--H4) or generated by \tool{} and curated by a participant~(H2, H4).
Each harm was rated by three random raters out of seven raters in terms of likelihood and severity on a 4-point Likert scale (\textit{strongly agree}, \textit{agree}, \textit{disagree}, and \textit{strongly disagree} to statements ``this harm is likely to happen'' and ``this harm is severe'').}
\label{table:participant-harm}
\\
\textbf{AI Feature}                                             & \textbf{Who (could be harmed)}                                             & \textbf{How (the harm could happen)}                                                                                                                                                                                                             & \textbf{Likelihood} & \textbf{Severity} \\
\hline
\multirow[t]{14}{\featurewidth}{Email summarizer (H1)}     & AI company                                                                 & AI company’s reputation may be harmed due to inaccurate summary                                                                                                                                                                                  & 3, 3, 3             & 2, 2, 3           \\ \cline{2-5}
& Author                                                                     & The AI summary might lose context and information, critical context. It could cut out some crucial part of the email.                                                                                                                            & 4, 3, 3             & 3, 3, 3           \\ \cline{2-5}
& Owner of the email                                                         & Rewrite could leak data from other users’ email.                                                                                                                                                                                                 & 3, 1, 2             & 4, 4, 2           \\ \cline{2-5}
& People whose names / personal information is lost in the summarization     & Potential data loss. The email could be very long, and the rewrite could miss some context (e.g., full name). The rewrite can cause confusion to the receivers.                                                                                  & 3, 3, 3             & 2, 3, 3           \\ \cline{2-5}
& Product team, company                                                      & Prompt injection attack.                                                                                                                                                                                                                         & 3, 1, 3             & 2, 1, 3           \\ \cline{2-5}
& Receiver                                                                   & Omission of vital information and context. The feature can miss some very important information.                                                                                                                                                 & 3, 3, 3             & 3, 2, 3           \\ \cline{2-5}
& Receiver                                                                   & The initial rewrite is a bit open ended, which might cause anxiety.                                                                                                                                                                              & 2, 3, 2             & 1, 3, 2           \\ \cline{2-5}
& Recipients                                                                 & If users can control the temperature of the LLM, they can generate more inaccurate emails.                                                                                                                                                       & 1, 3, 1             & 4, 2, 1           \\ \cline{2-5}
& Recipients due to loss of productivity and insight into their orgs metrics & The email may miss some metrics from the email.                                                                                                                                                                                                  & 3, 3, 3             & 2, 2, 3           \\ \cline{2-5}
& Scam victims                                                               & Scam emails are very common. Mass emailing a lot of people with easy automation can cause harm to victims.                                                                                                                                       & 2, 3, 3             & 4, 3, 3           \\ \cline{2-5}
& Sender                                                                     & In a professional setting, if the feature is not sufficient to capture the keywords, the summary can sound unprofessional. It can harm the reputation of the sender.                                                                             & 2, 3, 3             & 2, 2, 2           \\ \cline{2-5}
& Sender                                                                     & The summary is inaccurate, e.g., having the wrong time, names in the rewrite. It can harm the sender’s reputation.                                                                                                                               & 3, 3, 3             & 2, 3, 3           \\ \cline{2-5}
& Sender, Recipient                                                          & The feature introduces incorrect information that is not in the original email.                                                                                                                                                                  & 3, 3, 3             & 3, 2, 2           \\ \cline{2-5}
& Sender, recipient                                                          & Quality of the communication. If the tone of the rewrite is different from the original email, it can cause harm to both sender and recipient. Miscommunication.                                                                                 & 3, 2, 3             & 2, 2, 3           \\ \cline{2-5}
& Sender, recipient                                                          & The delicacy in the original email may be removed in the shorter email. The tone, emotion, and human elements may be changed.                                                                                                                    & 4, 3, 3             & 3, 2, 3           \\ \cline{2-5}
& Sender, recipient                                                          & The summary can have a different voice / style from the email (more directly, less soft). It can be perceived as negative.                                                                                                                       & 4, 3, 4             & 3, 2, 4           \\ \cline{2-5}
& Sender, recipient, company relationships                                   & AI can be culturally blind. It might not follow some typical email norm in the rewrite.                                                                                                                                                          & 3, 3, 3             & 2, 2, 3           \\ \cline{2-5}
& Senders, receivers                                                         & The AI feature puts words on other people’s behalf. It can have different viewpoints from the email. It can have misrepresented views.                                                                                                           & 2, 3, 4             & 3, 3, 4           \\ \cline{2-5}
& Senders, recipient                                                         & The summary can lose some tone, voice of the email, becoming less personal.                                                                                                                                                                      & 3, 3, 4             & 2, 3, 3           \\ \cline{2-5}
& Stakeholders of the decision                                               & If the communication is about decision making and the rewrite contains error, it can harm the stakeholders of the decision.                                                                                                                      & 1, 4, 3             & 1, 4, 3           \\ \cline{2-5}
& Team building the feature                                                  & Damage to reputation, credibility to have a feature fail this way.                                                                                                                                                                               & 3, 2, 3             & 3, 1, 2           \\ \hline
\multirow[t]{14}{\featurewidth}{Toxicity classifier (H2)}
& Advertisers                                                                & If the AI feature fails, ads might be present next to toxic content, losing revenues.                                                                                                                                                            & 3, 3, 3             & 1, 2, 3           \\ \cline{2-5}
& Bullies                                                                    & Teachers use it to identify students being bullied. The AI may make the teachers be biased against the bullies.                                                                                                                                  & 2, 3, 2             & 3, 3, 2           \\ \cline{2-5}
& Company of the AI product                                                  & The company may lose users due to the presence of more moderation.                                                                                                                                                                               & 3, 2, 1             & 4, 2, 1           \\ \cline{2-5}
& Customer                                                                   & Customer service agents use it to identify abusive customers. Customers may lose the opportunity to get help from customer service agents.                                                                                                       & 3, 3, 3             & 3, 2, 4           \\ \cline{2-5}
& Customer service agents                                                    & If customer service uses it to screen people, false positives: exposing customer service agents to toxic customers.                                                                                                                              & 3, 1, 3             & 3, 1, 3           \\ \cline{2-5}
& Employees of the company                                                   & HR departments use it to screen job applicants for toxic behavior. Employees may be unfairly rejected for jobs due to AI-generated toxicity labels.                                                                                              & 1, 3, 3             & 1, 3, 4           \\ \cline{2-5}
& End user (children \& parents)                                             & Mislabeled outputs will not be blocked by parental controls.                                                                                                                                                                                     & 3, 3, 3             & 3, 4, 4           \\ \cline{2-5}
& Government                                                                 & Military uses it to identify potential insurgents. Governments may use AI to target and surveil innocent people, leading to human rights violations.                                                                                             & 3, 2, 3             & 4, 3, 3           \\ \cline{2-5}
& Hate group targets                                                         & Hate groups use it to identify potential recruits. Hate groups may be able to recruit more people due to the AI tool.                                                                                                                            & 3, 2, 3             & 3, 3, 3           \\ \cline{2-5}
& Job applicants                                                             & Reddit has a karma score. Similarly, if a social media uses this feature to prioritize non-toxic content, misclassification can lose job opportunities for social media users (e..g, tweet not seen by companies).                               & 3, 4, 1             & 3, 3, 1           \\ \cline{2-5}
& Online moderators                                                          & Online moderators may lose their jobs because the AI tool can already distinguish toxic vs non-toxic so there is no need for a moderator.                                                                                                        & 2, 1, 3             & 3, 1, 4           \\ \cline{2-5}
& Online moderators                                                          & Online moderators may miss some toxic content and post it on social media.                                                                                                                                                                       & 3, 3, 3             & 3, 4, 3           \\ \cline{2-5}
& People being targeted by law enforcement                                   & Law enforcement uses it to identify potential terrorists. Potential terrorists ([different name]) may be unfairly targeted by law enforcement due to AI-generated misclassifications.                                                            & 3, 1, 3             & 3, 1, 4           \\ \cline{2-5}
& People who are not members of hate groups                                  & Hate groups use it to identify potential recruits. People who are not members of hate groups may be affected by increased hate group activity.                                                                                                   & 3, 3, 4             & 3, 3, 4           \\ \cline{2-5}
& Sender, Receivers                                                          & False positive. Flagging non-toxic content as toxic, causing harm to legitimate receivers of the information (filtered). Sender cannot post due to the filter.                                                                                   & 3, 3, 4             & 3, 3, 4           \\ \cline{2-5}
& Social media users                                                         & Social media users may be exposed to toxic content that was mislabeled by the AI as non-toxic.                                                                                                                                                   & 3, 3, 3             & 3, 4, 3           \\ \cline{2-5}
& Social media users                                                         & Social media users may feel frustrated or anxious about the accuracy of the AI tool.                                                                                                                                                             & 3, 3, 3             & 2, 4, 2           \\ \cline{2-5}
& Social media users                                                         & Social media users who do not use this AI might be exposed to some distorted message.                                                                                                                                                            & 2, 1, 3             & 3, 1, 3           \\ \cline{2-5}
& Social media users                                                         & Unfairly flagged as toxic (false positive trigger).                                                                                                                                                                                              & 3, 3, 3             & 3, 2, 2           \\ \cline{2-5}
& Student                                                                    & Teachers use it to help students identify toxic language. Students may feel frustrated or anxious about the accuracy of the AI tool.                                                                                                             & 2, 1, 3             & 2, 1, 4           \\ \cline{2-5}
& Victims                                                                    & Scammers use it to identify potential victims. Victims of scams may lose money or personal information.                                                                                                                                          & 2, 2, 1             & 4, 3, 1           \\ \hline
\multirow[t]{14}{\featurewidth}{Article Sumamrizer (H3)} & Anyone reading the document                                                & If the document is sent to anyone else, the summary may not well-represent the document. The readers of the summary might mis-understand the original document.                                                                                  & 4, 3, 3             & 3, 2, 3           \\ \cline{2-5}
& Business                                                                   & Write marketing content, memo. Bad summary can lead to miscommunication, causing financial loss.                                                                                                                                                 & 4, 3, 3             & 3, 3, 4           \\ \cline{2-5}
& Company publishing the content (news paper)                                & If the summary is wrong, people would lose confidence in the company.                                                                                                                                                                            & 3, 4, 4             & 2, 4, 3           \\ \cline{2-5}
& Economic loss                                                              & The summary may contain misinformation about some critical article, causing economic loss.                                                                                                                                                       & 3, 3, 1             & 3, 3, 1           \\ \cline{2-5}
& Employees                                                                  & Employees can use it to summarize documents in a workplace. Wrong summary can harm the stakeholders.                                                                                                                                             & 3, 3, 3             & 4, 3, 3           \\ \cline{2-5}
& End user                                                                   & Change the meaning of the article.                                                                                                                                                                                                               & 3, 4, 4             & 3, 4, 4           \\ \cline{2-5}
& Journalists / writers                                                      & If the summary is wrong, journalists’ reputation might be harmed.                                                                                                                                                                                & 1, 4, 3             & 1, 3, 4           \\ \cline{2-5}
& Kids                                                                       & Kids use it cheat in school for assignments.                                                                                                                                                                                                     & 3, 2, 3             & 2, 2, 3           \\ \cline{2-5}
& Readers                                                                    & AI fails to pick up key information (nuances) in the article => the readers miss the points from the article, miscommunication.                                                                                                                  & 3, 3, 3             & 2, 2, 2           \\ \cline{2-5}
& Readers, content providers, new                                            & If there are two facts in the article that are true independently, then it’s possible that the summary combines them where the statement is no longer true.                                                                                      & 3, 2, 3             & 2, 2, 3           \\ \cline{2-5}
& Readers, society                                                           & A lot of content is missed, making readers less educated overtime, causing societal loss (some events being lost in history).                                                                                                                    & 3, 2, 1             & 3, 3, 1           \\ \cline{2-5}
& Readers, writers                                                           & It would lose a lot of context and details.                                                                                                                                                                                                      & 3, 3, 3             & 2, 3, 2           \\ \cline{2-5}
& Social groups                                                              & Representational harm. It could stereotype some social groups in the output.                                                                                                                                                                     & 3, 3, 3             & 3, 3, 3           \\ \cline{2-5}
& Students                                                                   & Students can use this feature to summarize articles instead of reading for assignments.                                                                                                                                                          & 4, 3, 3             & 3, 3, 3           \\ \cline{2-5}
& Students                                                                   & Students can use this tool to cheat and lose opportunities to learn.                                                                                                                                                                             & 4, 1, 3             & 3, 1, 4           \\ \cline{2-5}
& Students                                                                   & Students misuse this feature to cheat on school assignments. They might lose learning opportunities.                                                                                                                                             & 4, 2, 3             & 3, 2, 3           \\ \cline{2-5}
& Users                                                                      & The context can be manipulated.                                                                                                                                                                                                                  & 2, 4, 3             & 2, 4, 3           \\ \cline{2-5}
& Users                                                                      & The summary can lose some information and context from the article.                                                                                                                                                                              & 3, 4, 3             & 3, 4, 2           \\ \cline{2-5}
& Users                                                                      & The tool makes mistakes (e.g., wrong summary). It can harm the users of the tool by missing information in the article.                                                                                                                          & 3, 3, 3             & 3, 3, 3           \\ \cline{2-5}
& Users                                                                      & Users may lose trust if the summary is not right. Users would become more stressful as well.                                                                                                                                                     & 3, 3, 4             & 2, 2, 3           \\ \cline{2-5}
& Writers                                                                    & If someone uses the tool to write, it may change people’s perception about how you write.                                                                                                                                                        & 3, 1, 1             & 2, 1, 2           \\ \hline
\multirow[t]{14}{\featurewidth}{Math tutor (H4)}          & Company of the AI product                                                  & Engineers use it to explain to non-technical stakeholders. The company may face increased legal liability due to AI-generated explanations being inaccurate or misleading.                                                                       & 2, 1, 2             & 1, 1, 2           \\ \cline{2-5}
& Editors                                                                    & Journal readers use it to get explanations of equations in inferential statistics sections. Journal editors may worry that the quality of their journal declines because the AI feature makes too many errors. Readers get frustrated over time. & 2, 1, 2             & 2, 1, 3           \\ \cline{2-5}
& Marginalized population                                                    & People with less resources are less likely to access this tool, losing opportunity to learn.                                                                                                                                                     & 4, 3, 3             & 3, 3, 3           \\ \cline{2-5}
& Minority social groups                                                     & If you ask the math tool about important social groups, it can refuse to answer => ignoring important questions, marginalizing some social groups.                                                                                               & 3, 3, 1             & 3, 3, 1           \\ \cline{2-5}
& Minority social groups                                                     & There are different ways to phrase things differently based on some social groups. If the user asks a question in non-profession English or non-English, it can cause alienation.                                                                & 4, 3, 3             & 3, 3, 3           \\ \cline{2-5}
& Non-technical stakeholder                                                  & Engineers use it to explain complex math concepts to non-technical stakeholders. Non-technical stakeholders may be misled by AI-generated explanations of complex mathematical models.                                                           & 3, 2, 3             & 3, 2, 3           \\ \cline{2-5}
& Parents of students                                                        & Students use it to learn math. Parents may feel frustrated or anxious about their child's education.                                                                                                                                             & 2, 2, 3             & 2, 2, 3           \\ \cline{2-5}
& Public                                                                     & The public may be misled by AI-generated explanations of scientific concepts.                                                                                                                                                                    & 3, 3, 3             & 3, 3, 3           \\ \cline{2-5}
& Rocket engineers                                                           & Wrong answers can harm the task.                                                                                                                                                                                                                 & 3, 3, 3             & 3, 2, 3           \\ \cline{2-5}
& Student                                                                    & Tutors use it to help students understand math concepts. Students may lose the opportunity to learn the material in a way that is tailored to their individual needs.                                                                            & 4, 2, 4             & 4, 2, 4           \\ \cline{2-5}
& Students                                                                   & If the students use this tool to cheat, they would have economic loss (not performing well on their jobs).                                                                                                                                       & 2, 1, 3             & 2, 1, 3           \\ \cline{2-5}
& Students                                                                   & Incorrect math answers harm the user’s learning process.                                                                                                                                                                                         & 3, 3, 3             & 2, 3, 3           \\ \cline{2-5}
& Students                                                                   & Students use it to cheat, losing opportunities to learn.                                                                                                                                                                                         & 4, 1, 3             & 3, 1, 3           \\ \cline{2-5}
& Students                                                                   & Students use it to learn math. Students may feel like they are losing their ability to learn about math problems.                                                                                                                                & 2, 1, 2             & 2, 1, 2           \\ \cline{2-5}
& Students                                                                   & Students use it to learn math. Students who do not use this AI product may feel like they are not getting the same quality of education as their peers.                                                                                          & 3, 3, 3             & 2, 3, 3           \\ \cline{2-5}
& Students                                                                   & Wrong solution, less optimized solution, wrong information, students learn the wrong thing.                                                                                                                                                      & 3, 3, 3             & 3, 2, 3           \\ \cline{2-5}
& Students and teachers                                                      & Students can use this app to cheat. Students do not learn.                                                                                                                                                                                       & 4, 3, 3             & 3, 2, 4           \\ \cline{2-5}
& Students who do not use this AI product                                    & Students use it to learn math. Students who do not use this AI product may feel left behind by their peers who do.                                                                                                                               & 3, 3, 4             & 3, 2, 4           \\ \cline{2-5}
& Students, society                                                          & If the answer is based on some context of the math problem, it can harm minority students.                                                                                                                                                       & 2, 3, 1             & 3, 3, 1           \\ \cline{2-5}
& Users                                                                      & If you ask tax rates / vote count / interest rate / wages, the wrong answer can cause political harm and social harm.                                                                                                                            & 3, 2, 3             & 3, 2, 3           \\ \cline{2-5}
& Users                                                                      & Not all questions are about numbers. Sometimes the app might refuse to answer hard math questions, and uses would feel upset.                                                                                                                    & 3, 3, 3             & 2, 2, 3           \\
\arrayrulecolor{grayIX}
\specialrule{1.2pt}{0pt}{0pt}
\end{longtable}}{{\begin{spacing}{1.25}
\small
\begin{longtable}[\linewidth]{T{\featurewidth}!{\color{grayIII}\vrule}T{\whowidth}!{\color{grayIII}\vrule}T{\howwidth}!{\color{grayIII}\vrule}T{\likelihoodwidth}!{\color{grayIII}\vrule}T{\severitywidth}}
\captionsetup{type=table}
\caption[]{A random subset~($n$=84) of 895 unique harms collected from 42 participants in our evaluation user study~(\autoref{sec:evaluation}).
This random subset includes 14 harms for each of the four AI features~(four prompts used in \hone{}--\hfour{}).
Depending on the experimental conditions, each harm was envisioned by a participant~(\hone{}--\hfour{}) or generated by \tool{} and curated by a participant~(\htwo{}, \hfour{}).
Each harm was rated by three random raters out of seven raters in terms of likelihood and severity on a 4-point Likert scale (\textit{strongly agree}, \textit{agree}, \textit{disagree}, and \textit{strongly disagree} to statements ``this harm is likely to happen'' and ``this harm is severe'').}
\Description{This table shows the harms collected from participants. The columns are "AI Feature", "Who (could be harmed)", "How (the harm could happend)", "Likelihood", and "Severity."}
\label{table:participant-harm}
\\
\specialrule{1.2pt}{0pt}{0pt}
\textbf{AI Feature}                                             & \textbf{Who (could be harmed)}                                             & \textbf{How (the harm could happen)}                                                                                                                                                                                                             & \textbf{Likelihood} & \textbf{Severity} \\
\arrayrulecolor{grayIII}
\hline
\endfirsthead
\arrayrulecolor{grayIX}
\specialrule{1.2pt}{1pt}{0pt}
\textbf{AI Feature}   & \textbf{Who (could be harmed)}                                             & \textbf{How (the harm could happen)}                                                                                                                                                                                                             & \textbf{Likelihood} & \textbf{Severity} \\
\arrayrulecolor{grayIII}
\hline
\endhead
\multirow[t]{14}{\featurewidth}{Email summarizer (\hone{})}     & AI company                                                                 & AI company’s reputation may be harmed due to inaccurate summary                                                                                                                                                                                  & 3, 3, 3             & 2, 2, 3           \\ \cline{2-5}
& Author                                                                     & The AI summary might lose context and information, critical context. It could cut out some crucial part of the email.                                                                                                                            & 4, 3, 3             & 3, 3, 3           \\ \cline{2-5}
& Owner of the email                                                         & Rewrite could leak data from other users’ email.                                                                                                                                                                                                 & 3, 1, 2             & 4, 4, 2           \\ \cline{2-5}
& People whose names / personal information is lost in the summarization     & Potential data loss. The email could be very long, and the rewrite could miss some context (e.g., full name). The rewrite can cause confusion to the receivers.                                                                                  & 3, 3, 3             & 2, 3, 3           \\ \cline{2-5}
& Product team, company                                                      & Prompt injection attack.                                                                                                                                                                                                                         & 3, 1, 3             & 2, 1, 3           \\ \cline{2-5}
& Receiver                                                                   & Omission of vital information and context. The feature can miss some very important information.                                                                                                                                                 & 3, 3, 3             & 3, 2, 3           \\ \cline{2-5}
& Receiver                                                                   & The initial rewrite is a bit open ended, which might cause anxiety.                                                                                                                                                                              & 2, 3, 2             & 1, 3, 2           \\ \cline{2-5}
& Recipients                                                                 & If users can control the temperature of the LLM, they can generate more inaccurate emails.                                                                                                                                                       & 1, 3, 1             & 4, 2, 1           \\ \cline{2-5}
& Recipients due to loss of productivity and insight into their orgs metrics & The email may miss some metrics from the email.                                                                                                                                                                                                  & 3, 3, 3             & 2, 2, 3           \\ \cline{2-5}
& Scam victims                                                               & Scam emails are very common. Mass emailing a lot of people with easy automation can cause harm to victims.                                                                                                                                       & 2, 3, 3             & 4, 3, 3           \\ \cline{2-5}
& Sender                                                                     & In a professional setting, if the feature is not sufficient to capture the keywords, the summary can sound unprofessional. It can harm the reputation of the sender.                                                                             & 2, 3, 3             & 2, 2, 2           \\ \cline{2-5}
& Sender                                                                     & The summary is inaccurate, e.g., having the wrong time, names in the rewrite. It can harm the sender’s reputation.                                                                                                                               & 3, 3, 3             & 2, 3, 3           \\ \cline{2-5}
& Sender, Recipient                                                          & The feature introduces incorrect information that is not in the original email.                                                                                                                                                                  & 3, 3, 3             & 3, 2, 2           \\ \cline{2-5}
& Sender, recipient                                                          & Quality of the communication. If the tone of the rewrite is different from the original email, it can cause harm to both sender and recipient. Miscommunication.                                                                                 & 3, 2, 3             & 2, 2, 3           \\ \cline{2-5}
& Sender, recipient                                                          & The delicacy in the original email may be removed in the shorter email. The tone, emotion, and human elements may be changed.                                                                                                                    & 4, 3, 3             & 3, 2, 3           \\ \cline{2-5}
& Sender, recipient                                                          & The summary can have a different voice / style from the email (more directly, less soft). It can be perceived as negative.                                                                                                                       & 4, 3, 4             & 3, 2, 4           \\ \cline{2-5}
& Sender, recipient, company relationships                                   & AI can be culturally blind. It might not follow some typical email norm in the rewrite.                                                                                                                                                          & 3, 3, 3             & 2, 2, 3           \\ \cline{2-5}
& Senders, receivers                                                         & The AI feature puts words on other people’s behalf. It can have different viewpoints from the email. It can have misrepresented views.                                                                                                           & 2, 3, 4             & 3, 3, 4           \\ \cline{2-5}
& Senders, recipient                                                         & The summary can lose some tone, voice of the email, becoming less personal.                                                                                                                                                                      & 3, 3, 4             & 2, 3, 3           \\ \cline{2-5}
& Stakeholders of the decision                                               & If the communication is about decision making and the rewrite contains error, it can harm the stakeholders of the decision.                                                                                                                      & 1, 4, 3             & 1, 4, 3           \\ \cline{2-5}
& Team building the feature                                                  & Damage to reputation, credibility to have a feature fail this way.                                                                                                                                                                               & 3, 2, 3             & 3, 1, 2           \\ \hline
\multirow[t]{14}{\featurewidth}{Toxicity classifier (\htwo{})}
& Advertisers                                                                & If the AI feature fails, ads might be present next to toxic content, losing revenues.                                                                                                                                                            & 3, 3, 3             & 1, 2, 3           \\ \cline{2-5}
& Bullies                                                                    & Teachers use it to identify students being bullied. The AI may make the teachers be biased against the bullies.                                                                                                                                  & 2, 3, 2             & 3, 3, 2           \\ \cline{2-5}
& Company of the AI product                                                  & The company may lose users due to the presence of more moderation.                                                                                                                                                                               & 3, 2, 1             & 4, 2, 1           \\ \cline{2-5}
& Customer                                                                   & Customer service agents use it to identify abusive customers. Customers may lose the opportunity to get help from customer service agents.                                                                                                       & 3, 3, 3             & 3, 2, 4           \\ \cline{2-5}
& Customer service agents                                                    & If customer service uses it to screen people, false positives: exposing customer service agents to toxic customers.                                                                                                                              & 3, 1, 3             & 3, 1, 3           \\ \cline{2-5}
& Employees of the company                                                   & HR departments use it to screen job applicants for toxic behavior. Employees may be unfairly rejected for jobs due to AI-generated toxicity labels.                                                                                              & 1, 3, 3             & 1, 3, 4           \\ \cline{2-5}
& End user (children \& parents)                                             & Mislabeled outputs will not be blocked by parental controls.                                                                                                                                                                                     & 3, 3, 3             & 3, 4, 4           \\ \cline{2-5}
& Government                                                                 & Military uses it to identify potential insurgents. Governments may use AI to target and surveil innocent people, leading to human rights violations.                                                                                             & 3, 2, 3             & 4, 3, 3           \\ \cline{2-5}
& Hate group targets                                                         & Hate groups use it to identify potential recruits. Hate groups may be able to recruit more people due to the AI tool.                                                                                                                            & 3, 2, 3             & 3, 3, 3           \\ \cline{2-5}
& Job applicants                                                             & Reddit has a karma score. Similarly, if a social media uses this feature to prioritize non-toxic content, misclassification can lose job opportunities for social media users (e..g, tweet not seen by companies).                               & 3, 4, 1             & 3, 3, 1           \\ \cline{2-5}
& Online moderators                                                          & Online moderators may lose their jobs because the AI tool can already distinguish toxic vs non-toxic so there is no need for a moderator.                                                                                                        & 2, 1, 3             & 3, 1, 4           \\ \cline{2-5}
& Online moderators                                                          & Online moderators may miss some toxic content and post it on social media.                                                                                                                                                                       & 3, 3, 3             & 3, 4, 3           \\ \cline{2-5}
& People being targeted by law enforcement                                   & Law enforcement uses it to identify potential terrorists. Potential terrorists ([different name]) may be unfairly targeted by law enforcement due to AI-generated misclassifications.                                                            & 3, 1, 3             & 3, 1, 4           \\ \cline{2-5}
& People who are not members of hate groups                                  & Hate groups use it to identify potential recruits. People who are not members of hate groups may be affected by increased hate group activity.                                                                                                   & 3, 3, 4             & 3, 3, 4           \\ \cline{2-5}
& Sender, Receivers                                                          & False positive. Flagging non-toxic content as toxic, causing harm to legitimate receivers of the information (filtered). Sender cannot post due to the filter.                                                                                   & 3, 3, 4             & 3, 3, 4           \\ \cline{2-5}
& Social media users                                                         & Social media users may be exposed to toxic content that was mislabeled by the AI as non-toxic.                                                                                                                                                   & 3, 3, 3             & 3, 4, 3           \\ \cline{2-5}
& Social media users                                                         & Social media users may feel frustrated or anxious about the accuracy of the AI tool.                                                                                                                                                             & 3, 3, 3             & 2, 4, 2           \\ \cline{2-5}
& Social media users                                                         & Social media users who do not use this AI might be exposed to some distorted message.                                                                                                                                                            & 2, 1, 3             & 3, 1, 3           \\ \cline{2-5}
& Social media users                                                         & Unfairly flagged as toxic (false positive trigger).                                                                                                                                                                                              & 3, 3, 3             & 3, 2, 2           \\ \cline{2-5}
& Student                                                                    & Teachers use it to help students identify toxic language. Students may feel frustrated or anxious about the accuracy of the AI tool.                                                                                                             & 2, 1, 3             & 2, 1, 4           \\ \cline{2-5}
& Victims                                                                    & Scammers use it to identify potential victims. Victims of scams may lose money or personal information.                                                                                                                                          & 2, 2, 1             & 4, 3, 1           \\ \hline
\multirow[t]{14}{\featurewidth}{Article Sumamrizer (\hthree{})} & Anyone reading the document                                                & If the document is sent to anyone else, the summary may not well-represent the document. The readers of the summary might mis-understand the original document.                                                                                  & 4, 3, 3             & 3, 2, 3           \\ \cline{2-5}
& Business                                                                   & Write marketing content, memo. Bad summary can lead to miscommunication, causing financial loss.                                                                                                                                                 & 4, 3, 3             & 3, 3, 4           \\ \cline{2-5}
& Company publishing the content (news paper)                                & If the summary is wrong, people would lose confidence in the company.                                                                                                                                                                            & 3, 4, 4             & 2, 4, 3           \\ \cline{2-5}
& Economic loss                                                              & The summary may contain misinformation about some critical article, causing economic loss.                                                                                                                                                       & 3, 3, 1             & 3, 3, 1           \\ \cline{2-5}
& Employees                                                                  & Employees can use it to summarize documents in a workplace. Wrong summary can harm the stakeholders.                                                                                                                                             & 3, 3, 3             & 4, 3, 3           \\ \cline{2-5}
& End user                                                                   & Change the meaning of the article.                                                                                                                                                                                                               & 3, 4, 4             & 3, 4, 4           \\ \cline{2-5}
& Journalists / writers                                                      & If the summary is wrong, journalists’ reputation might be harmed.                                                                                                                                                                                & 1, 4, 3             & 1, 3, 4           \\ \cline{2-5}
& Kids                                                                       & Kids use it cheat in school for assignments.                                                                                                                                                                                                     & 3, 2, 3             & 2, 2, 3           \\ \cline{2-5}
& Readers                                                                    & AI fails to pick up key information (nuances) in the article => the readers miss the points from the article, miscommunication.                                                                                                                  & 3, 3, 3             & 2, 2, 2           \\ \cline{2-5}
& Readers, content providers, new                                            & If there are two facts in the article that are true independently, then it’s possible that the summary combines them where the statement is no longer true.                                                                                      & 3, 2, 3             & 2, 2, 3           \\ \cline{2-5}
& Readers, society                                                           & A lot of content is missed, making readers less educated overtime, causing societal loss (some events being lost in history).                                                                                                                    & 3, 2, 1             & 3, 3, 1           \\ \cline{2-5}
& Readers, writers                                                           & It would lose a lot of context and details.                                                                                                                                                                                                      & 3, 3, 3             & 2, 3, 2           \\ \cline{2-5}
& Social groups                                                              & Representational harm. It could stereotype some social groups in the output.                                                                                                                                                                     & 3, 3, 3             & 3, 3, 3           \\ \cline{2-5}
& Students                                                                   & Students can use this feature to summarize articles instead of reading for assignments.                                                                                                                                                          & 4, 3, 3             & 3, 3, 3           \\ \cline{2-5}
& Students                                                                   & Students can use this tool to cheat and lose opportunities to learn.                                                                                                                                                                             & 4, 1, 3             & 3, 1, 4           \\ \cline{2-5}
& Students                                                                   & Students misuse this feature to cheat on school assignments. They might lose learning opportunities.                                                                                                                                             & 4, 2, 3             & 3, 2, 3           \\ \cline{2-5}
& Users                                                                      & The context can be manipulated.                                                                                                                                                                                                                  & 2, 4, 3             & 2, 4, 3           \\ \cline{2-5}
& Users                                                                      & The summary can lose some information and context from the article.                                                                                                                                                                              & 3, 4, 3             & 3, 4, 2           \\ \cline{2-5}
& Users                                                                      & The tool makes mistakes (e.g., wrong summary). It can harm the users of the tool by missing information in the article.                                                                                                                          & 3, 3, 3             & 3, 3, 3           \\ \cline{2-5}
& Users                                                                      & Users may lose trust if the summary is not right. Users would become more stressful as well.                                                                                                                                                     & 3, 3, 4             & 2, 2, 3           \\ \cline{2-5}
& Writers                                                                    & If someone uses the tool to write, it may change people’s perception about how you write.                                                                                                                                                        & 3, 1, 1             & 2, 1, 2           \\ \hline
\multirow[t]{14}{\featurewidth}{Math tutor (\hfour{})}          & Company of the AI product                                                  & Engineers use it to explain to non-technical stakeholders. The company may face increased legal liability due to AI-generated explanations being inaccurate or misleading.                                                                       & 2, 1, 2             & 1, 1, 2           \\ \cline{2-5}
& Editors                                                                    & Journal readers use it to get explanations of equations in inferential statistics sections. Journal editors may worry that the quality of their journal declines because the AI feature makes too many errors. Readers get frustrated over time. & 2, 1, 2             & 2, 1, 3           \\ \cline{2-5}
& Marginalized population                                                    & People with less resources are less likely to access this tool, losing opportunity to learn.                                                                                                                                                     & 4, 3, 3             & 3, 3, 3           \\ \cline{2-5}
& Minority social groups                                                     & If you ask the math tool about important social groups, it can refuse to answer => ignoring important questions, marginalizing some social groups.                                                                                               & 3, 3, 1             & 3, 3, 1           \\ \cline{2-5}
& Minority social groups                                                     & There are different ways to phrase things differently based on some social groups. If the user asks a question in non-profession English or non-English, it can cause alienation.                                                                & 4, 3, 3             & 3, 3, 3           \\ \cline{2-5}
& Non-technical stakeholder                                                  & Engineers use it to explain complex math concepts to non-technical stakeholders. Non-technical stakeholders may be misled by AI-generated explanations of complex mathematical models.                                                           & 3, 2, 3             & 3, 2, 3           \\ \cline{2-5}
& Parents of students                                                        & Students use it to learn math. Parents may feel frustrated or anxious about their child's education.                                                                                                                                             & 2, 2, 3             & 2, 2, 3           \\ \cline{2-5}
& Public                                                                     & The public may be misled by AI-generated explanations of scientific concepts.                                                                                                                                                                    & 3, 3, 3             & 3, 3, 3           \\ \cline{2-5}
& Rocket engineers                                                           & Wrong answers can harm the task.                                                                                                                                                                                                                 & 3, 3, 3             & 3, 2, 3           \\ \cline{2-5}
& Student                                                                    & Tutors use it to help students understand math concepts. Students may lose the opportunity to learn the material in a way that is tailored to their individual needs.                                                                            & 4, 2, 4             & 4, 2, 4           \\ \cline{2-5}
& Students                                                                   & If the students use this tool to cheat, they would have economic loss (not performing well on their jobs).                                                                                                                                       & 2, 1, 3             & 2, 1, 3           \\ \cline{2-5}
& Students                                                                   & Incorrect math answers harm the user’s learning process.                                                                                                                                                                                         & 3, 3, 3             & 2, 3, 3           \\ \cline{2-5}
& Students                                                                   & Students use it to cheat, losing opportunities to learn.                                                                                                                                                                                         & 4, 1, 3             & 3, 1, 3           \\ \cline{2-5}
& Students                                                                   & Students use it to learn math. Students may feel like they are losing their ability to learn about math problems.                                                                                                                                & 2, 1, 2             & 2, 1, 2           \\ \cline{2-5}
& Students                                                                   & Students use it to learn math. Students who do not use this AI product may feel like they are not getting the same quality of education as their peers.                                                                                          & 3, 3, 3             & 2, 3, 3           \\ \cline{2-5}
& Students                                                                   & Wrong solution, less optimized solution, wrong information, students learn the wrong thing.                                                                                                                                                      & 3, 3, 3             & 3, 2, 3           \\ \cline{2-5}
& Students and teachers                                                      & Students can use this app to cheat. Students do not learn.                                                                                                                                                                                       & 4, 3, 3             & 3, 2, 4           \\ \cline{2-5}
& Students who do not use this AI product                                    & Students use it to learn math. Students who do not use this AI product may feel left behind by their peers who do.                                                                                                                               & 3, 3, 4             & 3, 2, 4           \\ \cline{2-5}
& Students, society                                                          & If the answer is based on some context of the math problem, it can harm minority students.                                                                                                                                                       & 2, 3, 1             & 3, 3, 1           \\ \cline{2-5}
& Users                                                                      & If you ask tax rates / vote count / interest rate / wages, the wrong answer can cause political harm and social harm.                                                                                                                            & 3, 2, 3             & 3, 2, 3           \\ \cline{2-5}
& Users                                                                      & Not all questions are about numbers. Sometimes the app might refuse to answer hard math questions, and uses would feel upset.                                                                                                                    & 3, 3, 3             & 2, 2, 3           \\
\arrayrulecolor{grayIX}
\specialrule{1.2pt}{0pt}{0pt}
\end{longtable}\end{spacing}}}

\end{document}